\documentclass{aa}  

\usepackage{graphicx}
\usepackage{txfonts}
\usepackage{xcolor}
\usepackage{lipsum}
\usepackage[normalem]{ ulem }
\usepackage{soul}
\usepackage{amsmath}
\usepackage{upgreek}

\definecolor{mygray}{gray}{0.9}
\definecolor{bd}{HTML}{842121}
\newcommand{\aava}[1]{\textcolor{black}{#1}}
\newcommand{\Od}{\mathcal{R}}
\begin{document}

   \title{How do tidal waves interact with convective vortices\\ in rapidly-rotating planets and stars?}


   \author{V. Dandoy\inst{1,2}, J. Park\inst{3}, K. Augustson\inst{4,5}, A. Astoul\inst{6}, and S. Mathis\inst{1}
          }

   \institute{\inst{1} AIM, CEA, CNRS, Universit\'e Paris-Saclay, Universit\'e Paris Cit\'e, F-91191 Gif-sur-Yvette Cedex, France\\
    \inst{2} Institute for Theoretical Astroparticle Physics, Karlsruhe Institute of Technology (KIT), D-76021 Karlsruhe, Germany\\
    \inst{3} Centre for Fluid and Complex Systems, Coventry University, Coventry CV1 5FB, UK\\
    \email{junho.park@coventry.ac.uk}\\
    \inst{4} Department of Engineering Sciences and Applied Mathematics, Northwestern University, Evanston IL 60208, USA\\ 
    \inst{5} CIERA, Northwestern University, Evanston IL 60201, USA\\
    \inst{6} Department of Applied Mathematics, School of Mathematics, University of Leeds, Leeds, LS2 9JT, UK
                           }

   \date{}

 
  \abstract
   {The dissipation of tidal inertial waves in planetary and stellar convective regions is one of the key mechanisms that drive the evolution of star-planet/planet-moon systems. It is particularly efficient for young low-mass stars and gaseous giant planets, which are rapid rotators. In this context, the interaction between tidal inertial waves and turbulent convective flows must be modelled in a realistic and robust way. In the state-of-the-art simulations, the friction applied by convection on tidal waves is modelled most of the time by an effective eddy-viscosity. This approach may be valid when the characteristic length scales of convective eddies are smaller than those of tidal waves. However, it becomes highly questionable in the case where tidal waves interact with potentially stable large-scale vortices, as those observed at the pole of Jupiter and Saturn. They are potentially triggered by convection in rapidly-rotating bodies in which the Coriolis acceleration forms the flow in columnar vortical structures along the direction of the rotation axis.}
   {We investigate the complex interactions between 
   a tidal inertial wave and a columnar convective vortex.}
   {We use a quasi-geostrophic semi-analytical model of a convective columnar vortex, which is validated by numerical simulations. First, we carry out linear stability analysis using both numerical and asymptotic Wentzel-Kramers-Brillouin-Jeffreys (WKBJ) methods. Next, we conduct linear numerical simulations of the interactions between such a convective vortex and an incoming tidal inertial wave.}
   {The convective columnar vortex we consider is found to be centrifugally stable in the range $-\Omega_{p}\leq\Omega_{0}\leq3.62\Omega_{p}$ and unstable outside this range, where $\Omega_{0}$ is the local rotation rate of the vortex at its center and $\Omega_{p}$ is the global planetary (stellar) rotation rate.
   From the linear stability analysis, we found that this vortex is prone to centrifugal instability with perturbations with azimuthal wavenumbers $m=\left\{0,1,2\right\}$, which correspond potentially to eccentricity, obliquity and asynchronous tides, respectively. The modes with $m>2$ are found to be neutral or stable. The WKBJ analysis provides analytic expressions of the dispersion relations for neutral and unstable modes when the axial (vertical) wavenumber is sufficiently large. We verify that in the unstable regime, an incoming tidal inertial wave triggers the growth of the most unstable mode of the vortex. This would lead to turbulent dissipation. 
  For stable convective columns, the wave-vortex interaction leads to the mixing of momentum for tidal inertial waves while it creates a low-velocity region around the vortex core and a new wave-like perturbation in the form of a progressive wave radiating in the far field. The emission of this secondary wave is the strongest when the wavelength of the incoming wave is close to the characteristic size (radius) of the vortex. Incoming tidal waves can also experience complex angular momentum exchanges locally at critical layers of stable vortices.
  }
   {The interaction between tidal inertial waves and large-scale coherent convective vortices in rapidly-rotating planets (stars) leads to turbulent dissipation in the unstable regime and complex behaviors such as mixing of momentum and radiation of new waves in the far field or wave-vortex angular momentum exchanges in the stable regime. 
   These phenomena cannot be modelled using a simple effective eddy-viscosity.}


   \maketitle
%
\section{Introduction}
Tidal star-planet and planet-moon interactions are one of the key physical mechanisms that drive the evolution of planetary systems \citep[e.g.][]{Laskar2012,Ahuir2021a,Lainey2020}. In the case of stars and fluid (gaseous/liquid) planetary layers, the tidal force trigger several flows. On the one hand, the tidal potential induces a large-scale non-wavelike flow associated with the tidal deformation \citep[e.g.][]{Zahn1966a,Remus2012,Ogilvie2013}. On the other hand, since this flow is not a solution of the complete hydrodynamical equations, the non-wavelike flow is completed by tidal waves, the so-called dynamical tide \citep[][]{Zahn1975,Ogilvie2004}. The kinetic and potential energies of these flows are then dissipated through different friction mechanisms like turbulent friction in convective layers and heat diffusion in stably stratified regions \citep[e.g.][]{Ogilvie2014,Mathis2019}. In this framework, the rate of tidal dissipation in stellar and planetary convective regions has broad consequences for the evolution of star-planet and planet-moon systems. In the case of star-planet systems, the dissipation of tidal inertial waves, which have the Coriolis acceleration as a restoring force, in the convective envelope of low-mass stars during their pre-main-sequence leads to a strong outward/inward migration if the planet is above/inside the co-rotation orbit \citep[e.g.][]{Bolmont2016}. This shapes the orbital distribution of short-period gaseous exoplanets \citep[e.g.][]{Benbakoura2019,Barker2020,Ahuir2021a}. The efficiency of this tidal friction is due to the rapid rotation of low-mass stars during this evolutionary stage \citep[e.g.][]{Gallet2013,Gallet2015}. It leads to an efficient excitation of tidal inertial waves. In addition, in our solar system, a revolution has occurred in our understanding of the tidal evolution of the Jupiter and Saturn systems. Both planets are the seat of a tidal dissipation stronger by one or several orders of magnitude than previous predictions based on scenarios for the formation of their moons \citep{Goldreich1966}. This strong dissipation is necessary to explain their rapid orbital migration, which has been discovered thanks to high-precision astrometric measurements \citep{Lainey2009,Lainey2012,Lainey2017,Lainey2020}. As in the case of rapidly rotating young stars, the dissipation of tidal inertial waves propagating in the convective regions of Jupiter and Saturn, which are also fast rotators, has been proposed as one of the possible mechanisms to explain the observed strong dissipation \citep[][]{Ogilvie2004,Wu2005,Goodman2009,Lainey2020}. Therefore, we have to address two fundamental questions: (i) how are tidal flows interacting with turbulent convective flows and how are they dissipated, and; (ii) what is the role of rapid rotation in these processes?\\

The interaction between tidal and turbulent convective flows is one of the most challenging and debated topics in the theory of stellar and fluid planetary tides. The first proposed modellings have all adopted the vision of convective small-scale eddies acting on tidal flows, which have the largest characteristic length scales. This led \cite{Zahn1966b}, \cite{Zahn1989} and \cite{GoldreichKeeley1977} to use the theory of the mixing length to derive an effective eddy viscosity which is applied to the large-scale non-wavelike/equilibrium tide \citep[e.g.][]{Zahn1966b,Barker2020} and to tidal inertial waves \citep[][]{Wu2005}. A long debate has occurred on the dependence of this eddy viscosity as a function of the ratio between the tidal period ($P_{\rm T}$) and the convective characteristic turnover time ($P_{\rm c}$)  \citep[][]{Goodman1997,Penev2007,Ogilvie2012}. The most recent state-of-the-art simulations \citep{Duguid2020a,Duguid2020b,Vidal2020a,Vidal2020b} favoured the \cite{GoldreichKeeley1977} proposition with a scaling proportional to $\left(P_{\rm T}/P_{\rm c}\right)^{2}$ for high values of the tidal frequency but with a more complex behaviour at intermediate values. Recent work by \cite{Terquem2021a} proposed a more refined formalism leading to promising results \citep{Terquem2021b} but they are challenged by direct numerical simulations \citep{Barker2021}. A key point on these studies is that they do not study specifically the action of rotation on turbulent convective flows.\\

A first attempt has been done by \cite{Mathis2016} using a mixing length theory for rotating convective flows \citep{Stevenson1979,Augustson2019}, which has been confirmed by direct numerical simulation \citep{Barker2014,Currie2020}. This theory allows us to derive characteristic convective velocity and length scales that take into account the action of the Coriolis acceleration on convective flows. It decreases the efficiency of the heat transport leading to a decreased effective turbulent eddy-viscosity applied on tidal flows. At the same time, breakthroughs occur in our knowledge of stellar and planetary rapidly rotating convective flows thanks to global nonlinear numerical simulations and space missions JUNO and Cassini exploring Jupiter and Saturn, respectively. First, stable vortices have been discovered at the poles of Jupiter and Saturn \citep[e.g.][]{Adriani2018,Godfrey1988,Sanchez2006,Dyudina2008,Fletcher2018}. Their formation can result from deep-seated rapidly rotating convection \citep{Yadav2020,Garcia2020,Cai2021}. At the same time \cite{Julien12} and \cite{Hindman2020} provide a systematic classification of convective flows as a function of the rotation rate. These studies lead to columnar convective turbulent structures aligned along the rotation axis direction. This alignment results from the Taylor-Proudman constraint applied by the Coriolis acceleration on convective flows. In the presence of a companion, that would lead to configurations where tidal inertial waves propagate in turbulent rapidly-rotating convective flows with such structured columnar vortices. In this complex configuration, convective eddies can thus have the largest scale than those assumed in the mixing length approach and it becomes necessary to study their complex interactions with tidal inertial waves. 

In this article, we will thus study the propagation of a monochromatic (tidal) inertial wave through a columnar turbulent convective vortex as a laboratory of this process. 
In Sect.~\ref{sec:Formulation}, we formulate the equations to understand the wave-vortex interaction problem. 
In this framework, we propose a semi-analytical model for the convective Taylor columnar vortex and describe the regime where centrifugal instability occurs.
We formulate the equations for the stability analysis and linear evolution of perturbations for a tidally-forced inertial wave interacting with the vortex. 
In Sect.~\ref{sec:Stability}, we study 
the stability of the convective column based on numerical stability computation and compare with asymptotic results from a detailed WKBJ analysis reported in Appendix.
In the unstable regime, we verify that an incoming radial tidal wave can, with other motions, trigger the most unstable mode. 
This may lead to turbulence.
In Sect.~\ref{sec:Interaction}, we {conduct numerical simulations to investigate how a tidal inertial wave interacts with a stable convective column in the linear regime. 
In Sect.~\ref{sec:nonlinear}, we propose possible scenarios of the nonlinear interactions between convective vortices and tidally-forced inertial waves.}
Finally, conclusions, discussions, and perspectives are provided in Sect.~\ref{sec:Conclusion}.

\section{Mathematical formulation for the interactions between a convective Taylor column and inertial modes}
\label{sec:Formulation}

\subsection{Governing equations}

  \begin{figure*}
   \centering
   \includegraphics[width=10cm]{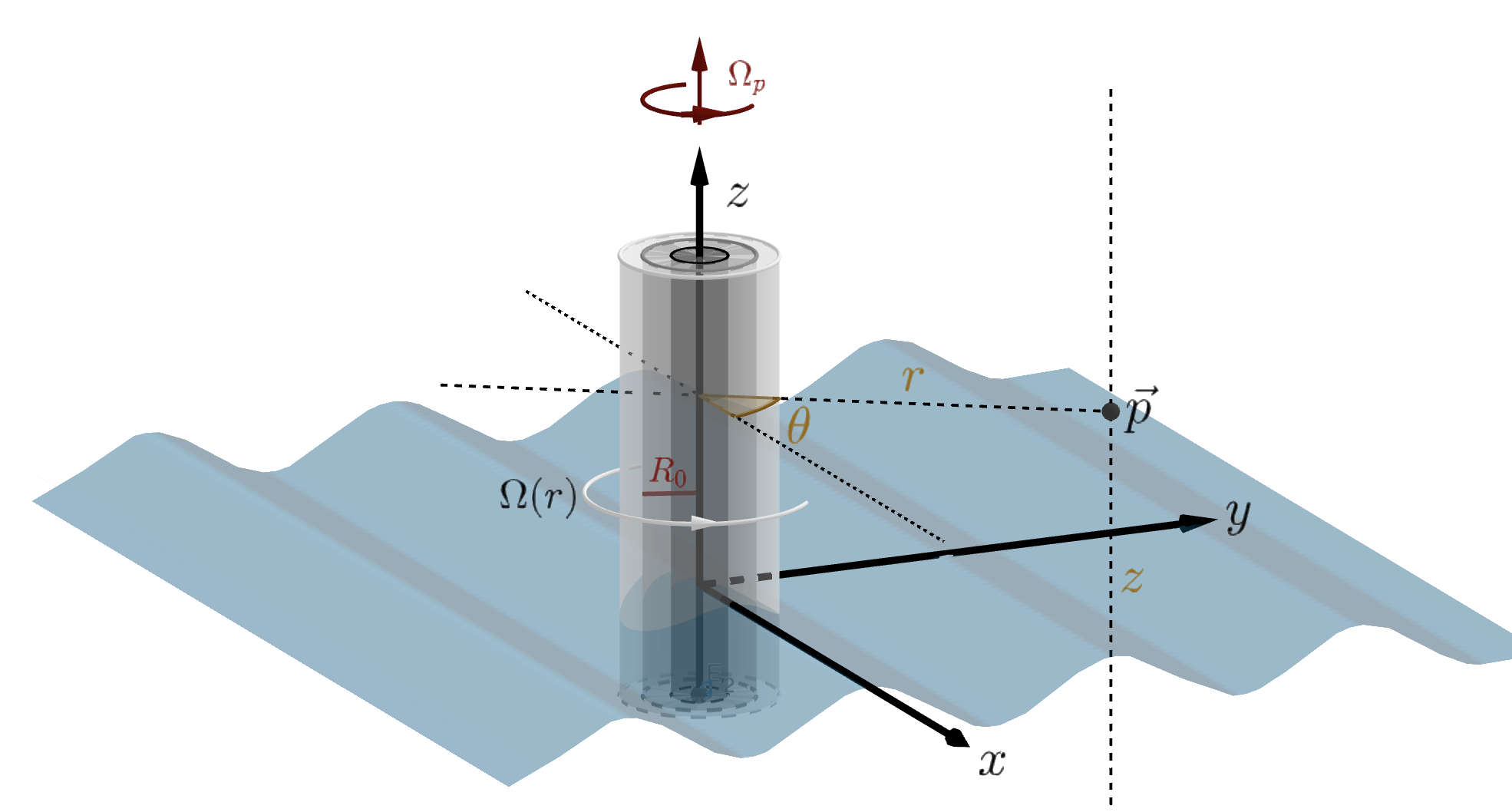}
 \includegraphics[height=5cm]{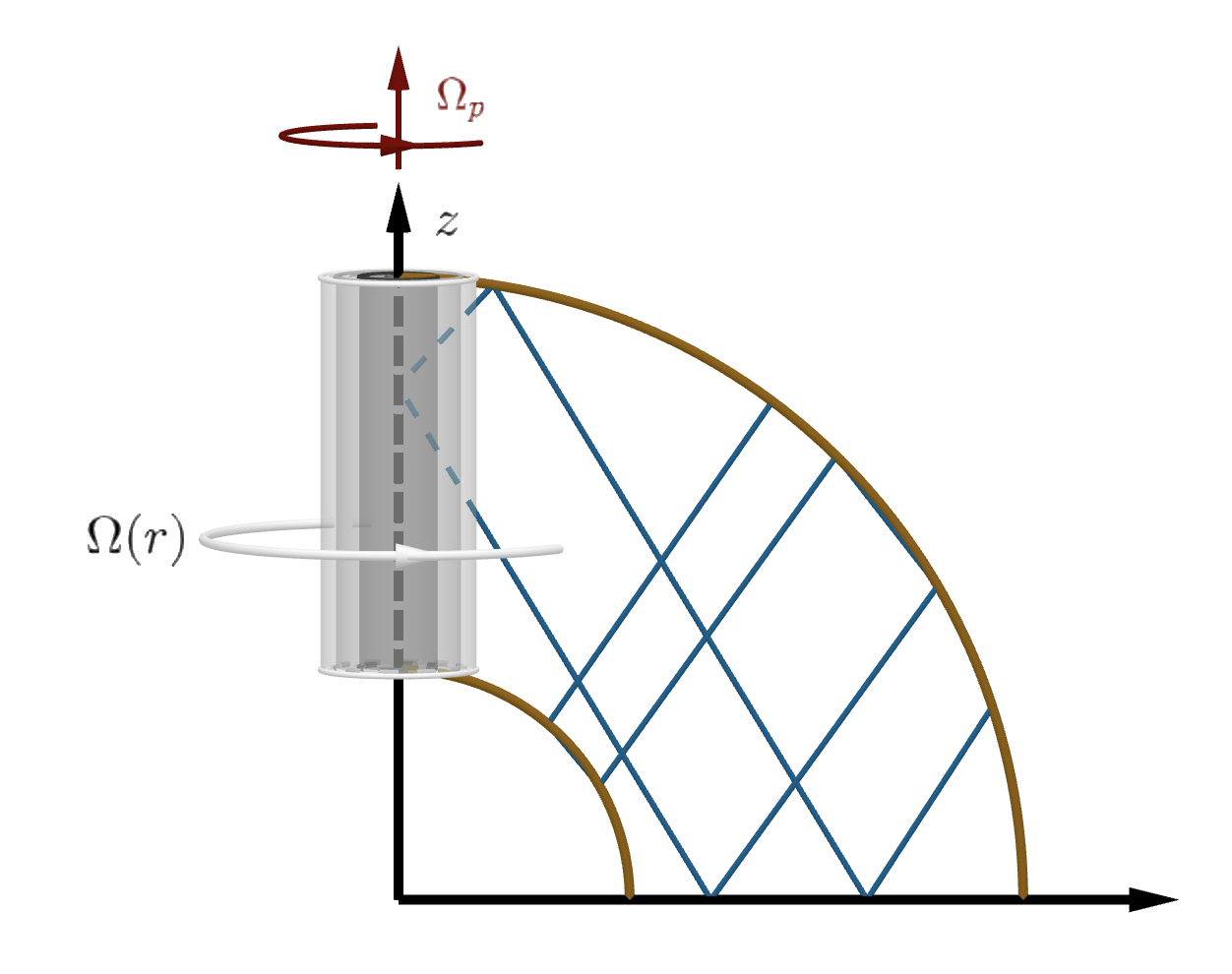}
     \caption{(Left) Schematic of the interaction between a tidal inertial wave and a single convective Taylor column in the $f$-plane. We introduce the cylindrical coordinates $\left(r,\theta,z \right)$.
     (Right) Schematic of a meridian cut of planets/stars showing propagation of a tidal inertial wave (blue lines) in a convective envelope and its interaction with a single Taylor column. 
     }
         \label{fig:column}
   \end{figure*}
To understand how a convective Taylor column interacts with {(tidal)} inertial waves (Fig.~\ref{fig:column}), we first investigate an intrinsic property of the convective structure: stability. 
The stability analysis allows us to examine how a convective column would respond to external forcing and when intense instability would occur because of its interaction with specific perturbations such as linear eigenmodes of basic states or the optimal perturbation represented as the sum of the eigenmodes \citep[][]{Schmid2001,Antkowiak2005}. 
In a local framework rotating with an angular speed $\Omega_{p}$, the global planetary (stellar) rotation rate, we consider the {continuity and momentum equations:}
\begin{equation}
\label{eq:continuity_general}
\frac{\partial\rho}{\partial t}+\nabla\cdot\left(\rho\vec{u}\right)=0,
\end{equation}
\begin{equation}
\label{eq:momentum_general}
\frac{\partial\vec{u}}{\partial t}+\left(\vec{u}\cdot\nabla\right)\vec{u}+\vec{f}\times\vec{u}+\frac{\vec{f}}{2}\times\left(\frac{\vec{f}}{2}\times\vec{r}\right)=-\frac{1}{\rho}\nabla P+\vec{g}+\upnu\nabla^{2}\vec{u}+\vec{F}_{e},
\end{equation}
where $\rho$ is the density, $\vec{u}$ is the velocity vector, $\vec{f}$ is the Coriolis vector where $\vec{f}=f\hat{\mathrm{e}}_{z}=2\Omega_{p}\hat{\mathrm{e}}_{z}$, $\vec{r}$ is the position vector, $P$ is the pressure, $\vec{g}$ is the gravity vector, 
$\upnu$ is the kinematic viscosity, 
$\nabla$ is the vector differential operator and $\vec{F}_{e}$ is the vector of external (tidal) forcing. 
We assume an incompressibility condition by considering the density $\rho$ as a constant and decompose the pressure $P$ into $P=P_{0}+p$ where $P_{0}$ satisfies the hydrostatic balance with the centrifugal acceleration taken into account as
\begin{equation}
\label{eq:pressurebalance}
\frac{1}{\rho}\nabla P_{0}=\vec{g}-\frac{\vec{f}}{2}\times\left(\frac{\vec{f}}{2}\times\vec{r}\right).
\end{equation}
Under these assumptions, Eqs.~(\ref{eq:continuity_general}) and (\ref{eq:momentum_general}) reduce to 
\begin{equation}
\label{eq:continuity}
\nabla\cdot\vec{u}=0,
\end{equation}
\begin{equation}
\label{eq:momentum}
\frac{\partial\vec{u}}{\partial t}+\left(\vec{u}\cdot\nabla\right)\vec{u}+\vec{f}\times\vec{u}=-\nabla \pi+\upnu\nabla^{2}\vec{u}+\vec{F}_{e},
\end{equation}
where $\pi=p/\rho$ is the normalized pressure.\\

Our study considers a local {polar} traditional $f$-plane centered around the rotation axis of the planet/star, where the rotation vector and the gravity are aligned, with an associated cylindrical coordinate system $(r,\theta,z)$. 
In this case, we have the following set of equations for the velocity $\vec{u}=(u,v,w)$, the normalized pressure $\pi$, and the forcing $\vec{F}_{e}=(F_{u},F_{v},F_{w})$:
\begin{equation}
\label{eq:continuity_cylinder}
\frac{1}{r}\frac{\partial(ru)}{\partial r}+\frac{1}{r}\frac{\partial v}{\partial \theta}+\frac{\partial w}{\partial z}=0,
\end{equation}
\begin{equation}
\label{eq:momentum_radial}
\begin{aligned}
&\frac{\partial u}{\partial t}+u\frac{\partial u}{\partial r}+\frac{v}{r}\frac{\partial u}{\partial \theta}+w\frac{\partial u}{\partial z}-\frac{v^{2}}{r}-fv
\\
&=-\frac{\partial\pi}{\partial r}+\upnu\left(\nabla^{2}u-\frac{u}{r^{2}}-\frac{2}{r}\frac{\partial v}{\partial \theta}\right)+F_{u},
\end{aligned}
\end{equation}
\begin{equation}
\label{eq:momentum_azimuthal}
\begin{aligned}
&\frac{\partial v}{\partial t}+u\frac{\partial v}{\partial r}+\frac{v}{r}\frac{\partial v}{\partial \theta}+w\frac{\partial v}{\partial z}+\frac{uv}{r}+fu\\
&=-\frac{1}{r}\frac{\partial\pi}{\partial \theta}+\upnu\left(\nabla^{2}v-\frac{v}{r^{2}}+\frac{2}{r}\frac{\partial u}{\partial \theta}\right)+F_{v},
\end{aligned}
\end{equation}
\begin{equation}
\label{eq:momentum_vertical}
\frac{\partial w}{\partial t}+u\frac{\partial w}{\partial r}+\frac{v}{r}\frac{\partial w}{\partial \theta}+w\frac{\partial w}{\partial z}=-\frac{\partial\pi}{\partial z}+\upnu\nabla^{2}w+F_{w},
\end{equation}
where the vertical component of the Coriolis acceleration vanishes. Being in the polar traditional $f$-plane allows us to treat a case where the equations for (tidal) inertial waves propagating within a vortex with a radial structure are separable. 

While the vortex has here a polar nature, the inertial waves can also propagate in a Cartesian, planar manner, as depicted by a schematic in Fig.~\ref{fig:column} (left panel) of the wave-vortex interaction in the Cartesian coordinate system. 
To analyze this planar wave propagation, we need to consider the set of equations for the velocity $\vec{u}=(u_{x},u_{y},w)$ in Cartesian coordinates $(x,y,z)$ on the traditional $f$-plane as follows:
\begin{equation}
\label{eq:continuity_cartesian}
\frac{\partial u_{x}}{\partial x}+\frac{\partial u_{y}}{\partial y}+\frac{\partial w}{\partial z}=0,
\end{equation}
\begin{equation}
\label{eq:momentum_x}
\frac{\partial u_{x}}{\partial t}+u_{x}\frac{\partial u_{x}}{\partial x}+u_{y}\frac{\partial u_{x}}{\partial y}+w\frac{\partial u_{x}}{\partial z}-fu_{y}=-\frac{\partial \pi}{\partial x}+\upnu\nabla^{2}u_{x}+F_{u_{x}},
\end{equation}
\begin{equation}
\label{eq:momentum_y}
\frac{\partial u_{y}}{\partial t}+u_{x}\frac{\partial u_{y}}{\partial x}+u_{y}\frac{\partial u_{y}}{\partial y}+w\frac{\partial u_{y}}{\partial z}+fu_{x}=-\frac{\partial \pi}{\partial y}+\upnu\nabla^{2}u_{y}+F_{u_{y}},
\end{equation}
\begin{equation}
\label{eq:momentum_z}
\frac{\partial w}{\partial t}+u_{x}\frac{\partial w}{\partial x}+u_{y}\frac{\partial w}{\partial y}+w\frac{\partial w}{\partial z}=-\frac{\partial \pi}{\partial z}+\upnu\nabla^{2}w+F_{w},
\end{equation}
where the following relations among the horizontal velocity components are satisfied:
\begin{equation}
u_{x}=u\cos\theta-v\sin\theta,~~
u_{y}=u\sin\theta+v\cos\theta.
\end{equation}
This polar traditional study constitutes the first necessary step to understanding the complex wave-vortex interaction. It is for instance relevant to study the interactions between tidal inertial waves and large polar vortices as those observed in giant planets. However, one should keep in mind that, in a more general case away from the poles \citep[e.g. a case with the nontraditional $f$-plane approximation applicable at a general colatitude, see e.g.][]{Park2021}, the propagation equation for inertial waves is not separable anymore \citep{Gerkema2005}. This would lead to a more complex situation, where 2D inertial waves attractors \citep[e.g.][]{Maas2001,Rieutord2001} are interacting with a vertical vortex \citep[][see also Fig.~\ref{fig:column} right panel]{Duran-Matute2013,Boury2021}. We first focus here on the already complex polar problem while the fully 2D problem will be studied in a future study.

%
\subsection{Convective Taylor column}
\subsubsection{Basics of rapidly rotating flows}
Convection plays a fundamental role in stars and planets. 
Rotation fundamentally modifies the behavior of convection through the action of the Coriolis acceleration \citep[e.g.][]{Davidson2013} and changes the equilibrium structure of the body because of the centrifugal force \citep[e.g.][]{Wang2016}. This impacts the interactions and energy exchanges with (gravito-) inertial (tidal) waves \citep{Mathis2014,Augustson2020}.

In rotating convection, one of the fundamental flows that appear even in very turbulent regimes, are columnar structures that result from the interaction of buoyancy-driven convection and the Coriolis acceleration. Rapid rotation tends to cause those structures to align with the rotation axis of the body. Indeed, in an inviscid adiabatic fluid, a steady state solution obeys the Taylor-Proudman theorem where the fluid velocity becomes invariant along the rotation vector \citep[e.g.][]{Davidson2013}. However, the entropy gradient is in general slightly nonadiabatic with a density gradient leading to quasigeostrophic flow in rapidly rotating systems with net heat flux. Such flows can be modelled using asymptotic methods if the scale separation along the direction of the rotation axis and the orthogonal one is large enough. These dynamically rich nonlinear systems have been studied in great detail \citep{veronis59,gough75,julien98}. Their solutions exhibit the formation of columnar flow structures in both laminar and turbulent regimes in numerical and experimental settings \citep{stellmach14,aurnou15,cheng15}. Exact analytical solutions validated on numerical simulations can be found as well \citep{Grooms2010,grooms15}. These solutions are separable with a fixed horizontal structure and a varying vertical one. They have been discussed by \citet{julien98}, \citet{Sprague2006}, \citet{Grooms2010}, \citet{Julien12}, and \citet{grooms15}. The equations for the vertical structure can be condensed into a nonlinear boundary value problem. This provides a robust analytical description of the structure of the flow with which we will be able to investigate their stability and interactions with incoming tidal inertial waves.

\subsubsection{Semi-analytical quasigeostrophic models}
\begin{figure*}
   \centering
   \includegraphics[height=5cm]{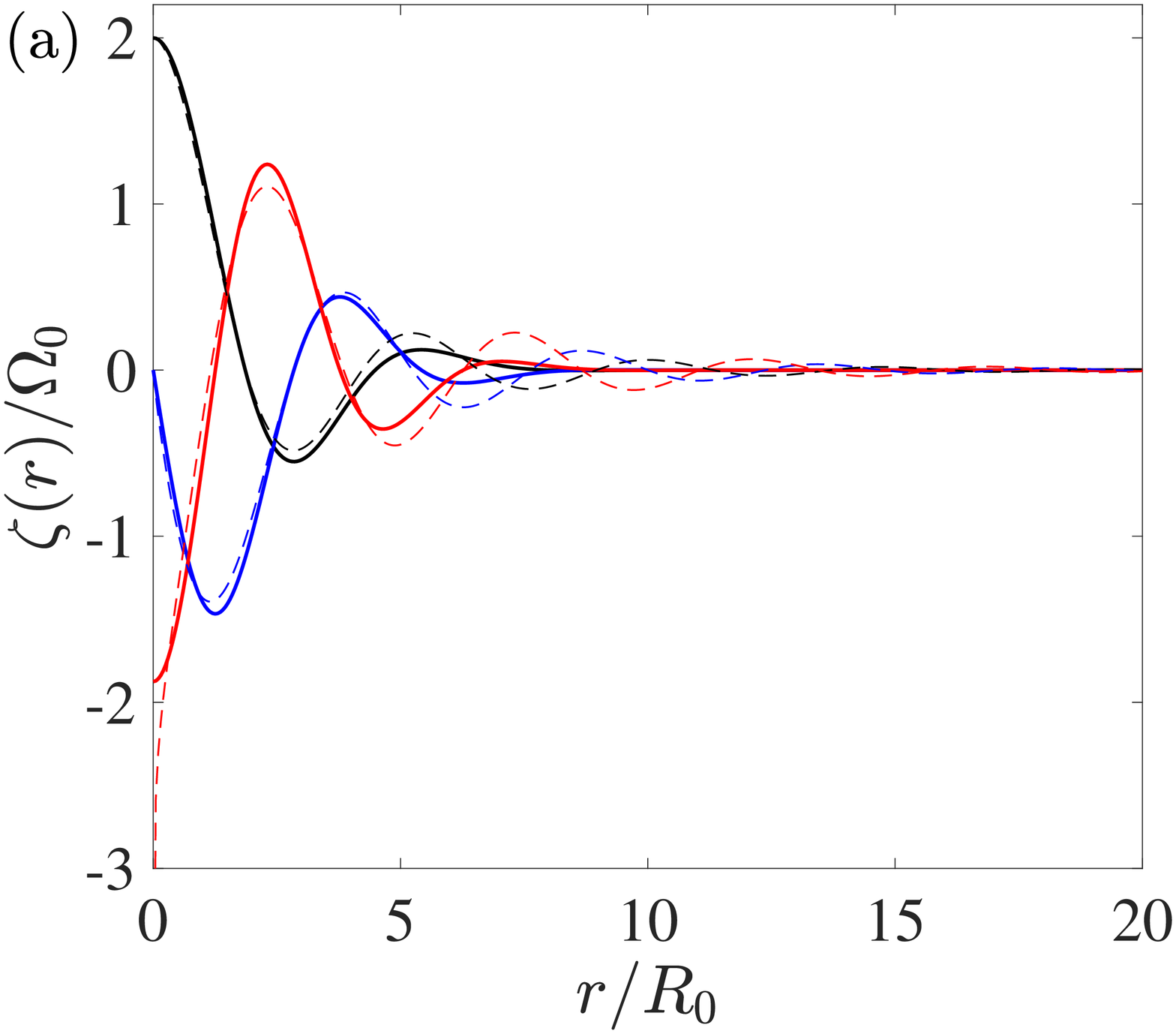}
   \includegraphics[height=5cm]{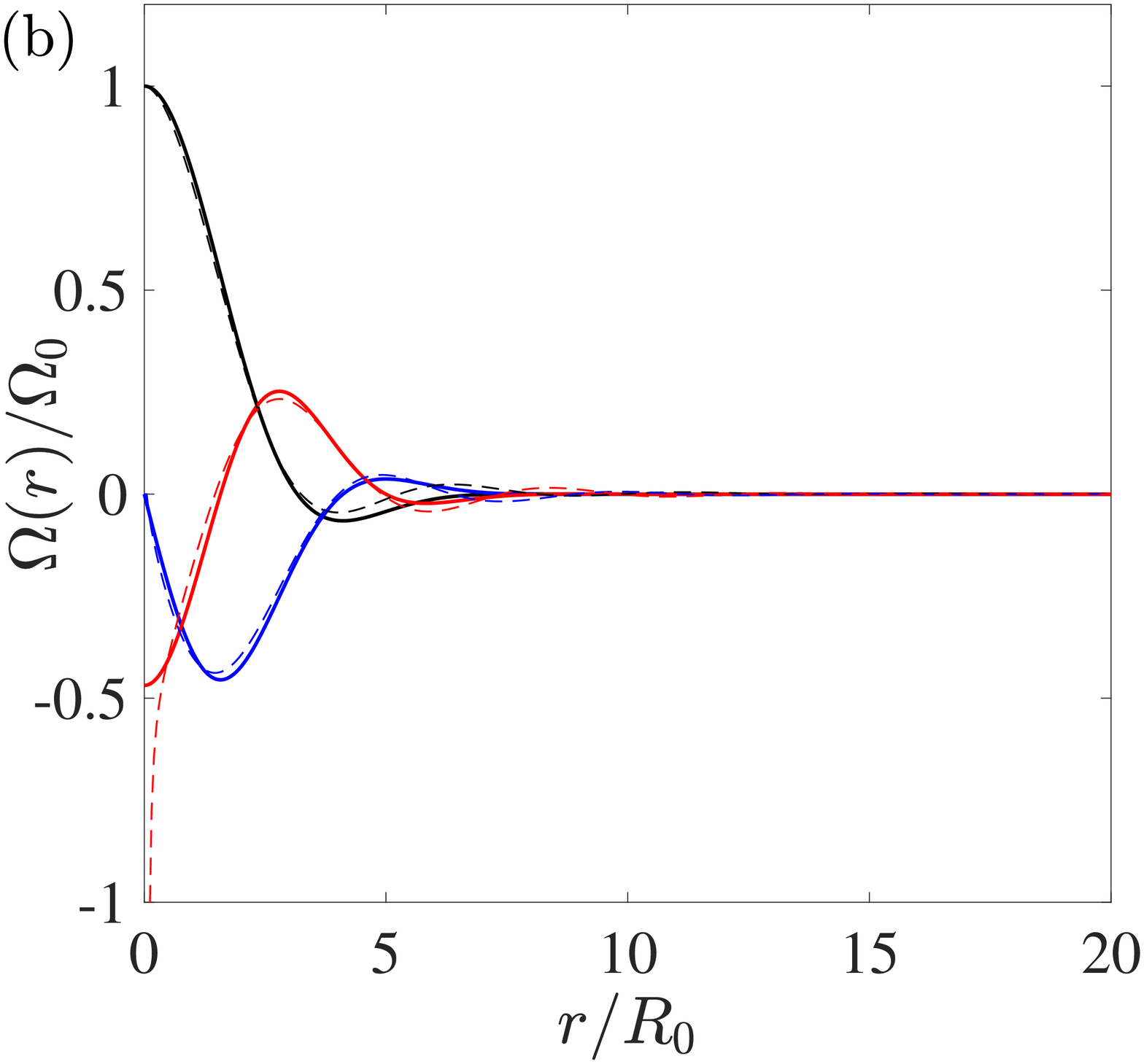}
   \includegraphics[height=5.1cm]{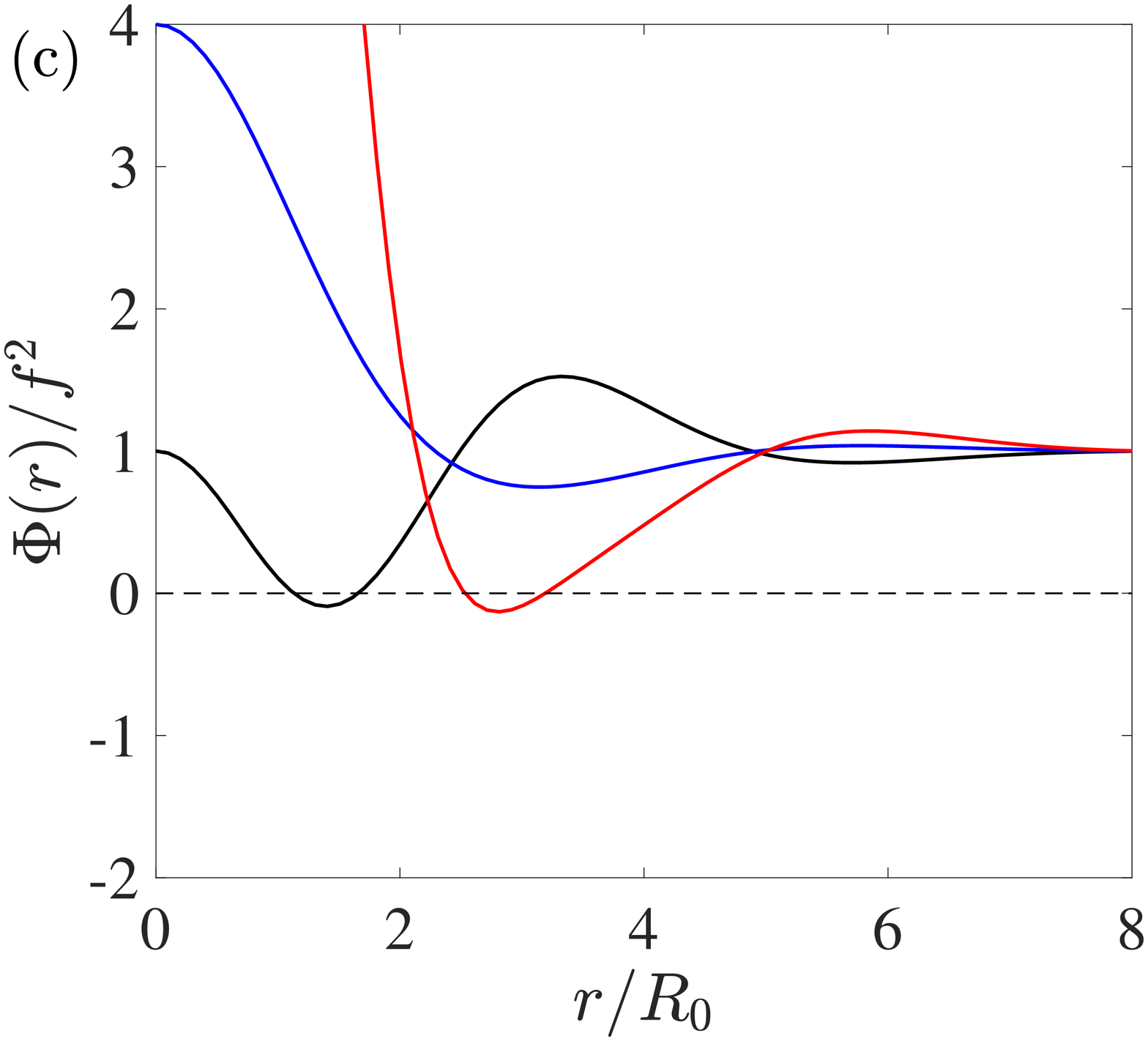}
     \caption{(a,b) Profiles of base vorticity $\zeta(r)$ and base angular velocity $\Omega(r)$ (black) and their first (blue) and second (red) derivatives for the current model (\ref{eq:base_Dandoy}) (solid lines) and (\ref{eq:base_Grooms}) from \citet{Grooms2010} (dashed lines). (c) Profiles of the Rayleigh discriminant $\Phi(r)$ for $\Omega_{0}/f=-1$ (black), $\Omega_{0}/f=0.5$ (blue) and $\Omega_{0}/f=2$ (red).
              }
         \label{Fig_base_flow}
\end{figure*}
Simulations of quasigeostrophic convective flows thus provide us guidance to model coherent structures that may exist in rapidly-rotating planets and stars. These simulations range from near the onset of instability to the most turbulent states that can be reached thanks to supercomputers, which are still far from the real planetary and stellar regimes. One characteristic that appears to bridge laminar and the most turbulent simulations are the formation of large-scale vortical structures that span the vertical domain. In the laminar regime, these vortices are long-lived (even persistent) and they only weakly interact with each other. In turbulent simulations, they are formed by the advection of small-scale vortices into domain-spanning larger-scale structures, representing an inverse cascade of energy into the largest-scale vortex which can be intermittent.  An analogy can be made between many of the features of the turbulent and the laminar structures, under the assumption that they weakly interact with each other.

Analytical solutions for such laminar and turbulent structures have been derived by \citet{Grooms2010} and \citet{grooms15}, respectively. They are solutions to a nonlinear eigenvalue problem that arises from a simplification of the full quasigeostrophic equations. First, it is assumed that nonlinear horizontal interactions between the columnar structures are weak while passive advection is still allowed. 
In this regard, each columnar structure is assumed to be axisymmetric and time-independent so that there is no horizontal self-interaction, which affects the structure by itself along its temporal evolution.
The equations resulting from these assumptions are Equations 9 and 10 in \citet{Grooms2010}.  One solution of these equations is of the separable form for the poloidal velocity potential $\phi(r,z) = \hat{\phi}(z) J_0(k r)$ that yields the vertical component of the vorticity $\zeta(z,r) = -\partial_z\phi$ with $J_0(k r)$ the zeroth-order Bessel function with a radial wavenumber $k$. This Bessel function solution is chosen over the better-fitting Hankel function solution examined in \citet{Grooms2010} because the latter diverges for $r\rightarrow 0$ {\citep[we refer the reader to][for the properties of these functions]{Abramowitz}}. However, as stated in \citet{Grooms2010}, the radial function must be truncated. We choose to do this exponentially so that an infinite radial domain may be studied when looking at the interaction with an incoming tidal inertial wave.  Moreover, the vertical structure is taken to be far from the bounding surfaces so that it can be locally approximated as a constant with $\hat{\phi}(z) = {\rm constant}$. In this case, we can define the local vorticity of the fluid such that:
\begin{equation}
\zeta=\frac{1}{r}\frac{\partial (r^{2}\Omega)}{\partial r},
\end{equation}
where $\Omega$ is the local angular velocity. A comparison of the approximate solution employed here (solid lines) versus the solution derived in \citet{Grooms2010} (dashed lines) is shown in Fig. \ref{Fig_base_flow}, where the divergence of the Hankel function can be seen as the radius approaches zero but the oscillations are damped requiring the exponential damping of the Bessel function.
More specifically, the radial profile of the vorticity $\zeta(r)$ in terms of the Hankel function is 
\begin{equation}
\label{eq:base_Grooms}
\zeta(r)\sim \mathrm{Re}\left[\mathrm{H}^{(1)}_{0}(k_{\rm{H}}r)\right],
\end{equation}
where $\mathrm{Re}$ denotes the real part, $\mathrm{H}^{(1)}_{0}$ denotes the Hankel function of the first kind at the zeroth order, and $k_{\mathrm{H}}=1.34\exp(0.153\mathrm{i})/R_{0}$ with $R_{0}$ the reference radial length scale of the structure. In contrast, the local angular velocity profile used here is instead 
\begin{equation}
\label{eq:base_Dandoy}
\Omega(r)=\Omega_{0}\exp\left(-\alpha r^{2}\right)\mathrm{J}_{0}(\beta r),
\end{equation}
where $\Omega_{0}$ is the local angular velocity of the convective column at $r=0$, $\alpha$ and $\beta$ are constants $\alpha=0.09/R_{0}^{2}$ and $\beta=0.76/R_{0}$ that are chosen to minimize the error between the vorticity (\ref{eq:base_Grooms}) proposed by \citet{Grooms2010} and our model.
As shown in Fig.~\ref{Fig_base_flow}(a,b), the two profiles (\ref{eq:base_Grooms}) and (\ref{eq:base_Dandoy}) match fairly well and there is no singularity at $r=0$ for the second derivative of the angular velocity (\ref{eq:base_Dandoy}) as well as the second derivative of the vorticity. 
Throughout the paper, we use the angular velocity profile given in Eq.~(\ref{eq:base_Dandoy}) to derive analytic expressions for the dispersion relations, which are crucial for understanding the interaction between a convective column and tidally-forced inertial waves.

\subsubsection{Conditions for the centrifugal instability}
Hydrodynamics theories on vortices and their instability have been well established, and they are easily applicable to understand the stability of general convective Taylor columns.
In rotating fluids, columnar convective structures are prone to centrifugal instability, which occurs due to an imbalance between the centrifugal acceleration and pressure gradient in the radial direction. 
The centrifugal instability can be predicted using the Rayleigh's criterion, initially proposed by \citet{Rayleigh1917} for an inviscid rotating flow with angular velocity $\Omega(r)$ and then generalized by \citet{Kloosterziel1991} for rotating flows on the traditional $f$-plane. 
The criterion states that an inviscid vortex on the $f$-plane can become centrifugally unstable if there exists a region where the Rayleigh's discriminant $\Phi(r)$ becomes negative, i.e.,  
\begin{equation}
\label{eq:Rayleigh_criterion}
    \Phi=(f+2\Omega)(f+\zeta)<0.
\end{equation}
The necessary and sufficient conditions for the centrifugal instability were originally derived for axisymmetric perturbations with $m=0$ \citep[][]{Synge1933}, but the follow-up study by \citet{Billant2005} extended the criterion for general perturbations with $m\geq0$ by deriving the dispersion relation using the Wentzel-Kramers-Brillouin-Jeffreys (WKBJ) approximation. 
The Rayleigh's criterion (\ref{eq:Rayleigh_criterion}) is also applicable for the centrifugal instability of vortices in rotating and stratified fluids \citep[][]{Park2013PoF} and for the inertial instability of shear flows in rotating and stratified fluids \citep[][]{Arobone2012,Park2020}.

For the convective column with the angular velocity profile (\ref{eq:base_Dandoy}), examples of $\Phi(r)$ normalized by $f^{2}$ are shown in Fig.~\ref{Fig_base_flow}c for different values of the ratio $\Omega_{0}/f$. 
For instance, for $\Omega_{0}=-f$ or $\Omega_{0}=2f$, there are regions where $\Phi(r)$ becomes negative, thus the convective column can become centrifugally unstable to incoming perturbations like tidal inertial waves. 
At $\Omega_{0}=0.5f$, $\Phi(r)$ is always positive for all $r$ so it is centrifugally stable. 
In this regime, incoming tidal inertial waves will interact with stable vortices and this interaction can promote new radiating waves. This configuration will be investigated in detail in section 4. 

\begin{figure}
   \centering
   \includegraphics[width=6cm]{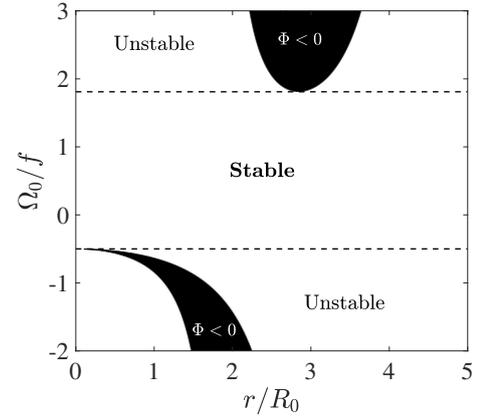}
     \caption{Sign of the Rayleigh discriminant $\Phi(r)$ in the parameter space $(r/R_{0},\Omega_{0}/f)$ where black regions indicate $\Phi<0$ and white regions indicate $\Phi\geq0$. Upper and lower dashed lines represent the upper limit ($\Omega_{0}/f=1.81$) and the lower limit ($\Omega_{0}/f=-0.5$) that bound the centrifugally stable region. 
              }
         \label{Fig_Rayleigh_contours}
   \end{figure}
Figure \ref{Fig_Rayleigh_contours} displays the sign of $\Phi$ in the parameter space $(r/R_{0},\Omega_{0}/f)$. 
We see that there are regions of the negative $\Phi$ for $\Omega_{0}<-0.5f$ and $\Omega_{0}>1.81f$.
In terms of the global planetary rotation rate $\Omega_{p}$, we can divide the regime of $\Omega_{0}$ into two regimes as
\begin{eqnarray}
\label{eq:Rayleigh_stable}
&-\Omega_{p}\leq\Omega_{0}\leq3.62\Omega_{p}&:~~\mathrm{centrifugally~stable},\nonumber\\
&\Omega_{0}>3.62\Omega_{p}~~\mathrm{or}~~\Omega_{0}<-\Omega_{p}&:~~\mathrm{centrifugally~unstable}.
\end{eqnarray}

\subsection{Linear stability equations}
While the regime (\ref{eq:Rayleigh_stable}) identifies where the base flow is centrifugally stable or unstable, we need to perform the linear stability analysis to compute quantitatively the growth rate of the centrifugal instability. 
To our knowledge, this is the first time that this analysis is performed in the case of the model of quasi-geostrophic columnar convective vortex studied here.

In the cylindrical coordinate system as presented in Fig.~\ref{fig:column} (right panel),  we consider perturbations subject to the base velocity $\vec{U}=(0,r\Omega(r),0)$ and pressure $\Pi$ as follows:
\begin{equation}
\label{eq:perturbation}
\acute{\vec{u}}=\vec{u}-\vec{U},~~~
\acute{\pi}=\pi-\Pi,
\end{equation}
where $\acute{\vec{u}}=(\acute{u},\acute{v},\acute{w})$ is the velocity perturbation and $\acute{\pi}$ is the normalised pressure perturbation.
When these perturbations are infinitesimal, we obtain from Eqs.~(\ref{eq:continuity_cylinder})-(\ref{eq:momentum_vertical}) the following linearized equations
\begin{equation}
\label{eq:continuity_ptb}
\frac{1}{r}\frac{\partial(r\acute{u})}{\partial r}+\frac{1}{r}\frac{\partial \acute{v}}{\partial \theta}+\frac{\partial \acute{w}}{\partial z}=0,
\end{equation}
\begin{equation}
\label{eq:momentum_radial_ptb}
\frac{\partial \acute{u}}{\partial t} +\Omega\frac{\partial \acute{u}}{\partial \theta}-(f+2\Omega)\acute{v}+\frac{\partial \acute{\pi}}{\partial r}=\upnu\left(\nabla^{2}\acute{u}-\frac{\acute{u}}{r^{2}}-\frac{2}{r}\frac{\partial \acute{v}}{\partial \theta}\right)+F_{u},
\end{equation}
\begin{equation}
\label{eq:momentum_azimuthal_ptb}
\frac{\partial \acute{v}}{\partial t} +\Omega\frac{\partial \acute{v}}{\partial \theta}+(f+\zeta)\acute{u}+\frac{1}{r}\frac{\partial \acute{\pi}}{\partial \theta}=\upnu\left(\nabla^{2}\acute{v}-\frac{\acute{v}}{r^{2}}+\frac{2}{r}\frac{\partial \acute{u}}{\partial \theta}\right)+F_{v},
\end{equation}
\begin{equation}
\label{eq:momentum_vertical_ptb}
\frac{\partial \acute{w}}{\partial t} +\Omega\frac{\partial \acute{w}}{\partial \theta}+\frac{\partial \acute{\pi}}{\partial z}=\upnu\nabla^{2}\acute{w}+F_{w}.
\end{equation}
First, we will examine the intrinsic stability of the convective Taylor column by considering no external tidal forcing (i.e., $\vec{F}_{e}=0$).
This problem is closely related to the case with free waves in a sheared region \citep[][]{Astoul2021}, where the inertial wave enters and perturbs the base shear flow.
The response of the base flow to (tidal) inertial waves strongly depends on its stability.
Therefore, it is crucial to investigate the stability beforehand to understand the wave-vortex interaction.
In the stability analysis, we apply the formulation of the normal mode to perturbations as follows:
\begin{equation}
\label{eq:normal_mode}
\left(
\begin{array}{c}
\acute{u}\\
\acute{v}\\
\acute{w}\\
\acute{\pi}
\end{array}
\right)=
\left(
\begin{array}{c}
\hat{u}(r)\\
\hat{v}(r)\\
\hat{w}(r)\\
\hat{p}(r)
\end{array}
\right)\exp\left[\mathrm{i}(k_{z}z+m\theta-\omega t)\right]+c.c.,
\end{equation}
where $\hat{\vec{u}}=(\hat{u},\hat{v},\hat{w})$ and $\hat{p}$ are the mode shapes of velocity and pressure, $k_{z}$ is the vertical wavenumber, $m$ is the azimuthal wavenumber, $c.c.$ denotes the complex conjugate, and $\omega=\omega_{r}+\mathrm{i}\omega_{i}$ is the complex frequency where the real part $\omega_{r}$ denotes the frequency and the imaginary part $\omega_{i}$ denotes the temporal growth rate, respectively. 
The use of the normal mode leads to the following linear stability equations:
\begin{equation}
\label{eq:continuity_mode}
\frac{1}{r}\frac{\mathrm{d}(r\hat{u})}{\mathrm{d}r}+\frac{\mathrm{i}m\hat{v}}{r}+\mathrm{i}k_{z}\hat{w}=0,
\end{equation}
\begin{equation}
\label{eq:momentum_radial_mode}
\mathrm{i}s\hat{u}-(f+2\Omega)\hat{v}+\frac{\mathrm{d}\hat{p}}{\mathrm{d}r}=\upnu\left(\hat{\nabla}^{2}\hat{u}-\frac{\hat{u}}{r^{2}}-\frac{2\mathrm{i}m\hat{v}}{r}\right),
\end{equation}
\begin{equation}
\label{eq:momentum_azimuthal_mode}
\mathrm{i}s\hat{v}+(f+\zeta)\hat{u}+\frac{\mathrm{i}m\hat{p}}{r}=\upnu\left(\hat{\nabla}^{2}\hat{v}-\frac{\hat{v}}{r^{2}}+\frac{2\mathrm{i}m\hat{u}}{r}\right),
\end{equation}
\begin{equation}
\label{eq:momentum_vertical_mode}
\mathrm{i}s\hat{w}+\mathrm{i}k_{z}\hat{p}=\upnu\hat{\nabla}^{2}\hat{w},
\end{equation}
where $s=-\omega+m\Omega$ is the Doppler-shifted frequency and $\hat{\nabla}^{2}=\mathrm{d}^{2}/\mathrm{d}r^{2}+(1/r)\mathrm{d}/\mathrm{d}r-k_{z}^{2}-m^{2}/r^{2}$ is the Laplacian operator. 
Due to the symmetry: $\omega(m,k_{z})=\omega(m,-k_{z})=-\omega^{*}(-m,-k_{z})$ \citep[][]{Park2013PoF} where $^*$ denotes the complex conjugate, we consider hereafter only the non-negative wavenumbers ($m\geq0$ and $k_{z}\geq0$).  
The linear stability equations (\ref{eq:continuity_mode})-(\ref{eq:momentum_vertical_mode}) can be solved numerically with the following simplified eigenvalue problem
\begin{equation}
\label{eq:eigenvalue_problem}
-\mathrm{i}\omega\mathcal{A}\hat{\textbf{q}}=\mathcal{B}\hat{\textbf{q}},
\end{equation}
where $\hat{\textbf{q}}=(\hat{u},\hat{v})$ and $\mathcal{A}$ and $\mathcal{B}$ are the operator matrices as 
\begin{equation}
\label{eq:Matrices_AB}
\mathcal{A}=
\left[
\begin{array}{cc}
\mathcal{A}_{11} & \mathcal{A}_{12}\\
\mathcal{A}_{21} & \mathcal{A}_{22}
\end{array}
\right]
,~~
\mathcal{B}=\left[
\begin{array}{cc}
\mathcal{B}_{11} & \mathcal{B}_{12}\\
\mathcal{B}_{21} & \mathcal{B}_{22}
\end{array}
\right]
\end{equation}
where
\begin{equation}
\mathcal{A}_{11}=1-\frac{1}{k_{z}^{2}}\frac{\mathrm{d}}{\mathrm{d}r}\left(\frac{1}{r}+\frac{\mathrm{d}}{\mathrm{d}r}\right),~
\mathcal{A}_{12}=-\frac{\mathrm{i}m}{rk_{z}^{2}}\left(\frac{\mathrm{d}}{\mathrm{d}r}-\frac{1}{r}\right),
\end{equation}
\begin{equation}
\mathcal{A}_{21}=-\frac{\mathrm{i}m}{rk_{z}^{2}}\left(\frac{1}{r}+\frac{\mathrm{d}}{\mathrm{d}r}\right),~
\mathcal{A}_{22}=1+\frac{m^{2}}{r^{2}k_{z}^{2}},
\end{equation}
\begin{equation}
\begin{aligned}
\mathcal{B}_{11}=&-\mathrm{i}m\left[\Omega-\frac{\mathrm{d}}{\mathrm{d}r}\left\{\frac{\Omega}{k_{z}^{2}}\left(\frac{1}{r}+\frac{\mathrm{d}}{\mathrm{d}r}\right)\right\}\right]\\
&+\upnu\left[\hat{\nabla}^{2}-\frac{1}{r^{2}}-\frac{1}{k_{z}^{2}}\frac{\mathrm{d}}{\mathrm{d}r}\left\{\hat{\nabla}^{2}\left(\frac{1}{r}+\frac{\mathrm{d}}{\mathrm{d}r}\right)\right\}\right],
\end{aligned}
\end{equation}
\begin{equation}
\mathcal{B}_{12}=(f+2\Omega)-\frac{m^{2}}{k_{z}^{2}}\frac{\mathrm{d}}{\mathrm{d}r}\left(\frac{\Omega}{r}\right)-\mathrm{i}m\upnu\left[\frac{2}{r^{2}}+\frac{1}{k_{z}^{2}}\frac{\mathrm{d}}{\mathrm{d}r}\left\{\hat{\nabla}^{2}\left(\frac{1}{r}\right)\right\}\right],
\end{equation}
\begin{equation}
\mathcal{B}_{21}=-(f+\zeta)-\frac{m^{2}}{rk_{z}^{2}}\left(\frac{1}{r}+\frac{\mathrm{d}}{\mathrm{d}r}\right)+\frac{\mathrm{i}m\upnu}{r}\left[\frac{2}{r}-\frac{1}{k_{z}^{2}}\hat{\nabla}^{2}\left(\frac{1}{r}+\frac{\mathrm{d}}{\mathrm{d}r}\right)\right],
\end{equation}
\begin{equation}
\mathcal{B}_{22}=-\mathrm{i}m\Omega\left(1+\frac{m^{2}}{r^{2}k_{z}^{2}}\right)+\upnu\left[\hat{\nabla}^{2}-\frac{1}{r^{2}}+\frac{m^{2}}{rk_{z}^{2}}\hat{\nabla}^{2}\left(\frac{1}{r}\right)\right].
\end{equation}
In this eigenvalue problem, we use the Chebyshev spectral method for a discretization with collocation points in the radial direction \citep[][]{Antkowiak2005,Park2012} to obtain a complete eigenvalue spectrum from the eigenvalue problem (\ref{eq:eigenvalue_problem}). 

{In the inviscid limit $\upnu=0$, equations (\ref{eq:continuity_mode})-(\ref{eq:momentum_vertical_mode}) can be further simplified into a single second-order ordinary differential equation for $\hat{u}$:
\begin{equation}
\label{eq:2ODE_u}
\begin{aligned}
\frac{\mathrm{d}^{2}\hat{u}}{\mathrm{d}r^{2}}&+\left(\frac{1}{r}-\frac{Q'}{Q}\right)\frac{\mathrm{d}\hat{u}}{\mathrm{d}r}\\
&+\left[-k_{z}^{2}\Delta-\frac{m^{2}}{r^{2}}-\frac{mrQ}{s}\left(\frac{f+\zeta}{r^{2}Q}\right)'+Q\left(\frac{1}{rQ}\right)'\right]\hat{u}=0,
\end{aligned}
\end{equation}
where prime denotes the radial derivative, $Q$ and $\Delta$ are the functions defined as
\begin{equation}
\label{eq:Q}
Q(r)=\frac{m^{2}}{r^{2}}+k_{z}^{2},~~
\Delta(r)=1-\frac{\Phi}{s^{2}}.
\end{equation}
In solving the 2nd-order ODE (\ref{eq:2ODE_u}), we consider the boundary conditions in the two limits $r\rightarrow\infty$ and $r\rightarrow0$ with analytic asymptotic solutions.
For instance, we can find an asymptotic solution for (\ref{eq:2ODE_u}) in the limit $r\rightarrow\infty$ as
\begin{equation}
\label{eq:2ODE_infinity_solution}
\hat{u}(r)\sim A_{1}\mathrm{H}_{0}^{(1)}\left(k_{r}r\right),
\end{equation}
where $A_{1}$ is the constant amplitude, $\mathrm{H}_{0}^{(1)}$ is the Hankel function of the first kind and $k_{r}$ is the radial wavenumber of the solution in the far field defined as
\begin{equation}
k_{r}=k_{z}\sqrt{\frac{f^{2}}{\omega^{2}}-1}.
\end{equation} 
The }solution (\ref{eq:2ODE_infinity_solution}) in the limit $r\rightarrow\infty$ behaves asymptotically {as an inertial wave} solution with an exponential decay if $|f|>|\omega |$ while the solution is evanescent if $|f|<|\omega|$. 
For the solution around the center $r=0$, we apply the Taylor expansion on (\ref{eq:continuity_mode})-(\ref{eq:momentum_vertical_mode}) to have the following asymptotic solutions:
\begin{equation}
\label{eq:solution1_center}
\hat{u}\sim O(r),~~
\hat{v}\sim O(r),~~
\hat{w}\sim O(1),~~
\hat{p}\sim O(1),
\end{equation} 
if $m=0$, or
\begin{equation}
\label{eq:solution2_center}
\hat{u}\sim O(r^{|m|-1}),~~
\hat{v}\sim O(r^{|m|-1}),~~
\hat{w}\sim O(r^{|m|}),~~
\hat{p}\sim O(r^{|m|}),
\end{equation} 
if $|m|\geq1$ \citep[see also,][]{Saffman1992}. 

\section{Stability of the convective column}
\label{sec:Stability}
This section will focus on analyzing the stability of the convective column model (\ref{eq:base_Dandoy}) that is prone to centrifugal instability.
The stability analysis is the first step to understand how a convective Taylor column responds to external perturbations.  
Using numerical and asymptotic methods, we compute the dispersion relation (i.e. the expression of the growth rate and frequency in terms of the wavenumbers) of the modes of the convective column. 
Such analyses can identify the most unstable mode, which is most likely to be observed in the interaction process between the convective column and inertial waves, as well as its characteristic temporal and length scales. 
In this section, we will consider the inviscid case where $\upnu=0$. This allows us to perform an asymptotic analysis with the WKBJ method for the centrifugal instability in the large-$k_{z}$ limit.
Here, we present the stability analysis results for our convective column model and the asymptotic expressions of the dispersion relations for neutral and unstable modes. 
The details of the WKBJ method are presented in Appendix \ref{sec:WKBJ}.
We also note that the viscosity stabilizes modes of the convective column, but it is expected that the dynamics of unstable or stable modes will not change significantly in the presence of the viscosity, which is weak in the interiors of stars and giant planets.
The linear evolution of neutral or unstable modes will be discussed in the later part of the section.

\subsection{Linear stability analysis results}
   \begin{figure*}
   \centering
   \includegraphics[height=4.8cm]{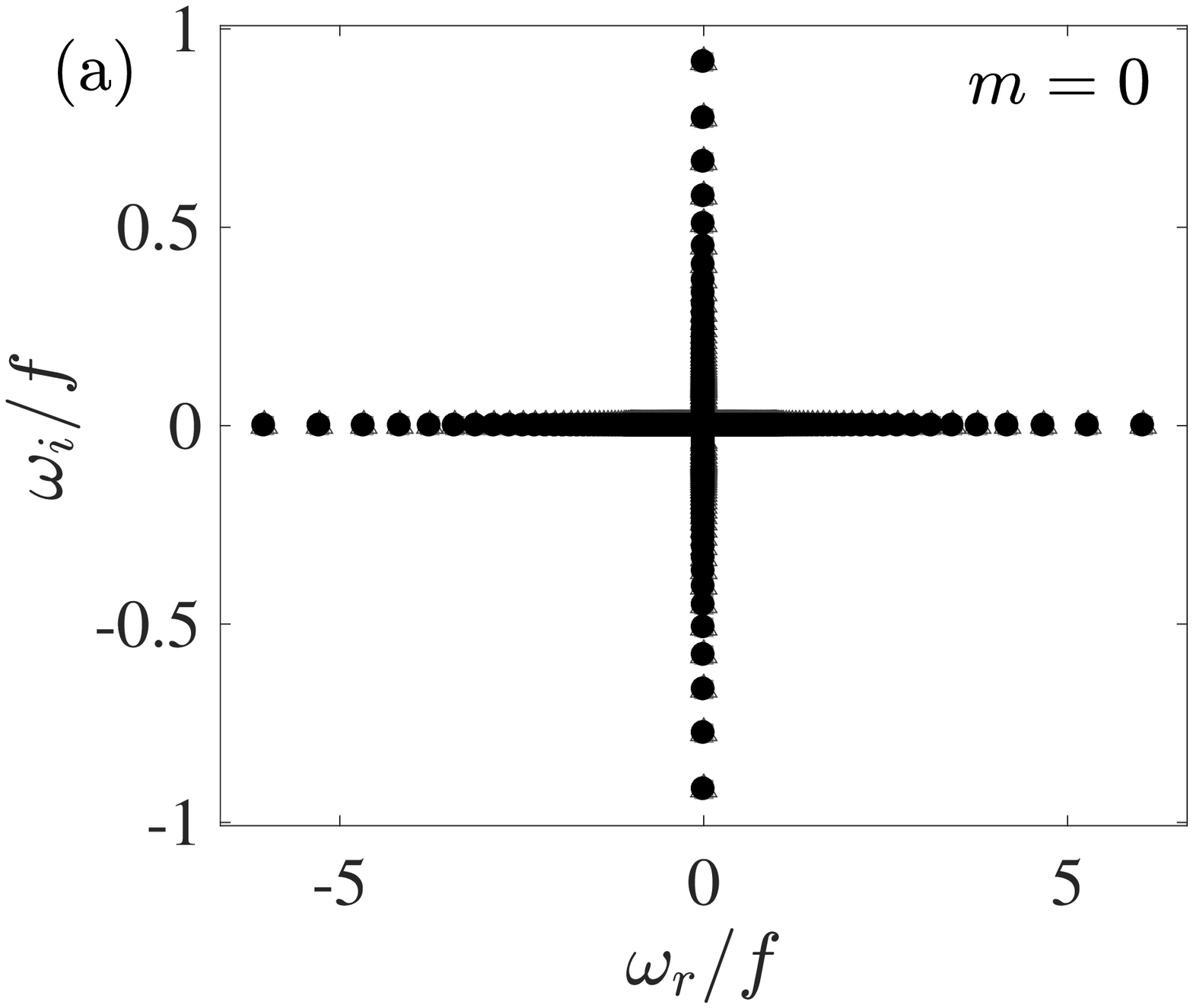}
   \includegraphics[height=4.8cm]{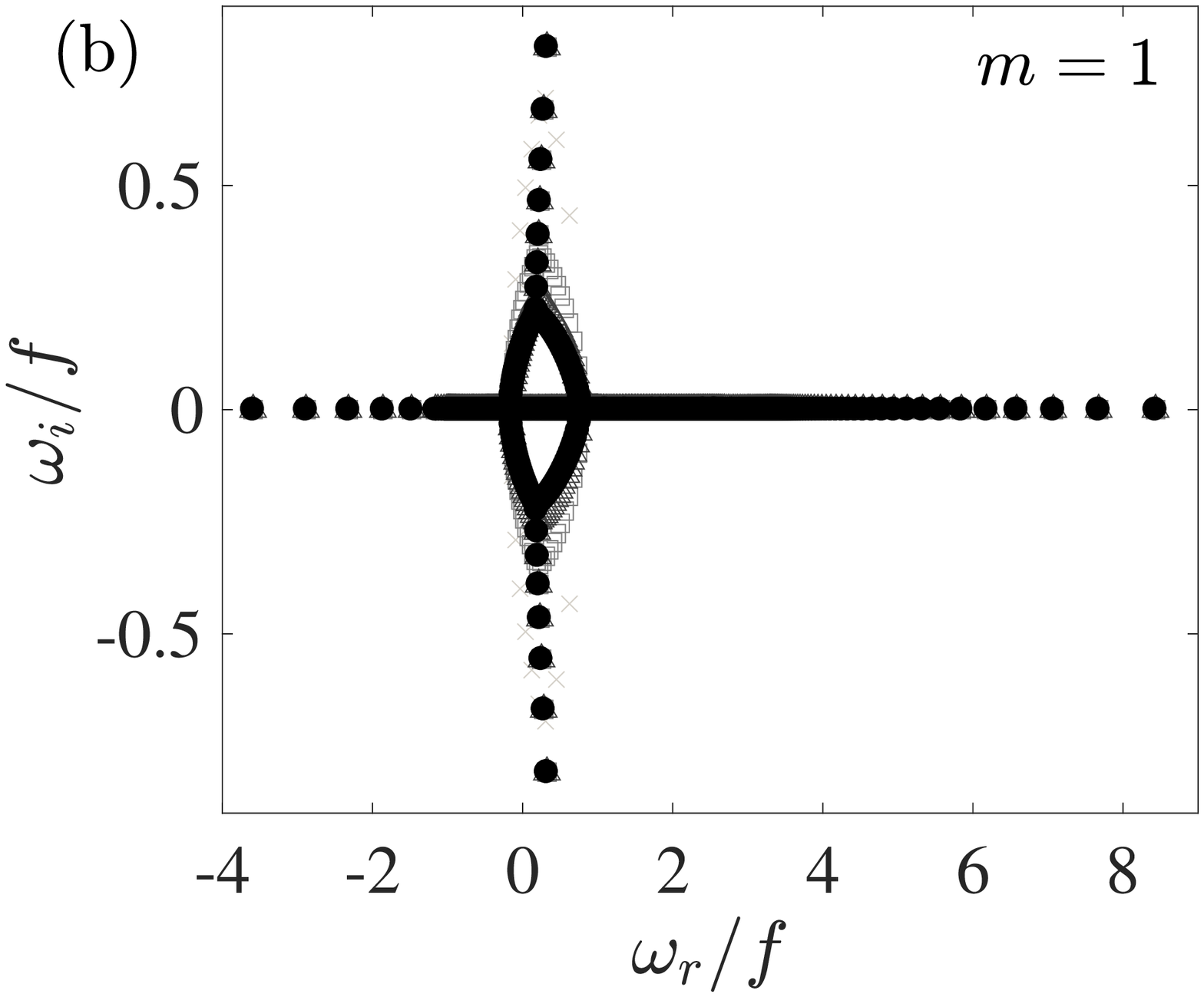}
   \includegraphics[height=4.8cm]{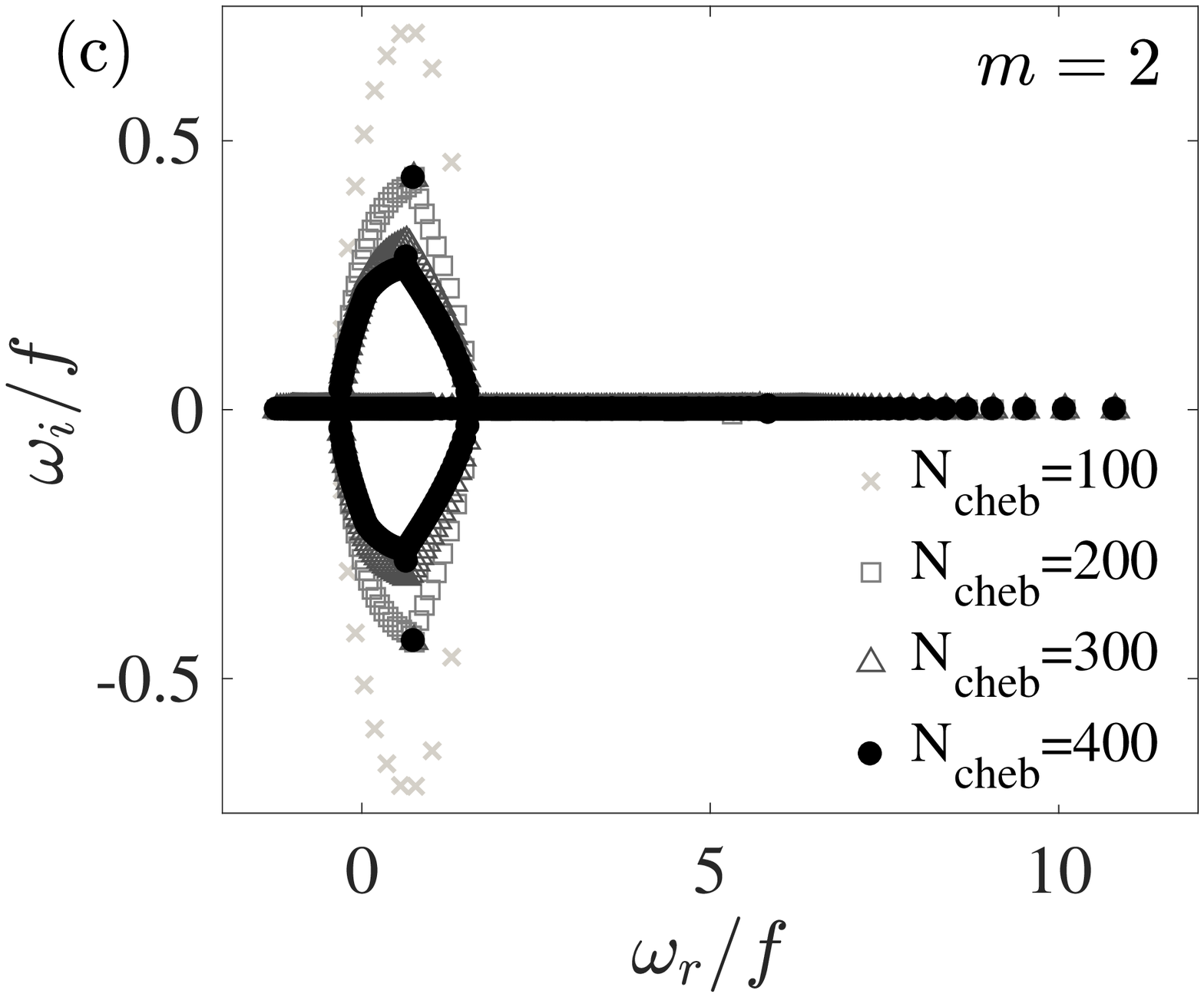}
     \caption{Eigenvalue spectra for an unstable case at $\Omega_{0}/f=3$ and $k_{z}R_{0}=10$ for (a) $m=0$, (b) $m=1$, and (c) $m=2$. Different numbers of the Chebyshev collocation points $\mathrm{N}_{\mathrm{cheb}}$ are tested for convergence. 
              }
         \label{Fig_eigenvalue_spectra}
   \end{figure*}
Figure \ref{Fig_eigenvalue_spectra} displays spectra of the eigenvalues $\omega=\omega_{r}+\mathrm{i}\omega_{i}$ at $\Omega_{0}/f=3$ and $k_{z}R_{0}=10$ for various azimuthal wavenumbers $m$. 
The $m=\{0,1,2\}$ modes correspond to the eccentricity, obliquity, and asynchronous tides, respectively. 
Dynamical tides with $m>0$ are particularly interesting to see how they {can potentially} destabilize a convective column leading to turbulence and dissipation. 
For the axisymmetric case ($m=0$), eigenvalues are symmetrically distributed with respect to the origin and we have neutral modes with the zero growth rate (i.e., $\omega_{i}=0$) in the frequency range $|\omega_{r}/f|<6$, and stable/unstable modes with the zero frequency (i.e., $\omega_{r}=0$) in the growth-rate range $|\omega_{i}/f|<1$.
These eigenvalues are numerically well converged at a relatively low resolution $N_{\mathrm{cheb}}=100$. 
In Fig.~\ref{Fig_eigenvalue_spectra}b, we see that eigenvalues for $m=1$ are no longer symmetric with respect to the line $\omega_{r}=0$ and we have neutral modes in the frequency range $-4<\omega_{r}/f<8.5$ and stable/unstable modes in the growth-rate range $|\omega_{i}/f|<0.85$ with a non-zero frequency $\omega_{r}/f\neq0$.
While some of these modes are well converged at a resolution $N_{\mathrm{cheb}}\geq200$, some modes with a non-zero growth rate in the frequency range $-0.2<\omega_{r}/f<0.8$ are not well converged even at a high resolution $N_{\mathrm{cheb}}=400$. 
We verified that these unconverged modes possess singular points (e.g., the critical layers at which the incident wave exchanges its angular momentum with the convective column) and much higher resolutions are required to resolve these modes correctly \citep[see also the numerical treatment by][]{Astoul2021}. 
In this paper, we will not investigate the dynamics of these singular modes but we will focus on converged modes that have high frequency or growth rate. 
The spectral characteristics for $m=2$ shown in Fig.~\ref{Fig_eigenvalue_spectra}c are similar to those for $m=1$, but we only observed two unstable modes that are numerically converged at $N_{\mathrm{cheb}}=400$.
For the parameters $\Omega_{0}/f=3$ and $k_{z}R_{0}=10$ considered in Fig.~\ref{Fig_eigenvalue_spectra}, we found that the growth rate of the most unstable mode decreases with $m$ and no unstable mode is observed after $m\geq3$.  

%
   \begin{figure}
   \centering
   \includegraphics[width=7cm]{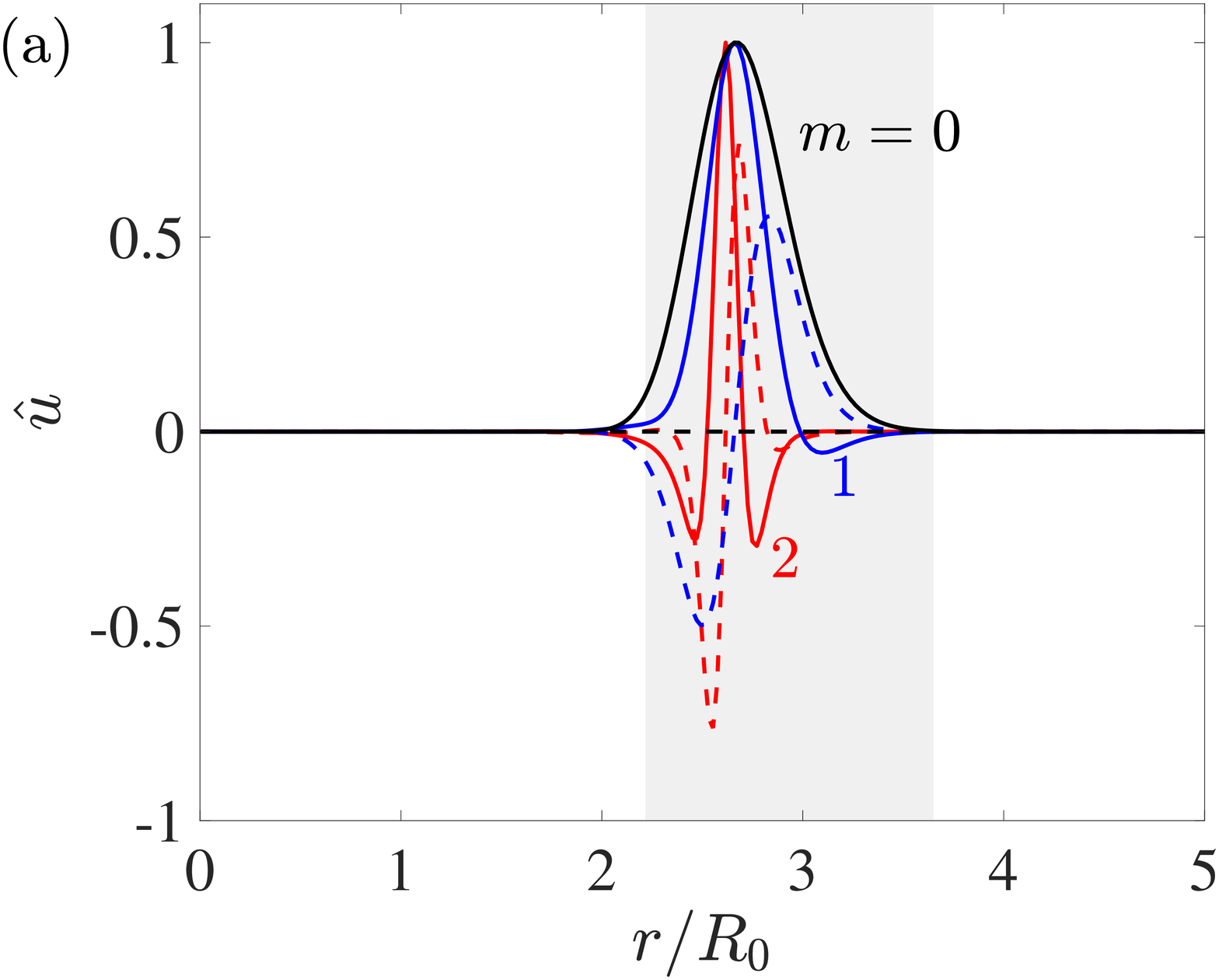}
      \includegraphics[width=7cm]{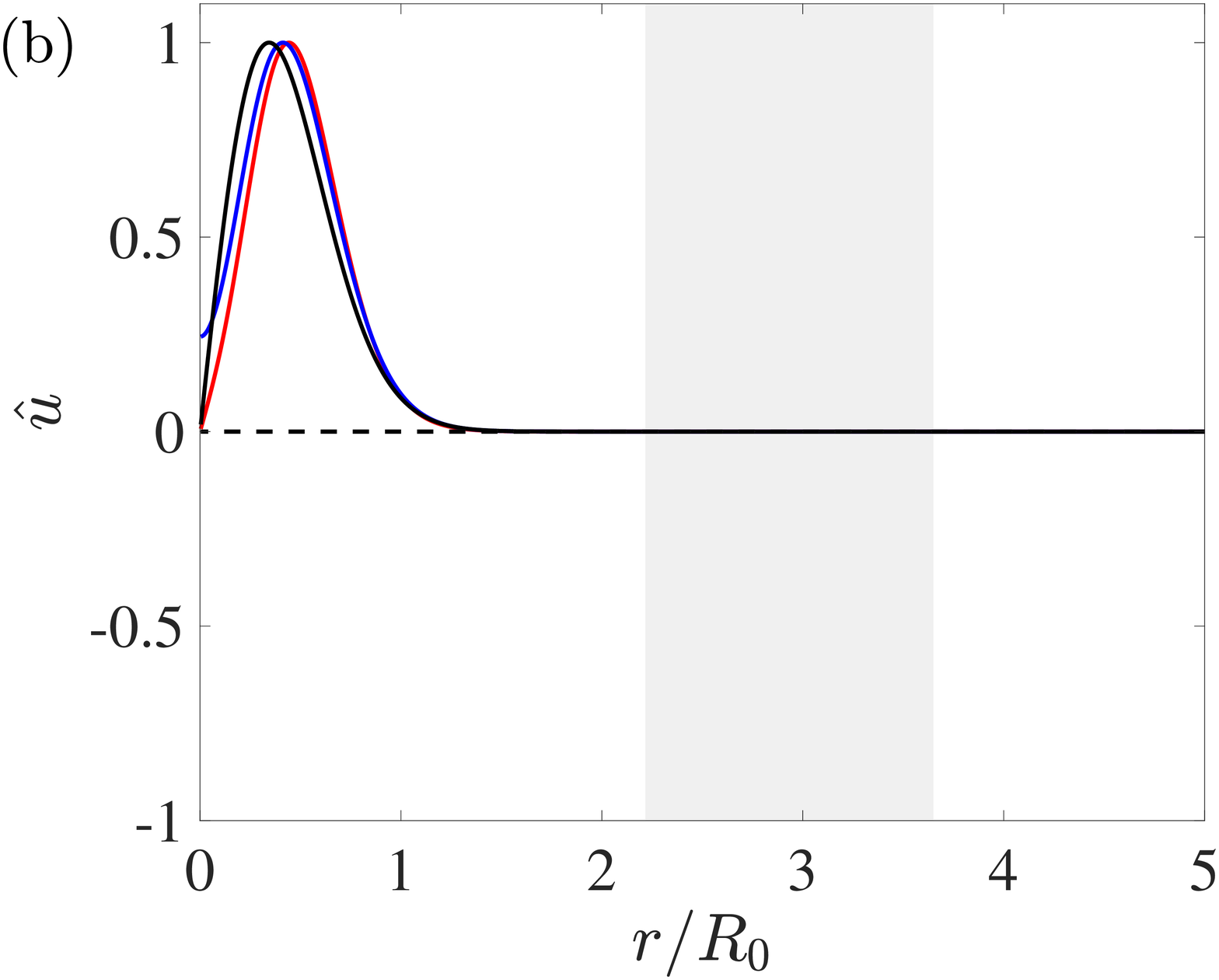}
     \caption{Real (solid) and imaginary (dashed) parts of (a) the most unstable mode and (b) the first neutral mode with the highest frequency at $\Omega_{0}/f=3$ and $k_{z}R_{0}=10$ for different azimuthal wavenumbers: $m=0$ (black), $m=1$ (blue), and $m=2$ (red). The gray-shaded region denotes where the Rayleigh discriminant $\Phi$ is negative. 
              }
         \label{Fig_eigenmodes}
   \end{figure}
Figure~\ref{Fig_eigenmodes} displays the mode shape $\hat{u}(r)$ of the most unstable modes and the first neutral modes with the highest frequency for various azimuthal wavenumbers $m$, which are picked up in the eigenvalue spectra in Fig.~\ref{Fig_eigenvalue_spectra}.
We also plot the gray-shaded area to demonstrate where the Rayleigh discriminant $\Phi(r)$ is negative. 
The unstable modes have an oscillatory shape inside the region $\Phi<0$ while they are evanescent such that they increase exponentially from $r=0$ and decrease exponentially as $r\rightarrow\infty$.
It is also found that the number of zero crossings for the unstable modes increases with $m$.
The neutral modes, on the other hand, have different features; for instance, they have a peak around $r\simeq0.5$ below the region $\Phi<0$ and decrease exponentially as $r\rightarrow\infty$.
From these solution behaviors, we can expect that the wavelike behavior is observed in the region $\Phi<0$ when it is centrifugally unstable. 
According to the definition by \citet{Ledizes2005}, a mode that has a wavelike behavior far from the core $r=0$ is called the ring mode, while a mode with the wavelike behavior around $r=0$ is called the core mode. 
Therefore, we can distinguish that the unstable modes in Fig.~\ref{Fig_eigenmodes}a are the ring mode while the neutral modes in Fig.~\ref{Fig_eigenmodes}b are the core mode. 

%
   \begin{figure*}
   \centering
     \includegraphics[height=3.7cm]{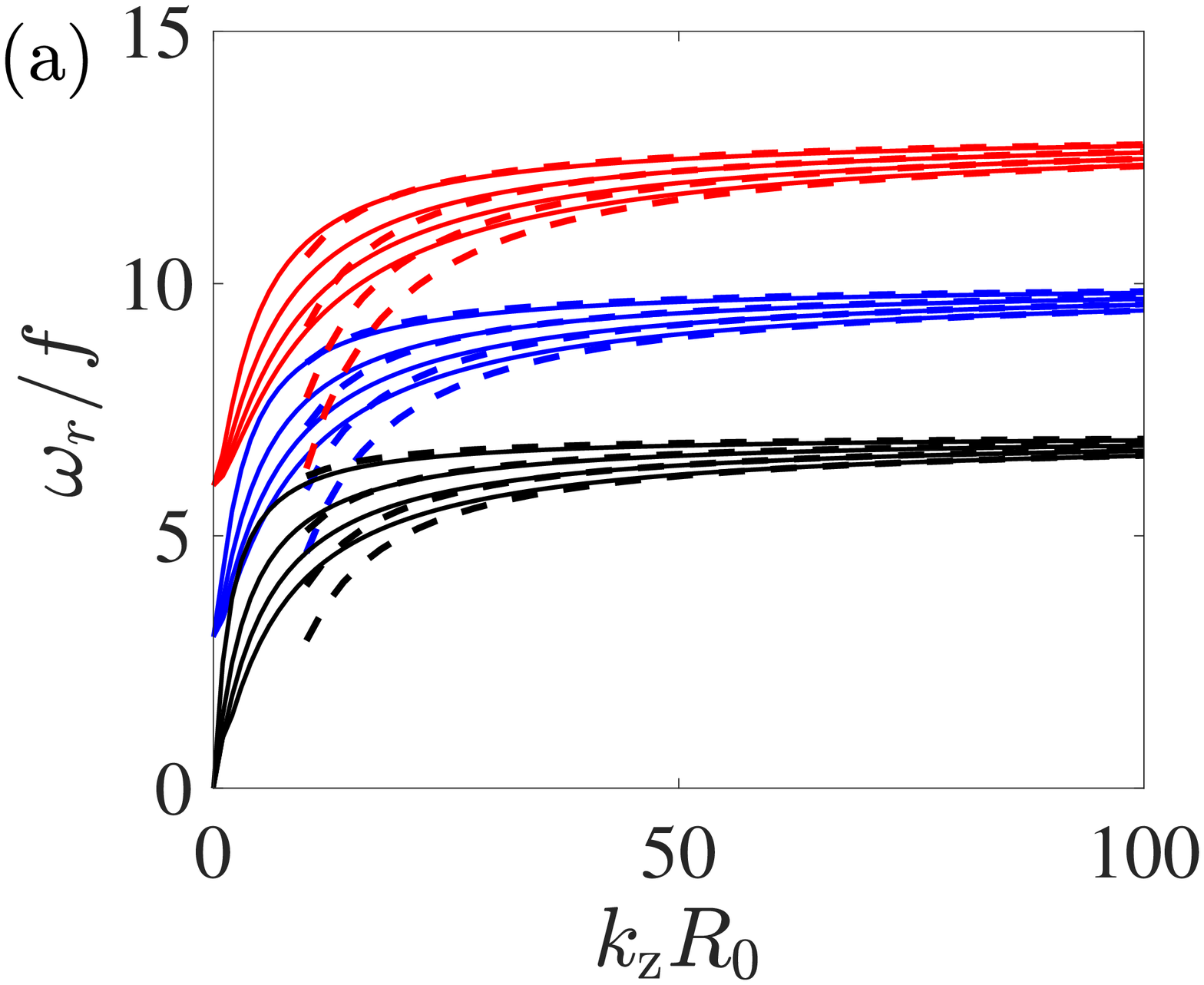}
     \includegraphics[height=3.7cm]{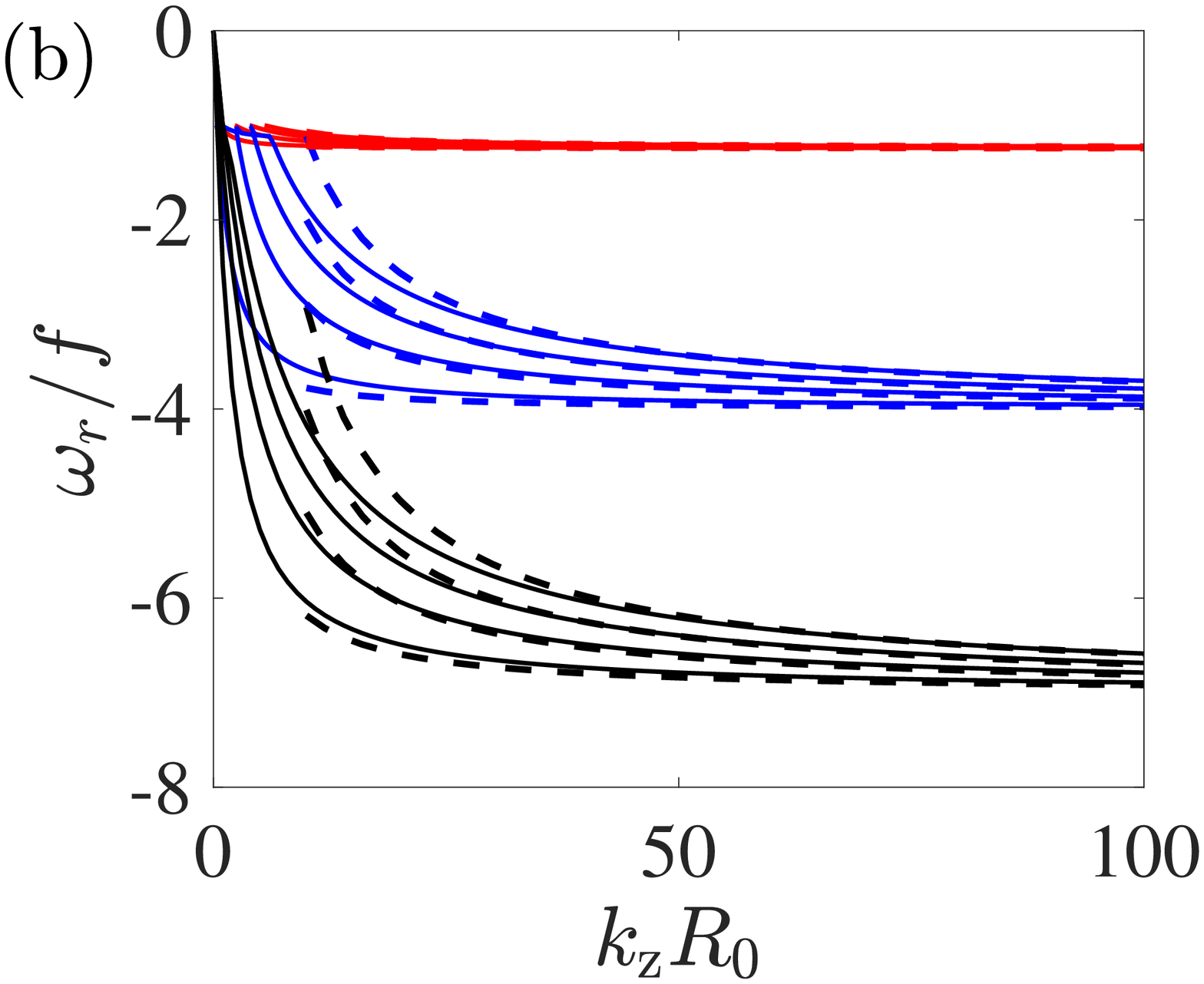}
     \includegraphics[height=3.7cm]{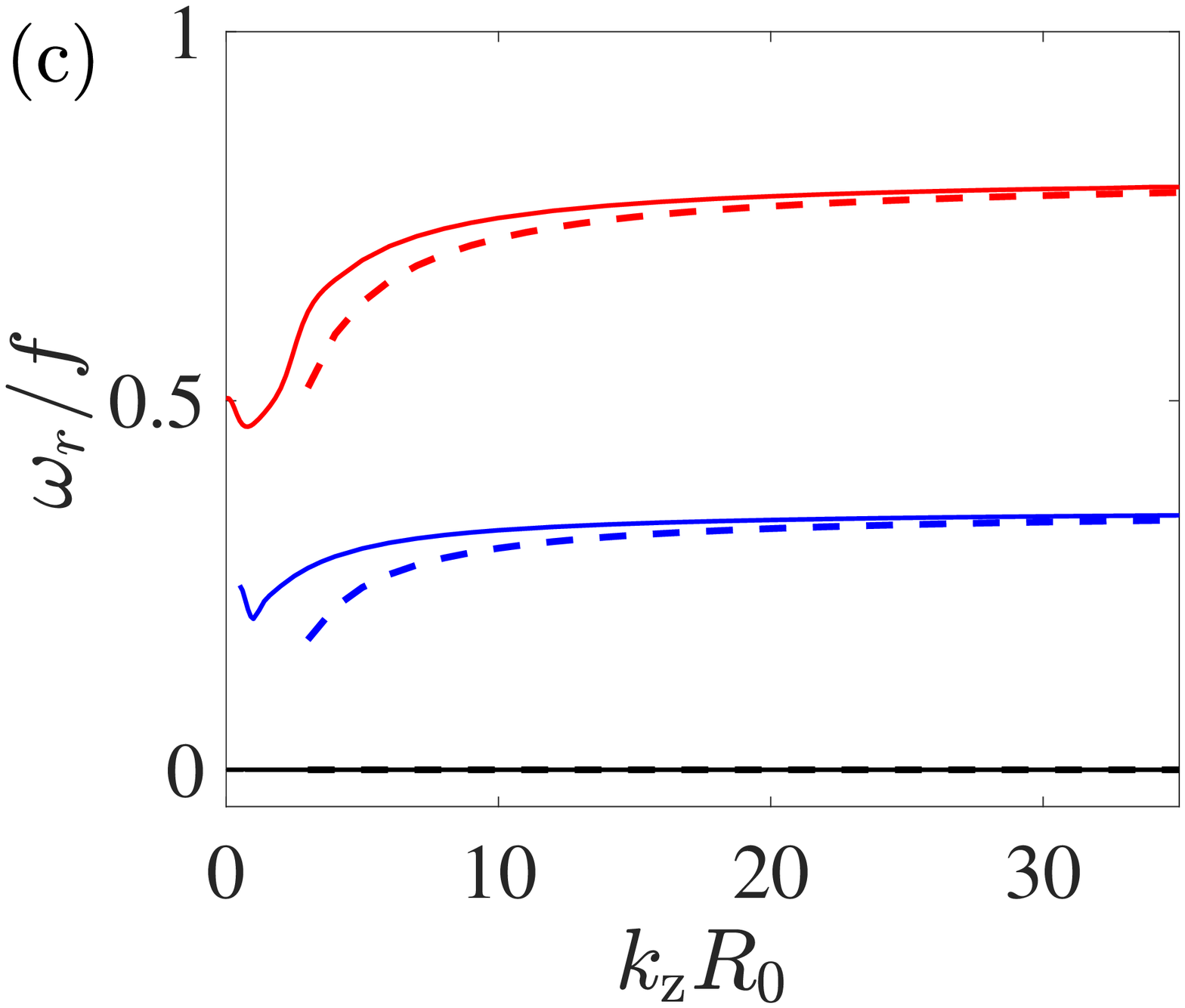}
     \includegraphics[height=3.7cm]{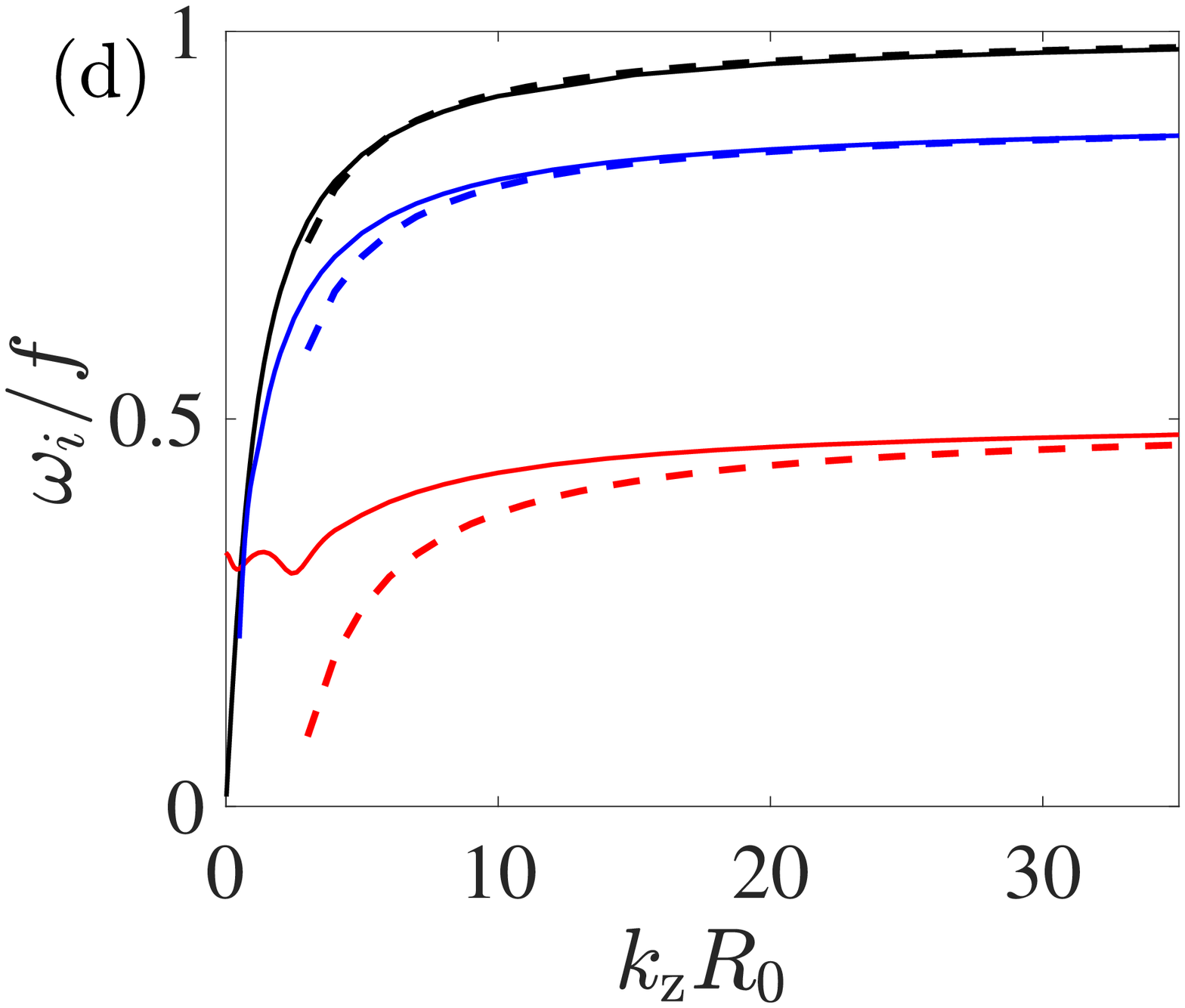}
     \caption{Panels a and b: frequency $\omega_{r}$ as a function of the vertical wavenumber $k_{z}$ of the neutral modes for (a) the four upper branches and (b) the four lower branches at $\Omega_{0}/f=3$ for $m=0$ (black), $m=1$ (blue), and $m=2$ (red). 
     Panels c and d: (c) frequency $\omega_{r}$ and (d) growth rate $\omega_{i}$ versus vertical wavenumber $k_{z}$ for the most unstable mode at $\Omega_{0}/f=3$ for $m=0$ (black), $m=1$ (blue), and $m=2$ (red).
          For all panels, solid lines are numerical results and dashed lines denote the WKBJ predictions for panel a: Eq.~(\ref{eq:dispersion_neutral_modes_core_upper}), panel b: Eqs.~(\ref{eq:dispersion_neutral_modes_core_lower}) for $m=0,1$ (core mode) and (\ref{eq:frequency_neutral_modes_ring_lower}) for $m=2$ (ring mode), and panels c,d: Eq.~(\ref{eq:dispersion_relation_complex_w}).
              }
         \label{Fig_eigenvalues_k}
   \end{figure*}
Figure \ref{Fig_eigenvalues_k} shows how eigenvalues change with the vertical wavenumber $k_{z}$ for neutral and unstable modes at $\Omega_{0}/f=3$.
For the neutral modes, there are countless branches depending on the number of zero-crossings but we only display the first four upper branches in Fig.~\ref{Fig_eigenvalues_k}(a,b) that have the highest frequencies and the four lower branches with the lowest frequencies. 
In Fig.~\ref{Fig_eigenvalues_k}a, we see that the frequency of the upper branches begins at $\omega_{r}=m\Omega_{0}$ at $k_{z}=0$, increases with $k_{z}$ and reaches an asymptote as $k_{z}\rightarrow\infty$. 
The asymptotic behaviors in the limit $k_{z}\rightarrow\infty$ can be understood by applying the Wentzel-Krammers-Brillouin-Jeffreys (WKBJ) approximation \citep[][]{Froman1965}.
Referring the readers to Appendix for detailed technical details at intermediate steps, we provide here the final analytic expressions of the dispersion relation for neutral and unstable modes obtained by using the WKBJ approximation.
For upper branches, we found the expressions of the dispersion relation for the neutral modes in the form of Taylor expansion with $1/k_{z}$ as
\begin{equation}
    \label{eq:dispersion_neutral_modes_core_upper}
\begin{aligned}
    &\omega=m\Omega_{0}+\sqrt{\Phi_{0}}-\frac{2}{k_{z}}\left(n-\frac{1}{4}\right) \sqrt{-\frac{\Phi_0''}{2} - m\Omega_0'' \sqrt{\Phi_0}},&\mathrm{if}~m\neq1,\\
    &\omega=m\Omega_{0}+\sqrt{\Phi_{0}}-\frac{2}{k_{z}}\left(n+\frac{1}{4}\right) \sqrt{-\frac{\Phi_0''}{2} - m\Omega_0'' \sqrt{\Phi_0}},&\mathrm{if}~m=1,\\
\end{aligned}
\end{equation}
where  $n$ denotes the branch number and the subscript 0 implies that the variables are evaluated at $r=0$. 
We see a good agreement between the numerical results and asymptotic predictions on the frequency of the upper branches, especially as $k_{z}\rightarrow0$. 
The frequency of the lower branches, on the other hand, decreases as $k_{z}$ increases and reaches an asymptote as $k_{z}\rightarrow\infty$.
From the WKBJ analysis, we found the analytical expressions of the dispersion relation for the lower branches as
\begin{equation}
    \label{eq:dispersion_neutral_modes_core_lower}
\begin{aligned}
    &\omega=m\Omega_{0}-\sqrt{\Phi_{0}}+\frac{2}{k_{z}}\left(n-\frac{1}{4}\right) \sqrt{-\frac{\Phi_0''}{2} + m\Omega_0'' \sqrt{\Phi_0}},&\mathrm{if}~m\neq1,\\
    &\omega=m\Omega_{0}-\sqrt{\Phi_{0}}+\frac{2}{k_{z}}\left(n+\frac{1}{4}\right) \sqrt{-\frac{\Phi_0''}{2} + m\Omega_0'' \sqrt{\Phi_0}},&\mathrm{if}~m=1.\\
\end{aligned}
\end{equation}
A good agreement is also obtained between the numerical and asymptotic results for the lower branches, as shown in Fig.~\ref{Fig_eigenvalues_k}(b).

In Fig.~\ref{Fig_eigenvalues_k}(c,d), we plot the frequency $\omega_{r}$ and growth rate $\omega_{i}$ of the most unstable mode versus $k_{z}$.
The axisymmetric mode ($m=0$) has zero frequency and its growth rate $\omega_{i}$ increases with $k_{z}$.
The frequency and growth rate of non-axisymmetric modes ($m\geq1$) increase with $k_{z}$ and they approach their asymptotes as $k_{z}\rightarrow\infty$.
From the WKBJ analysis, we found the following Taylor expansion of the complex frequency $\omega$:
\begin{equation}
\label{eq:dispersion_relation_complex_w}
\omega=\omega_{0}+\frac{\omega_{1}}{k_{z}}+O\left(\frac{1}{k_{z}^{2}}\right),
\end{equation}
where
\begin{equation}
\label{eq:dispersion_relation_w0}
\omega_{0}=m\Omega(r_{0})+\mathrm{i}\sqrt{-\Phi(r_{0})},
\end{equation}
\begin{equation}
\label{eq:dispersion_relation_w1}
\omega_{1}=\left.\frac{2n+1}{2\sqrt{2}\mathrm{i}}\sqrt{\Phi''-2m^{2}\Omega^{'2}+2\mathrm{i}m\Omega''\sqrt{-\Phi}}~\right|_{r=r_{0}},
\end{equation}
and $r_{0}$ is a double turning point where the radial derivative of $m\Omega(r)+\mathrm{i}\sqrt{-\Phi(r)}$ becomes zero \citep[for more details, see Appendix and][]{Billant2005}.
It is clearly shown in Fig.~\ref{Fig_eigenvalues_k} (c,d) that the numerical and asymptotic results are in good agreement. 
We verified numerically that there is no longer instability for higher azimuthal wavenumber $m\geq3$ for the convective column with the angular velocity profile given in Eq.~(\ref{eq:base_Dandoy}).
Furthermore, we found an instability at $k_{z}=0$ only for $m=2$. 
This zero-vertical-wavenumber instability is reminiscent of the vortex shear instability reported for other vortex profiles \citep[see e.g.,][]{Billant2005}.


\subsection{Linear evolution of unstable modes}
\label{sec:linear_cylindrical}
In addition to the stability analysis, we also investigate how inertial waves entering from the far field interact with a convective column. 
To understand the interaction, the first step is to perform linear simulations with a tidally-forced inertial wave incoming from the far field.
Considering the cylindrical coordinates, the mathematical formulation is similar to that of the stability analysis when the following ansatz is considered:
\begin{equation}
\label{eq:mode_shapes_time}
\left(
\begin{array}{c}
\acute{u}\\
\acute{v}\\
\acute{w}\\
\acute{\pi}
\end{array}
\right)=
\left(
\begin{array}{c}
\tilde{u}(r,t)\\
\tilde{v}(r,t)\\
\tilde{w}(r,t)\\
\tilde{p}(r,t)
\end{array}
\right)\exp\left[\mathrm{i}(k_{z}z+m\theta)\right]+c.c.,
\end{equation}
where $\tilde{\vec{u}}=(\tilde{u},\tilde{v},\tilde{w})$ and $\tilde{p}$ are the time-dependent mode shapes of velocity and normalised pressure. 
Applying the ansatz (\ref{eq:mode_shapes_time}) to Eqs.~(\ref{eq:continuity_ptb})-(\ref{eq:momentum_vertical_ptb}) leads to the following linear time-evolution equation
\begin{equation}
\label{eq:linear_evolution_equation}
    \mathcal{A}\frac{\partial\tilde{\textbf{q}}}{\partial t}=\mathcal{B}\tilde{\textbf{q}},
\end{equation}
where $\tilde{\textbf{q}}=(\tilde{u},\tilde{v})$ and $\mathcal{A}$ and $\mathcal{B}$ are the same operator matrices as in the eigenvalue problem (\ref{eq:eigenvalue_problem}).
To impose the boundary conditions around the center $r=0$, we consider the conditions (\ref{eq:solution1_center})-(\ref{eq:solution2_center}) and use the proper differentiation matrices to match the boundary conditions \citep[for details, we refer to][]{Antkowiak2005}.
To impose the boundary conditions in the far field, we consider the time-periodic solution as
\begin{equation}
\tilde{u}= A_{\infty}\exp\left(-\mathrm{i}\omega_{f}t\right),    
\end{equation}
where $A_{\infty}$ is the constant amplitude and $\omega_{f}$ is the forcing frequency.
For the time-marching of Eq.~(\ref{eq:linear_evolution_equation}), we use the implicit Euler method.
While the stability analysis uses the Chebyshev spectral method for discretization in the radial direction $r$ to facilitate the imposition of boundary conditions \citep[][]{Park2012}, we use the 4th-order finite difference method for the radial discretization to have a uniform spacing.
This allows us to have more collocation points in the far field and thus we can properly resolve the incoming inertial waves.\\ 

\begin{figure*}
   \centering
   \includegraphics[height=4.7cm]{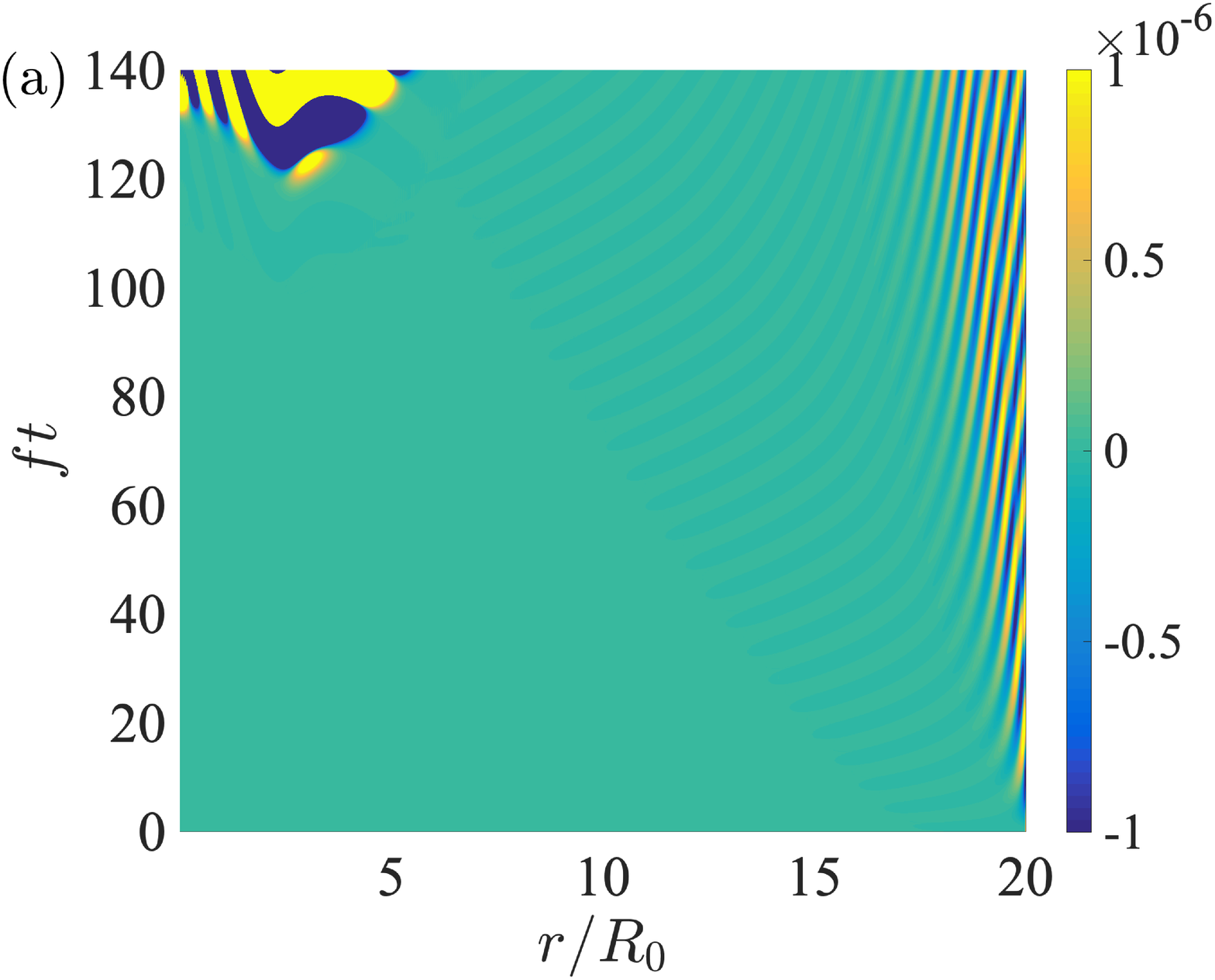}
   \includegraphics[height=4.7cm]{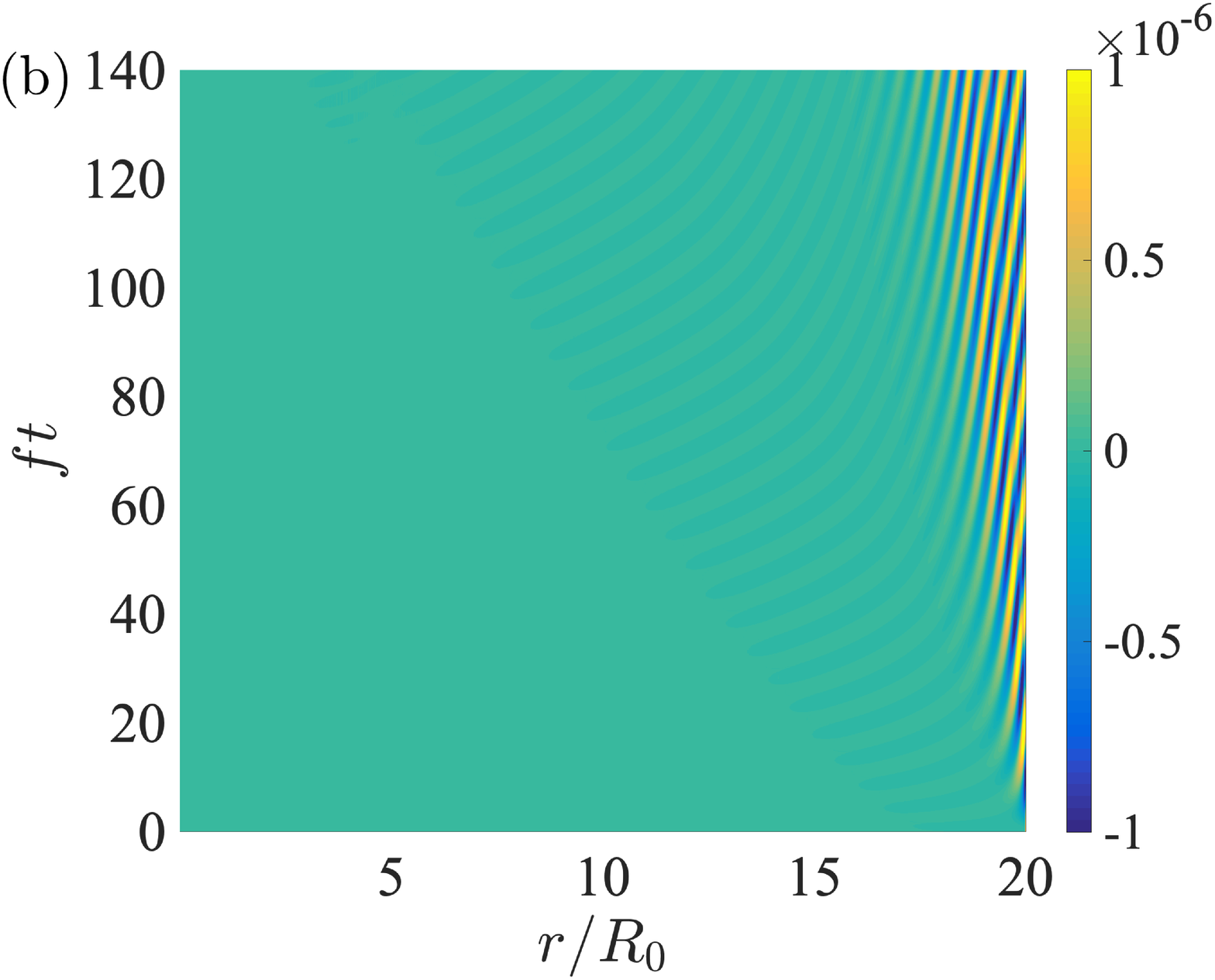}
   \includegraphics[height=4.7cm]{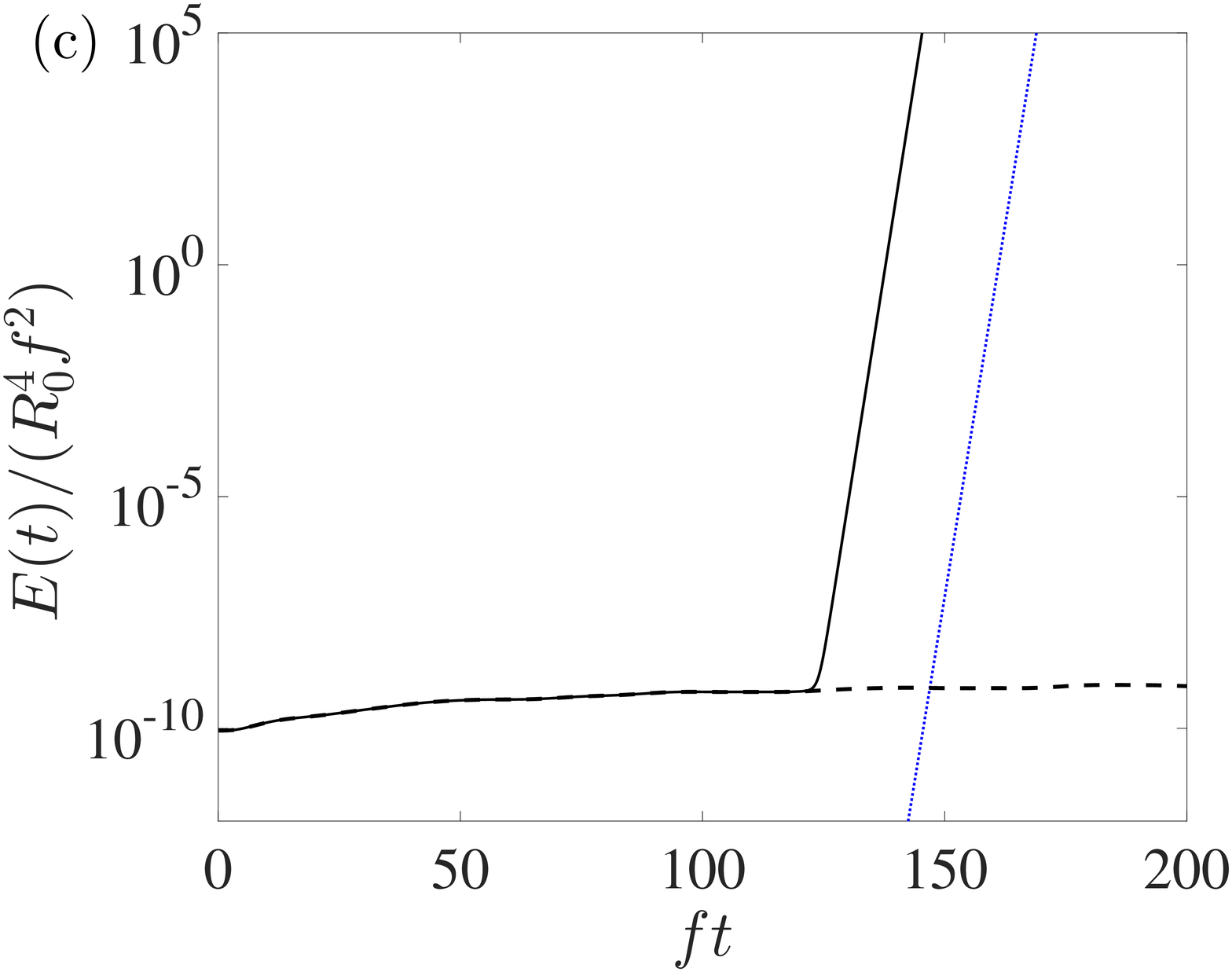}
     \caption{Panels a and b: Spatio-temporal diagrams in the space $(r,t)$ for inertial waves entering from $r/R_{0}=20$ at the forcing frequency $\omega_{f}=0.3$, $m=1$, $k_{z}R_{0}=5$, $\upnu=0$ for (a) an unstable column with $\Omega_{0}/f=3$ and (b) a stable column with $\Omega_{0}/f=0.5$. 
     Panel c: Time evolution of wave energy $E(t)$ for the unstable case (a) (black solid line) and stable case (b) (black dashed line). The blue line denotes the energy growth $C\exp(2\omega_{i} t)$ with a constant $C$ chosen for comparison with the unstable case. 
              }
         \label{Fig_waves}
   \end{figure*}
Panels a and b in Fig.~\ref{Fig_waves} display the spatio-temporal diagrams of the real part of the radial velocity $\mathrm{Re}\left[\tilde{u}_{r}(r,t)\right]$ as the inertial wave is forced at $r=20R_{0}$ with the amplitude $A_{\infty}=10^{-6}$, frequency $\omega_{f}=0.3f$, wavenumbers $(k_{z}R_{0},m)=(5,1)$ for two base flows: unstable and stable convective columns with $\Omega_{0}=3f$ and $\Omega_{0}=0.5f$, respectively.
On the one hand, the unstable case in Fig.~\ref{Fig_waves}a shows that the tidally-forced wave propagates inwards in the radial direction. 
This incoming cylindrical wave has a characteristic such that the front has a longer wavelength and a smaller amplitude than the rear part.
Although the front of the wave has a very small amplitude, it can still trigger the most unstable mode, which grows the fastest in time.  
If the nonlinearity is taken into account in this unstable case, strong interaction with the convective column can be induced and turbulence can be triggered as a consequence. 
On the other hand, we see in Fig.~\ref{Fig_waves}b for the stable case that the tidally-forced inertial wave propagates inwards but there is no sign of the strong interaction between the wave and the convective column. 
At a much later time $ft>140$, we verified indeed that no unstable mode is triggered for the stable case.  

To quantify the difference between the two {regimes}, we define the energy of perturbations:
\begin{equation}
\label{eq:energy_perturbations}
    E(t)=\int_{0}^{\infty}\left[|\tilde{u}_{r}|^{2}+|\tilde{u}_{\theta}|^{2}+|\tilde{u}_{z}|^{2}\right]r\mathrm{d}r.
\end{equation}
It is clearly shown in Fig.~\ref{Fig_waves}c that the perturbation energy for the unstable case increases significantly after $ft>120$ and grows exponentially in time, while the energy for the tidal inertial wave interacting with the stable convective column does not increase significantly in time.   
We found that the sudden increase of the perturbation energy for the unstable case corresponds to the emergence of the most unstable mode with the growth rate $\omega_{i,\max}$. 
This is verified by comparing the slope of the two curves: the unstable perturbation growth in black solid line and the exponential growth $\exp(2\omega_{i,\max})t$ of the most unstable mode in blue line. 
In comparing these two cases, we must consider the exponent $2\omega_{i,\max}$ which can be easily obtained by putting $\tilde{u}_{r}(r,t)=\hat{u}_{r}\exp(-\mathrm{i}\omega t)$ into the energy equation (\ref{eq:energy_perturbations}).
It is not shown here but we also verified that a similar appearance of the unstable mode occurs when the incoming inertial waves with different forcing frequencies are considered for unstable convective columns.

These results are coherent with what is expected from the stability analysis. 
Once the instability occurs and triggers turbulence, the vortex will be destroyed and turbulent dissipation will follow as a consequence 
\citep[see for instance the work by][for the case of the precession instability-driven turbulence]{Pizzi2022}. 
It is important here to underline that the trigger of unstable modes is not necessarily from tidal inertial waves but any low-amplitude noise is likely to cause the instability leading to turbulence.
Especially for the case where the tidal period scaled as $|\omega_{f}|^{-1}$ is larger than the characteristic time of the instability scaled as $|\omega_{i}|^{-1}$ (i.e., the case where $|\omega_{f}|<|\omega_{i}|$), one can expect that the instability develops fast before the incoming tidal wave starts to interact with the vortex, thus other types of low-amplitude perturbations would be the main trigger of the vortex instability.
A more important configuration is the one where tidal inertial waves interact with stable convective vortices as the numerous ones observed for instance at the surface of giant planets. 
What is less known is whether stable vortices would interact with the tidal waves. 
The current case only considers cylindrical inertial waves propagating in the radial direction.  
However, we must also study the general case where a large-scale stable vortex in planets or stars interacts with an incoming tidally-forced wave that goes through the center of the vortex as depicted in Fig.~\ref{fig:column} (left panel). To understand this regime, we examine in the following section the interaction between a stable vortex and a tidal inertial wave in the Cartesian coordinate system.

\section{Interaction between a planar inertial wave and a stable convective column}
\label{sec:Interaction}
In the previous section, we examine the interaction between the convective vortex and inertial wave, both of which have a cylindrical nature. 
A problem that draws more attention from astrophysicists working on tidal interaction and inertial-wave propagation \citep[see e.g.][]{Andre2017,Andre2019} is how the tidally-forced inertial wave interacts with complex fluid flow structures. In the case of convective vortices, this motivates us to consider inertial waves propagating in a plane and going through the vortex as described in Fig.~\ref{fig:column} (left panel). 
To investigate this, we formulate the problem in the Cartesian coordinate system and conduct two-dimensional linear simulations.
This allows us {to be able to examine} the vortex interaction with any incoming tidal waves, not necessarily restricted to the 1D cylindrical waves {as in the previous section}. A similar configuration is considered in \citet{McIntyre2019} who examines analytically the wave-vortex interaction by considering a vortex, which has an irrotational velocity field outside the vortex core, and a wave travelling through the vortex center.

We reformulate the base flow by considering the base velocity in the Cartesian coordinates as $\vec{U}=(U_{x},U_{y},0)$ where $U_{x}(x,y)=-r\Omega\sin\theta=-\Omega y$ and $U_{y}(x,y)=r\Omega\cos\theta=\Omega x$ are the base velocity in the $x$- and $y$-directions, respectively. 
Subject to this base flow, we obtain the linearized equations for perturbation velocity $\acute{\vec{u}}=(\acute{u}_{x},\acute{u}_{y},\acute{w})$ and pressure $\acute{\pi}$ as follows:  
\begin{equation}
\label{eq:continuity_cartesian_linear}
\frac{\partial \acute{u}_{x}}{\partial x}+\frac{\partial \acute{u}_{y}}{\partial y}+\frac{\partial \acute{w}}{\partial z}=0,
\end{equation}
\begin{equation}
\label{eq:momentum_x_linear}
\frac{\partial \acute{u}_{x}}{\partial t}+U_{x}\frac{\partial \acute{u}_{x}}{\partial x}+\frac{\partial U_{x}}{\partial x}\acute{u}_{x}+U_{y}\frac{\partial \acute{u}_{x}}{\partial y}+\frac{\partial U_{x}}{\partial y}\acute{u}_{y}-f\acute{u}_{y}=-\frac{\partial \acute{\pi}}{\partial x}+\upnu\nabla^{2}\acute{u}_{x},
\end{equation}
\begin{equation}
\label{eq:momentum_y_linear}
\frac{\partial \acute{u}_{y}}{\partial t}+U_{x}\frac{\partial \acute{u}_{y}}{\partial x}+\frac{\partial U_{y}}{\partial x}\acute{u}_{x}+U_{y}\frac{\partial \acute{u}_{y}}{\partial y}+\frac{\partial U_{y}}{\partial y}\acute{u}_{y}+f\acute{u}_{x}=-\frac{\partial \acute{\pi}}{\partial y}+\upnu\nabla^{2}\acute{u}_{y},
\end{equation}
\begin{equation}
\label{eq:momentum_z_linear}
\frac{\partial \acute{w}}{\partial t}+U_{x}\frac{\partial \acute{w}}{\partial x}+U_{y}\frac{\partial \acute{w}}{\partial y}=-\frac{\partial \acute{\pi}}{\partial z}+\upnu\nabla^{2}\acute{w}.
\end{equation}
As the base flow $\vec{U}(x,y)$ is homogeneous only in the $z$-direction, we consider the following ansatz for the perturbation in Cartesian coordinates:
\begin{equation}
\label{eq:ansatz_Cartesian}
\left(
\begin{array}{c}
\acute{u}_{x}\\
\acute{u}_{y}\\
\acute{w}\\
\acute{\pi}
\end{array}
\right)=
\left(
\begin{array}{c}
\check{u}(x,y,t)\\
\check{v}(x,y,t)\\
\check{w}(x,y,t)\\
\check{p}(x,y,t)
\end{array}
\right)\exp\left(\mathrm{i}k_{z}z\right)+c.c.,
\end{equation}
where $\check{\vec{u}}=(\check{u},\check{v},\check{w})$ and $\check{p}$ are the two-dimensional time-dependent mode shapes of velocity and normalised pressure, respectively.
Applying the ansatz (\ref{eq:ansatz_Cartesian}) to the linearized equations (\ref{eq:continuity_cartesian_linear})-(\ref{eq:momentum_z_linear}) leads to the following equations:
\begin{equation}
\label{eq:continuity_cartesian_linear_modal}
\frac{\partial \check{u}}{\partial x}+\frac{\partial \check{v}}{\partial y}+\mathrm{i}k_{z}\check{w}=0,
\end{equation}
\begin{equation}
\label{eq:momentum_x_linear_modal}
\frac{\partial \check{u}}{\partial t}+U_{x}\frac{\partial \check{u}}{\partial x}+\frac{\partial U_{x}}{\partial x}\check{u}+U_{y}\frac{\partial \check{u}}{\partial y}+\frac{\partial U_{x}}{\partial y}\check{v}-f\check{v}=-\frac{\partial \check{\pi}}{\partial x}+\upnu\check{\nabla}^{2}\check{u},
\end{equation}
\begin{equation}
\label{eq:momentum_y_linear_modal}
\frac{\partial \check{v}}{\partial t}+U_{x}\frac{\partial \check{v}}{\partial x}+\frac{\partial U_{y}}{\partial x}\check{u}+U_{y}\frac{\partial \check{v}}{\partial y}+\frac{\partial U_{y}}{\partial y}\check{v}+f\check{u}=-\frac{\partial \check{\pi}}{\partial y}+\upnu\check{\nabla}^{2}\check{v},
\end{equation}
\begin{equation}
\label{eq:momentum_z_linear_modal}
\frac{\partial \check{w}}{\partial t}+U_{x}\frac{\partial \check{w}}{\partial x}+U_{y}\frac{\partial \check{w}}{\partial y}=-\mathrm{i}k_{z}\check{\pi}+\upnu\check{\nabla}^{2}\check{w},
\end{equation}
where $\check{\nabla}^{2}=\partial^{2}/\partial x^{2}+\partial^{2}/\partial y^{2}-k_{z}^{2}$.
In an operator form, the above equations can be expressed as
\begin{equation}
\check{\nabla}\cdot\check{\vec{u}}=0,
\end{equation}
\begin{equation}
\frac{\partial\check{\vec{u}}}{\partial t}+\mathcal{L}_{\vec{U}}(\check{\vec{u}})=-\check{\nabla}\check{p},
\end{equation}
where $\mathcal{L}_{\vec{U}}$ denotes the linear operator applied to $\check{\vec{u}}$. 

To solve the two-dimensional linearized perturbation equations (\ref{eq:continuity_cartesian_linear_modal})-(\ref{eq:momentum_z_linear_modal}) and compute $\check{\vec{q}}=(\check{\vec{u}},\check{p})$ by numerical simulations, we use the fractional-step method for time marching and the Fourier spectral method for spatial discretization in the $x-$ and $y-$ coordinates to impose periodic boundary conditions. 
In the simulations, we consider as an initial condition an inertial wave propagating in the horizontal direction $x$ by imposing the wave with $k_{y}=0$ as shown in Fig.~\ref{Fig_planarwave_stablevortex} (upper left panel). 
There is no loss of generality in considering the inertial wave with $k_{y}=0$ as convective columns have a cylindrical geometry. 

The inertial-wave solution $\check{\vec{q}}_{iw}$ can be obtained by considering the zero base flow $\vec{U}=0$ in the equations (\ref{eq:continuity_cartesian_linear_modal})-(\ref{eq:momentum_z_linear_modal}) leading to the following equations:
\begin{equation}
\label{eq:continuity_cartesian_linear_IW}
\frac{\partial \check{u}_{iw}}{\partial x}+\mathrm{i}k_{z}\check{w}_{iw}=0,
\end{equation}
\begin{equation}
\label{eq:momentum_x_linear_IW}
\frac{\partial \check{u}_{iw}}{\partial t}-f\check{v}_{iw}=-\frac{\partial \check{\pi}_{iw}}{\partial x}+\upnu\check{\nabla}^{2}\check{u}_{iw},
\end{equation}
\begin{equation}
\label{eq:momentum_y_linear_IW}
\frac{\partial \check{v}_{iw}}{\partial t}+f\check{u}_{iw}=\upnu\check{\nabla}^{2}\check{v}_{iw},
\end{equation}
\begin{equation}
\label{eq:momentum_z_linear_IW}
\frac{\partial \check{w}_{iw}}{\partial t}=-\mathrm{i}k_{z}\check{\pi}_{iw}+\upnu\check{\nabla}^{2}\check{w}_{iw}.
\end{equation}
By considering the ansatz
\begin{equation}
\label{eq:ansatz_IW}
\check{\vec{q}}_{iw}=\left(
\begin{array}{c}
\check{u}_{iw}\\
\check{v}_{iw}\\
\check{w}_{iw}\\
\check{\pi}_{iw}
\end{array}
\right)=
\left(
\begin{array}{c}
\hat{u}_{iw}\\
\hat{v}_{iw}\\
\hat{w}_{iw}\\
\hat{p}_{iw}
\end{array}
\right)\exp\left(\mathrm{i}k_{\mathrm{x}}x-\mathrm{i}\omega t\right)+c.c.,
\end{equation}
we obtain the following dispersion relation for the inertial wave:
\begin{equation}
\label{eq:dispersion_IW}
\omega=\frac{k_{z}f}{\sqrt{k_{x}^{2}+k_{z}^{2}}}-\mathrm{i}\upnu k_{{xz}}^{2},
\end{equation}
where $k_{{xz}}^{2}=k_{x}^{2}+k_{z}^{2}$, and the corresponding polarization relations as follows:
\begin{equation}
\label{eq:amplitudes_IW}
\hat{v}_{iw}=\frac{f}{\mathrm{i}\omega-\upnu k_{{xz}}^{2}}\hat{u}_{iw},~~
\hat{w}_{iw}=-\frac{k_{x}}{k_{z}}\hat{u}_{iw},~~
\hat{p}_{iw}=-\frac{k_{x}\left(\omega+\mathrm{i}\upnu k_{{xz}}^{2}\right)}{k_{z}}\hat{u}_{iw},
\end{equation}
which are the classical solutions for viscous inertial waves when $k_{y}=0$. 
We note that the above inertial-wave solution $\check{\vec{q}}_{iw}$ is not a stationnary solution of the equations (\ref{eq:continuity_cartesian_linear_modal})-(\ref{eq:momentum_z_linear_modal}) as $\vec{U}\neq0$ in the equations, thus the perturbation $\check{\vec{q}}$ evolves in time as it interacts with the convective vortex. 

{
\begin{figure*}
   \centering
   \includegraphics[height=4.8cm]{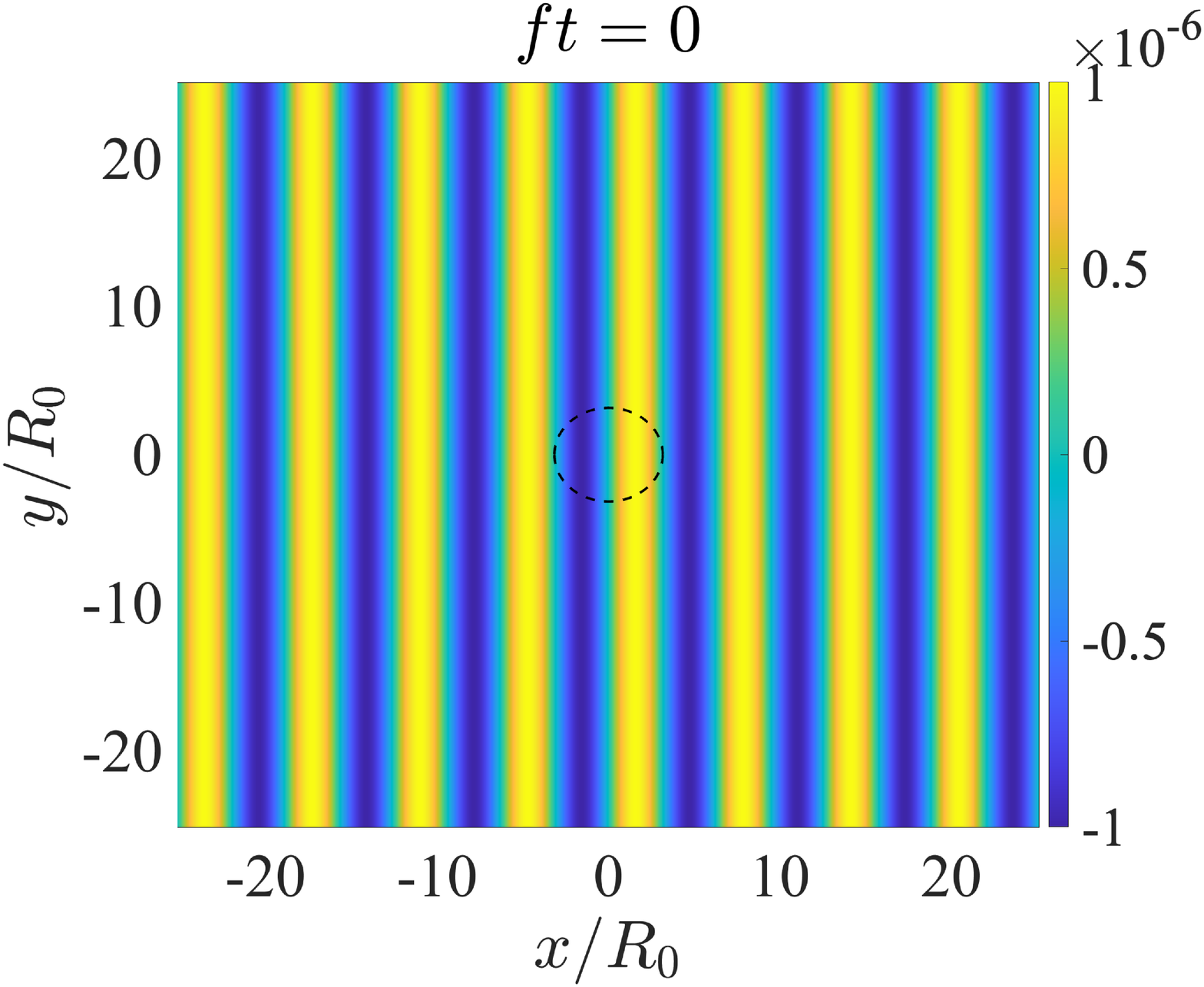}
   \includegraphics[height=4.8cm]{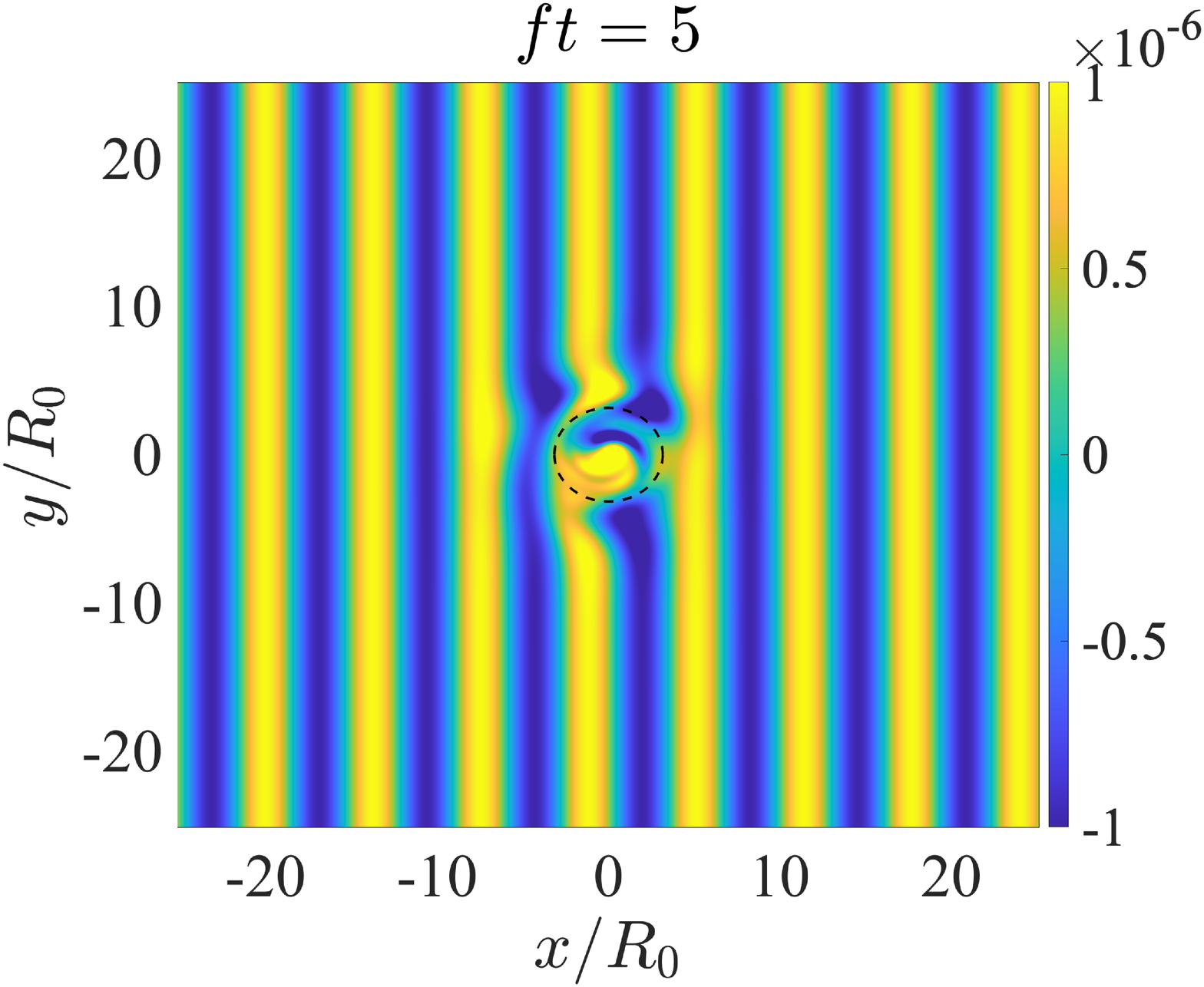}
   \includegraphics[height=4.8cm]{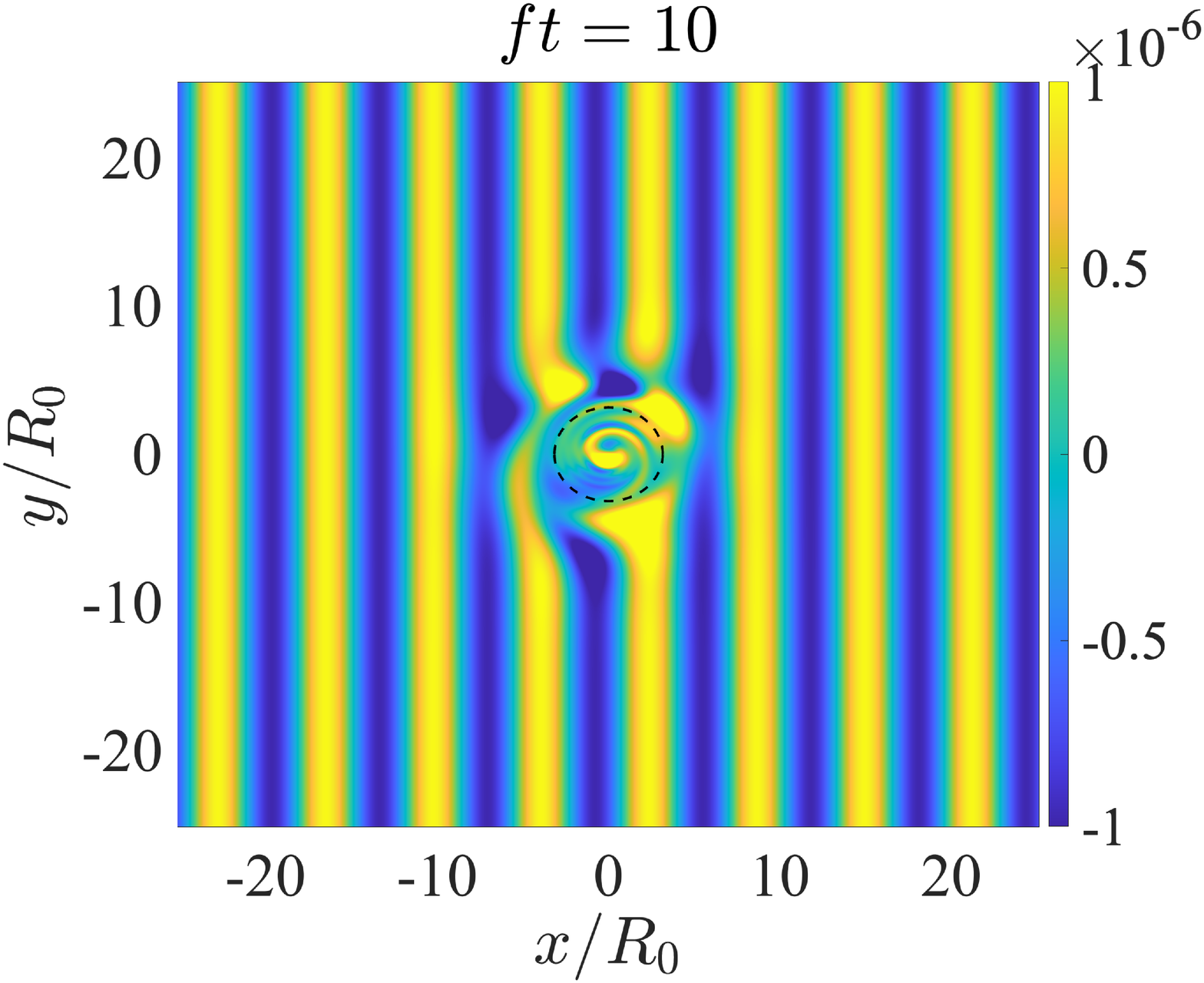}
   \includegraphics[height=4.8cm]{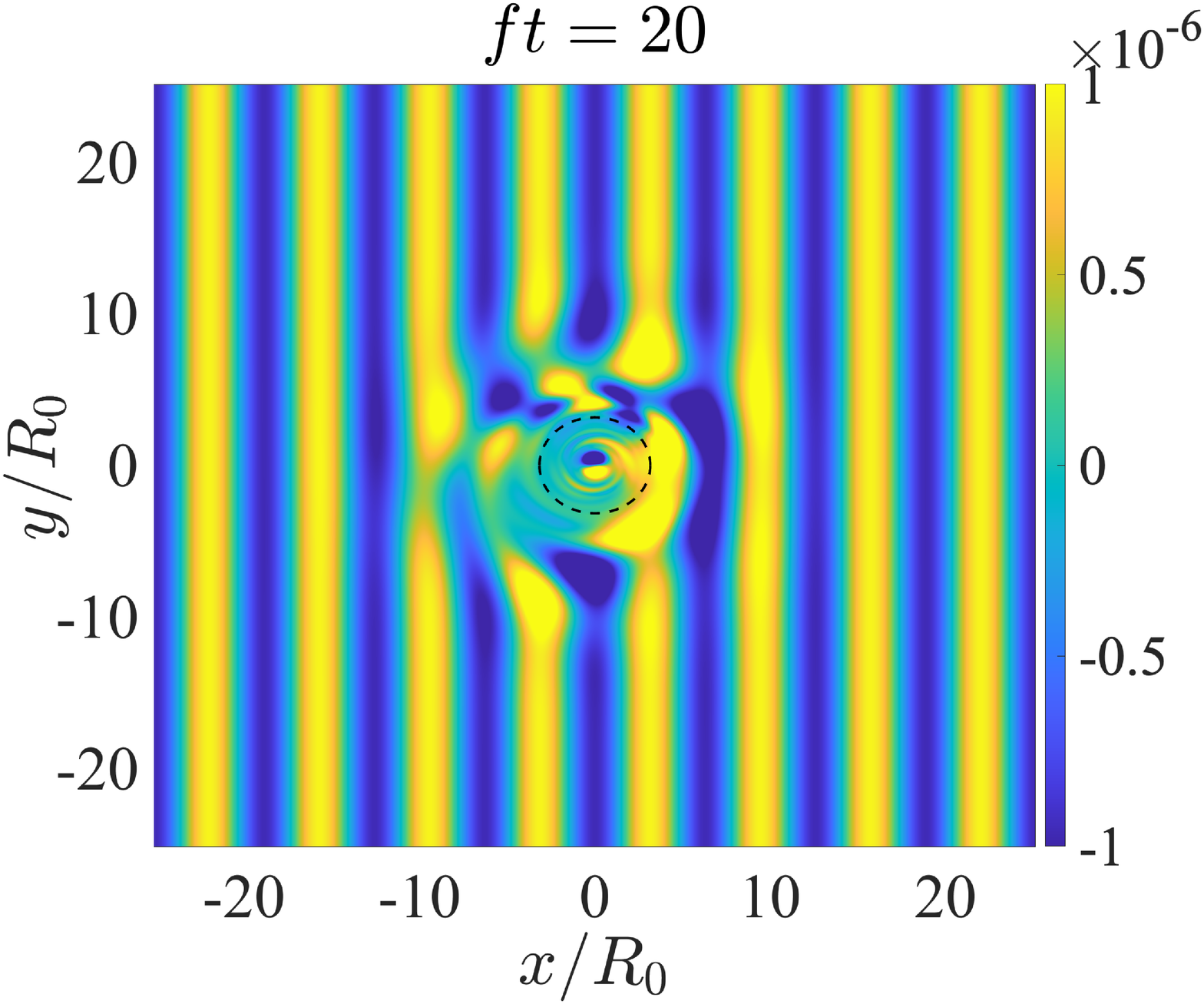}
   \includegraphics[height=4.8cm]{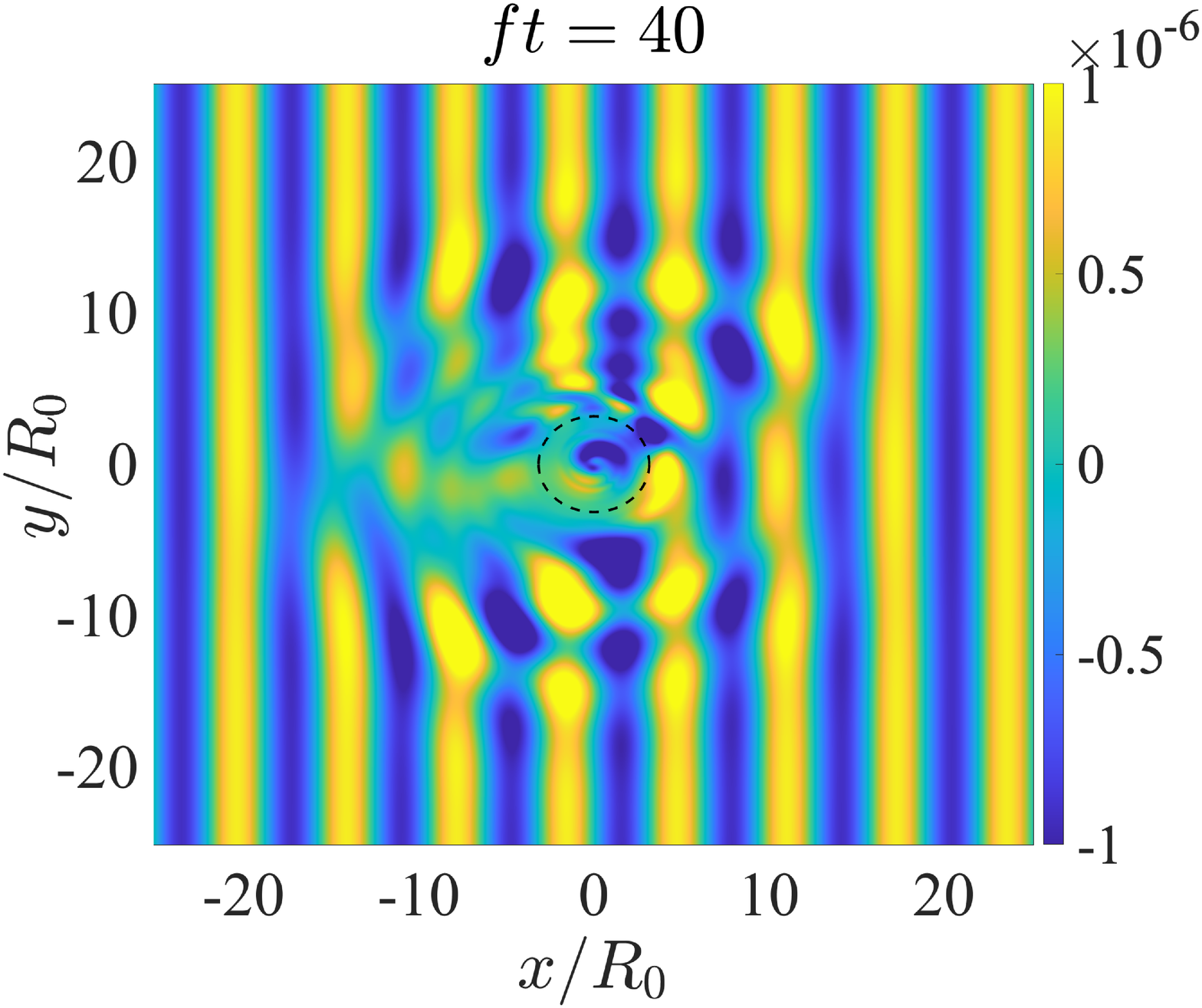}
   \includegraphics[height=4.8cm]{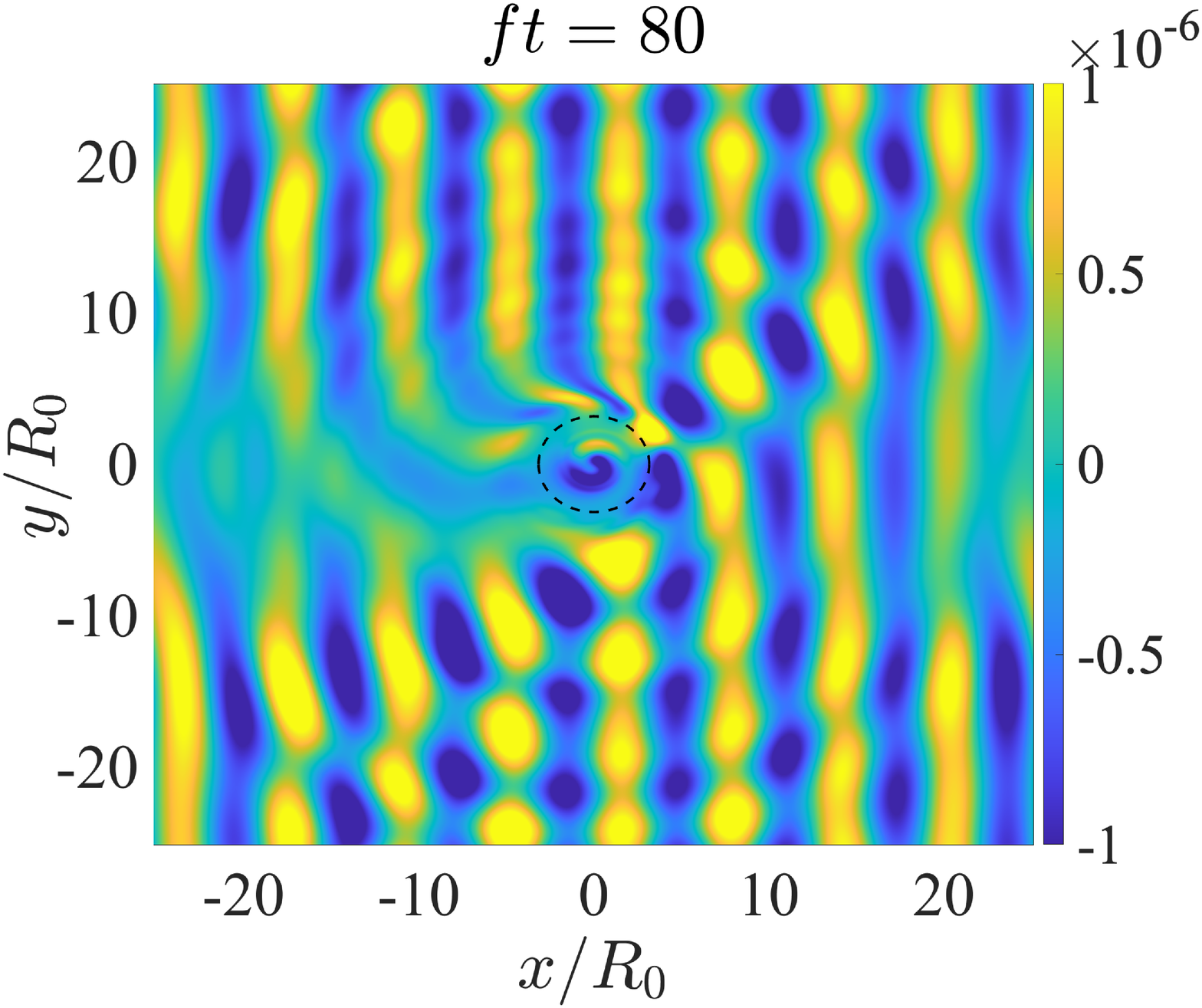}
     \caption{Time evolution of the perturbation velocity $\check{u}(x,y)$ illustrating the interaction between an inertial wave with $(k_{x}R_{0},k_{z}R_{0})=(1,1)$ and a stable vortex with $\Omega_{0}/f=1$ at $Ek=\upnu/(fR_{0}^{2})=10^{-3}$. The dashed line in each panel denotes $r/R_{0}=3.16$.
              }
         \label{Fig_planarwave_stablevortex}
   \end{figure*}
Figure \ref{Fig_planarwave_stablevortex} shows an example of the temporal evolution of perturbation interacting with a stable convective column at $\Omega_{0}/f=1$.
For indication of the vortex centre, we show by dashed lines a radius $r=3.16R_{0}$ at which the first zero-crossing of the angular velocity (i.e. $\Omega(r)=0$) occurs.
As the simulations consider linear equations, the wave-vortex interaction is one-way, i.e., inertial waves are affected by convective columns but not vice versa.  
We clearly see that the perturbation velocity $\check{u}$ of the inertial wave is mixed by the vortex at early stage. 
The mixing of momentum at this stage is likely due to the advection term $(\vec{U}\cdot\check{\nabla})\check{\vec{u}}$ as the perturbation is advected by the rotating base flow.
This mixing process by advection is similar to sweeping \citep[e.g.][]{Clarkdileoni2014,Campagne2015}, which occurs when the waves are advected by large-scale flows. 
However, we notify a difference from the sweeping process such that the wave-vortex interaction creates the low-velocity ring region around the vortex core as the mixing continues.}
{Inside the ring region, a confined perturbation with an azimuthal wavenumber $m=1$ appears as a result of the interaction. 
More interestingly, in the far field away from the vortex core, the interaction promotes radiation of a progressive wave.  
Although we impose the inertial wave in a planar, Cartesian way, it is noteworthy that the new radiating progressive wave has a cylindrical nature. 
These dynamical behaviors, mixing and radiation of a wavelike mode, are also observed for other wavenumber sets when either a long-wavelength, fast-moving wave (e.g., $k_{z}R_{0}=1$, $k_{x}R_{0}=0.25$ and phase speed $c_{x}=\mathrm{Re}(\omega/(fk_{x}R_{0}))=3.88$) or a short-wavelength, slow-moving wave (e.g., $k_{z}R_{0}=1$, $k_{x}R_{0}=2$, $c_{x}=0.22$) interact with stable convective columns.
We also note that such a new radiating wave is not observed for the cases with 1D cylindrical incoming waves studied in Sect.~\ref{sec:Stability}. 
These waves cannot penetrate the vortex center $r=0$ due to the boundary conditions (\ref{eq:boundary_condition_core}) while the 2D planar incoming waves are not restricted by any boundary conditions but they overlap the vortex core, a situation where a wave scattering \citep[][]{Stone2000} similar to what we see in Fig.~\ref{Fig_planarwave_stablevortex} or a wave-induced mean force \citep[][]{McIntyre2019} can be generated. 
}

\begin{figure*}
   \centering
   \includegraphics[height=4.8cm]{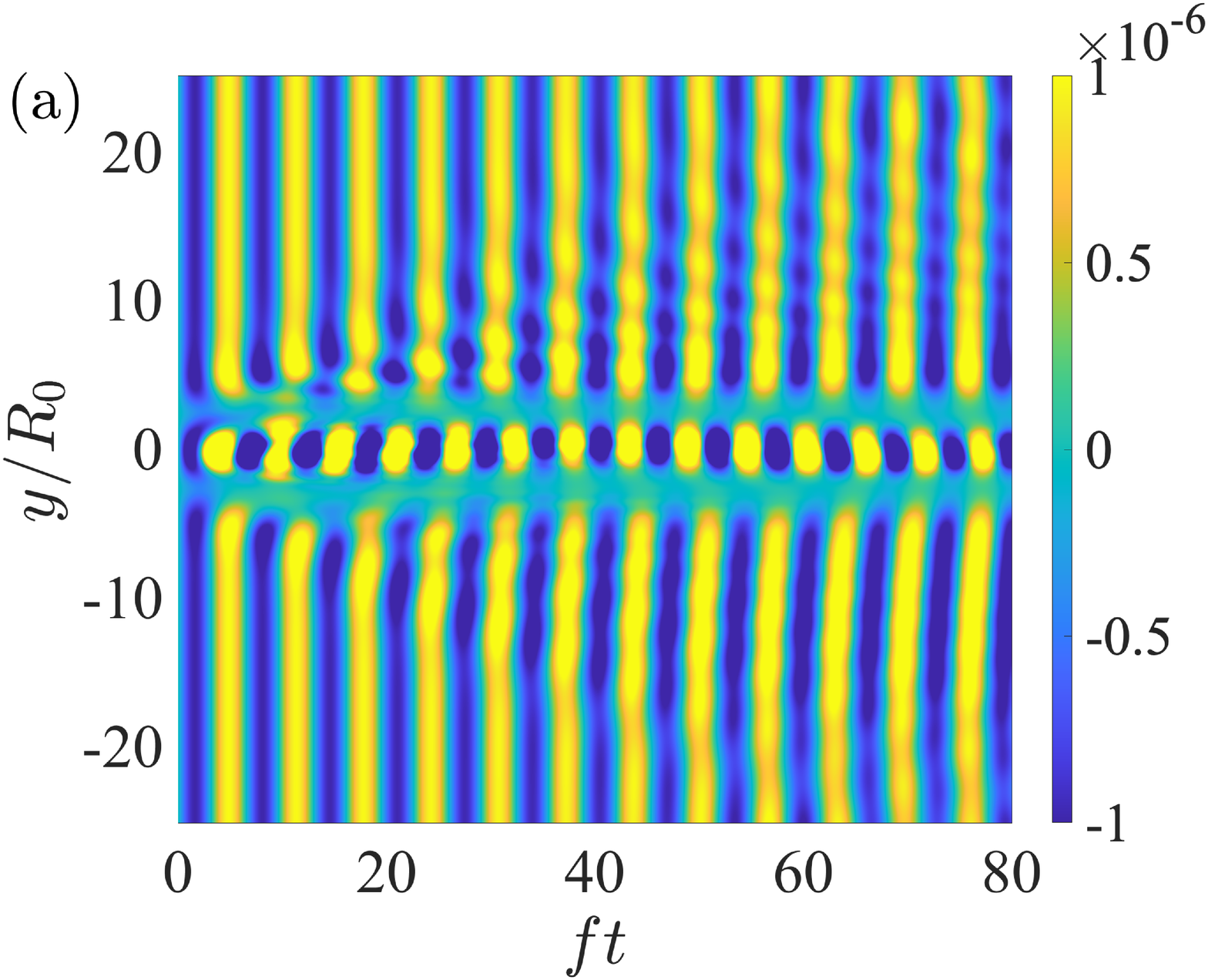}
   \includegraphics[height=4.8cm]{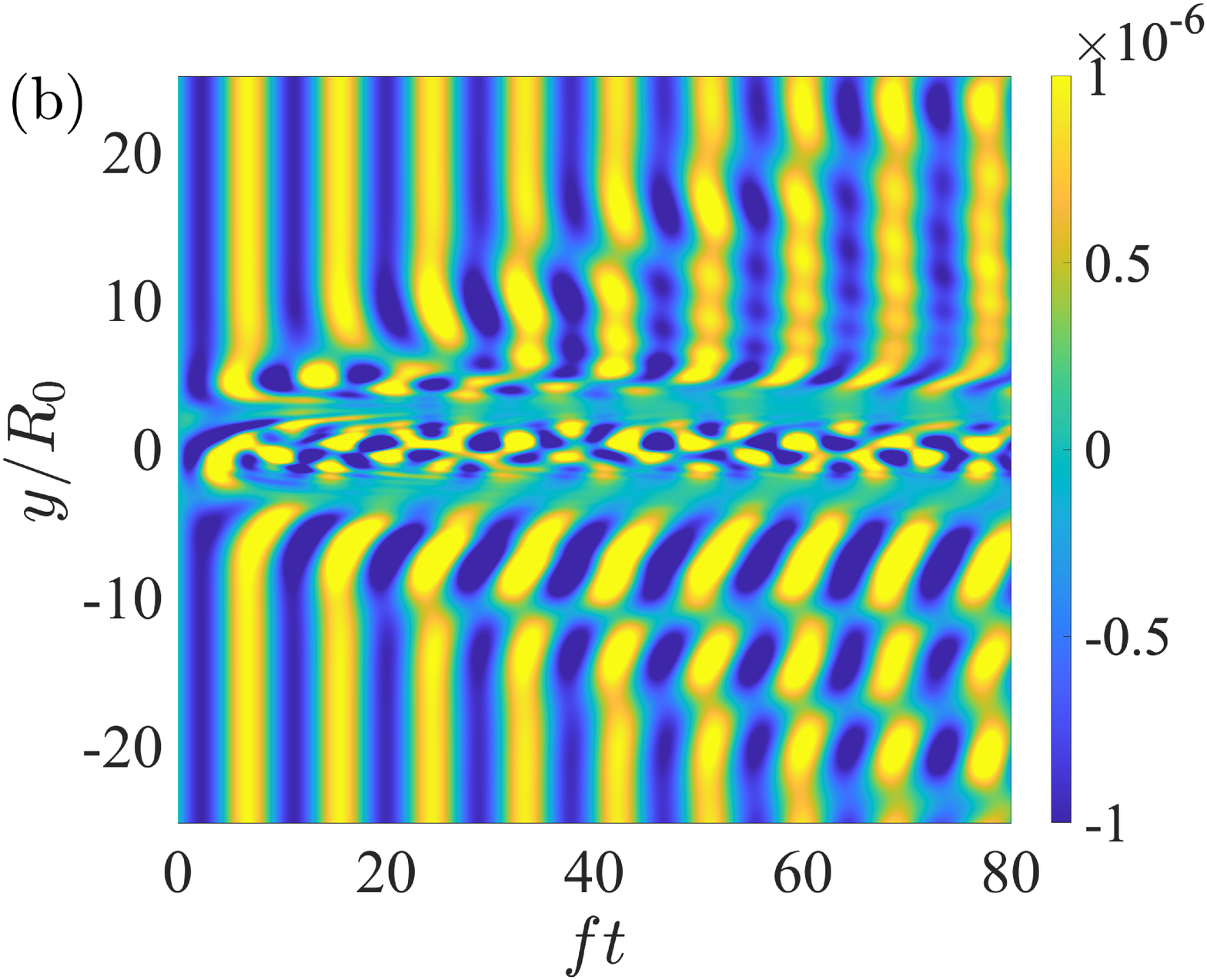}
   \includegraphics[height=4.8cm]{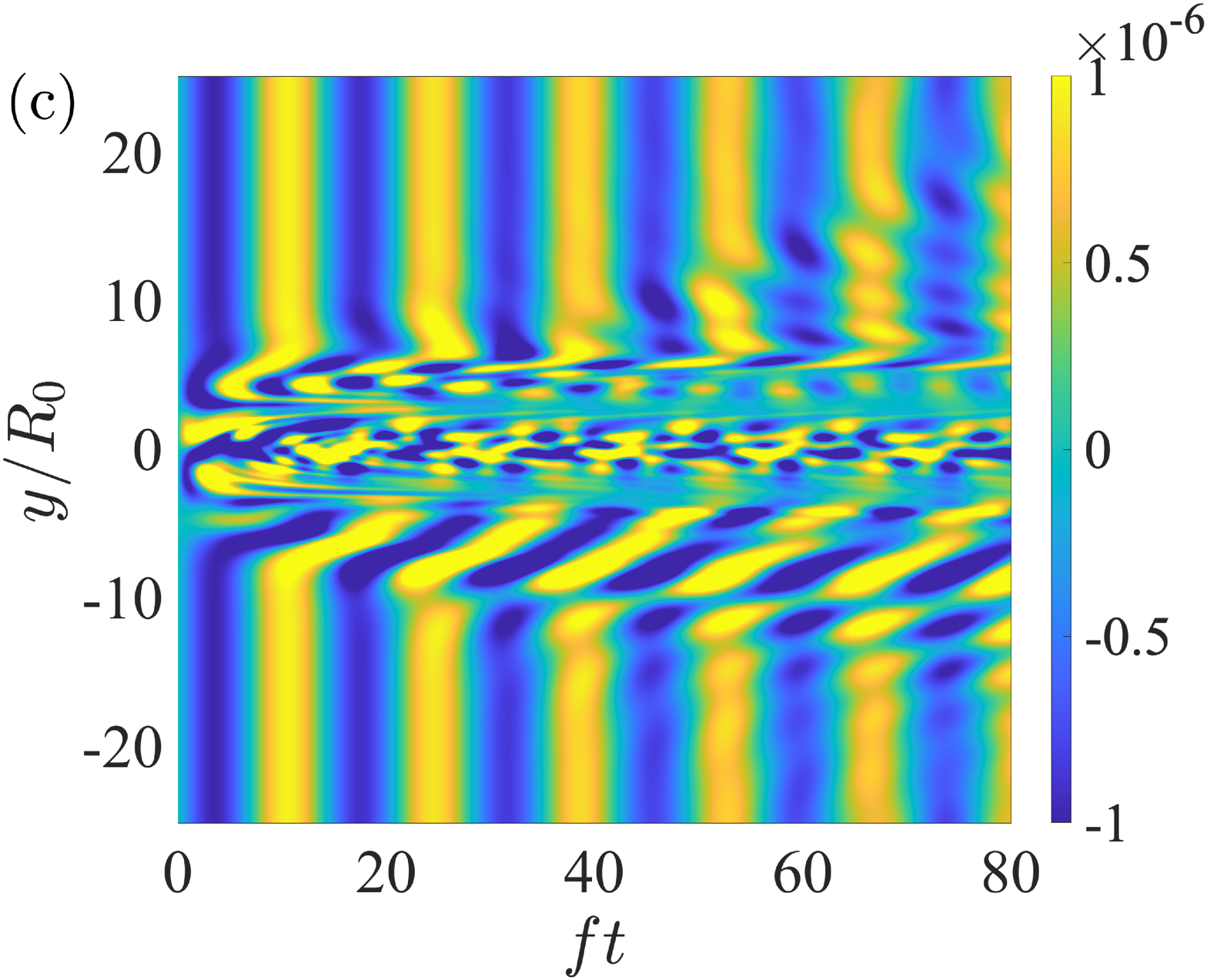}
     \caption{Spatio-temporal diagrams of $\check{u}$ in the space $(t,y)$ extracted at $x=0$ for the wavenumber sets $(k_{x}R_{0},k_{z}R_{0})$: (a) $(0.25,1)$, (b) $(1,1)$ and (c) $(2,1)$. 
              }
         \label{Fig_interaction_spatiotemporal}
   \end{figure*}
\begin{figure*}
   \centering
   \includegraphics[height=4.7cm]{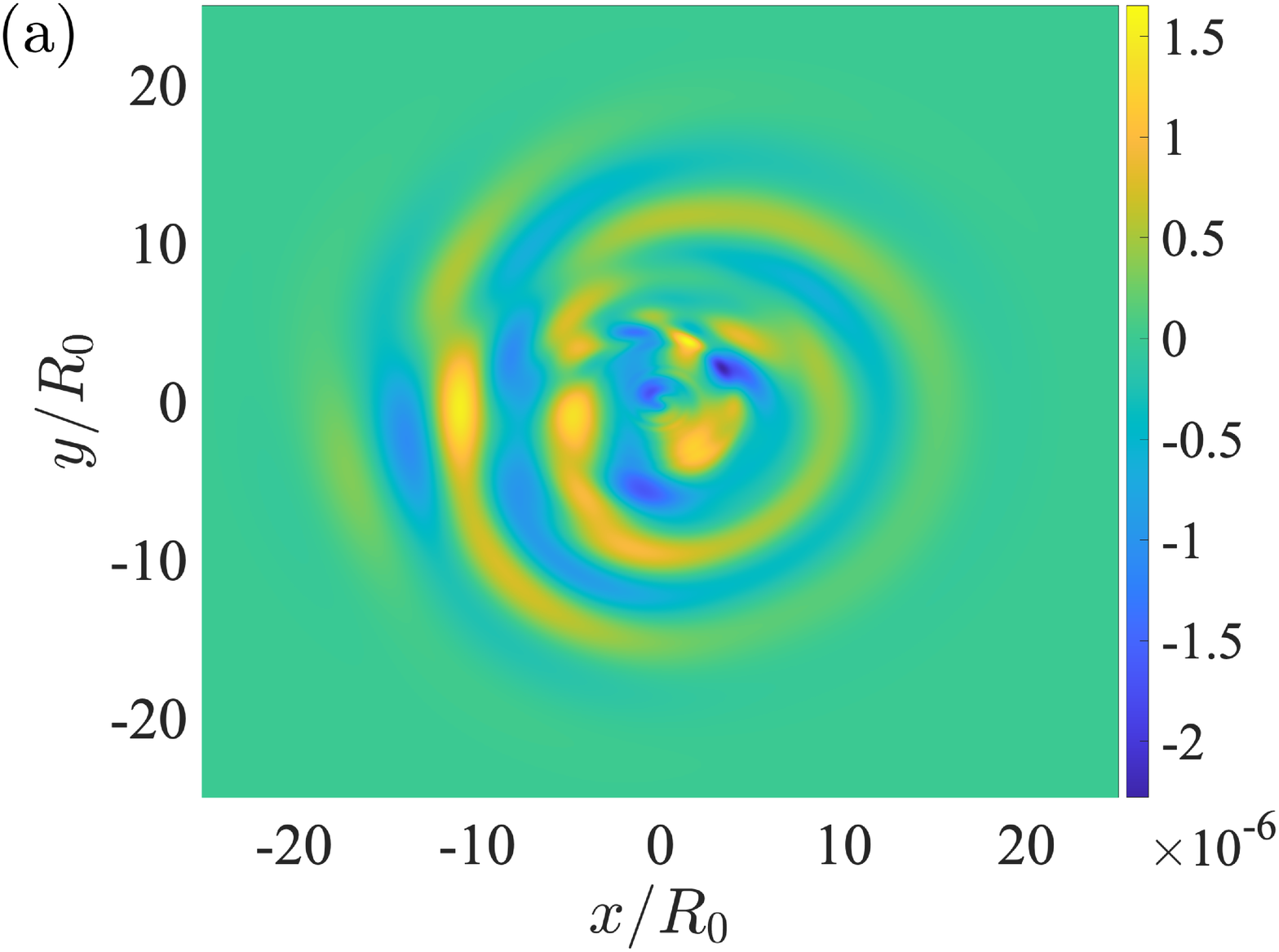}   
\includegraphics[height=4.7cm]{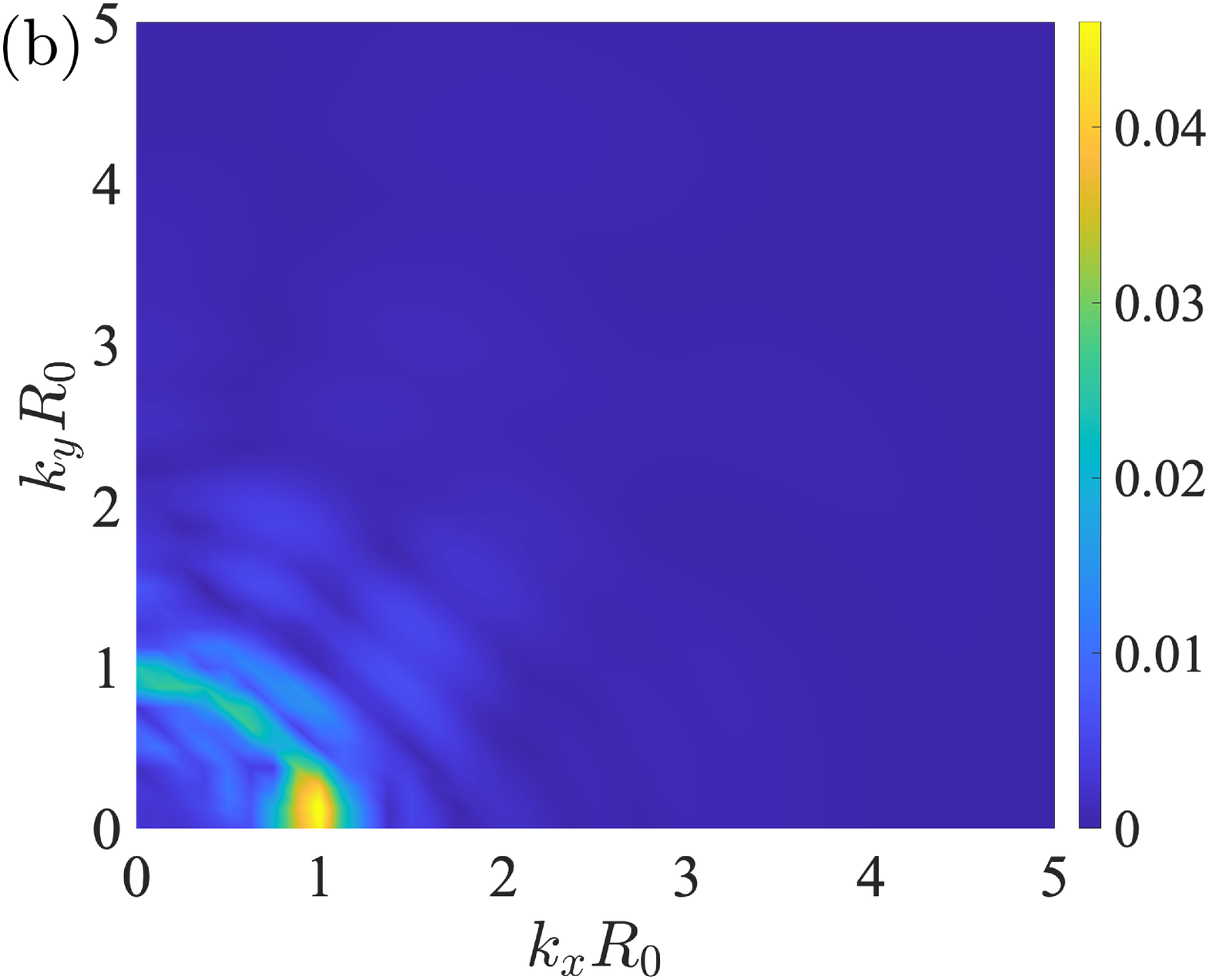}
\includegraphics[height=4.7cm]{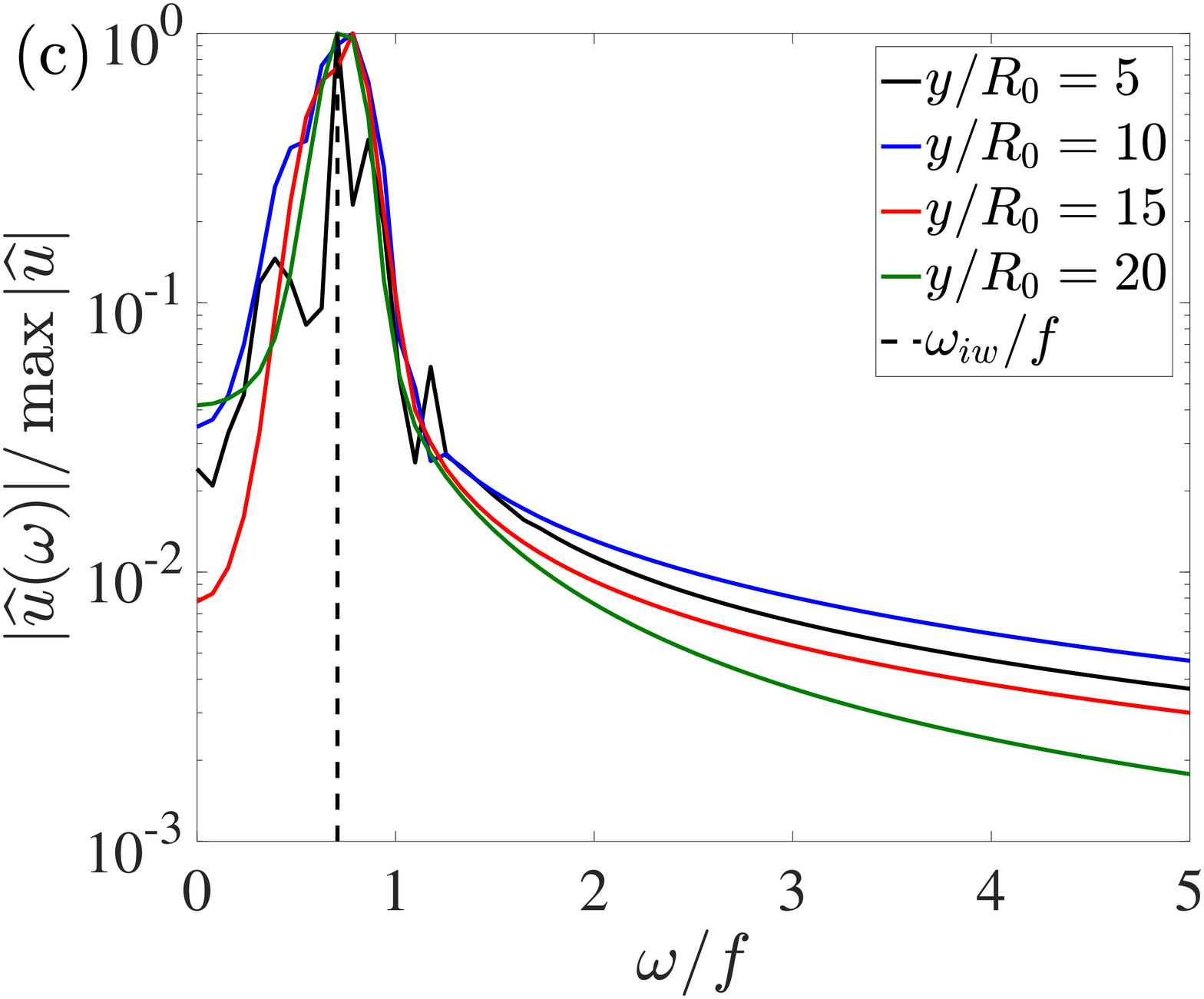}
\caption{(a) Perturbation velocity $\check{u}_{rm}(x,y)$ of the radiating mode extracted at $ft=40$ for the case in Fig.~\ref{Fig_planarwave_stablevortex}. (b) The corresponding wave amplitude $|\hat{u}(k_{x},k_{y},t)|$ in the wavenumber space $(k_{x},k_{y})$ at $ft=40$. 
(c) The frequency spectrums $\widehat{u}(\omega,x,y)$ for the perturbation velocity $\check{u}_{rm}$ extracted at $x=0$ for different $y$ locations. The spectrums are normalized by their maximums for comparision.  
}
         \label{Fig_new_rm}
   \end{figure*}
   \begin{figure}
   \centering
   \includegraphics[height=5.2cm]{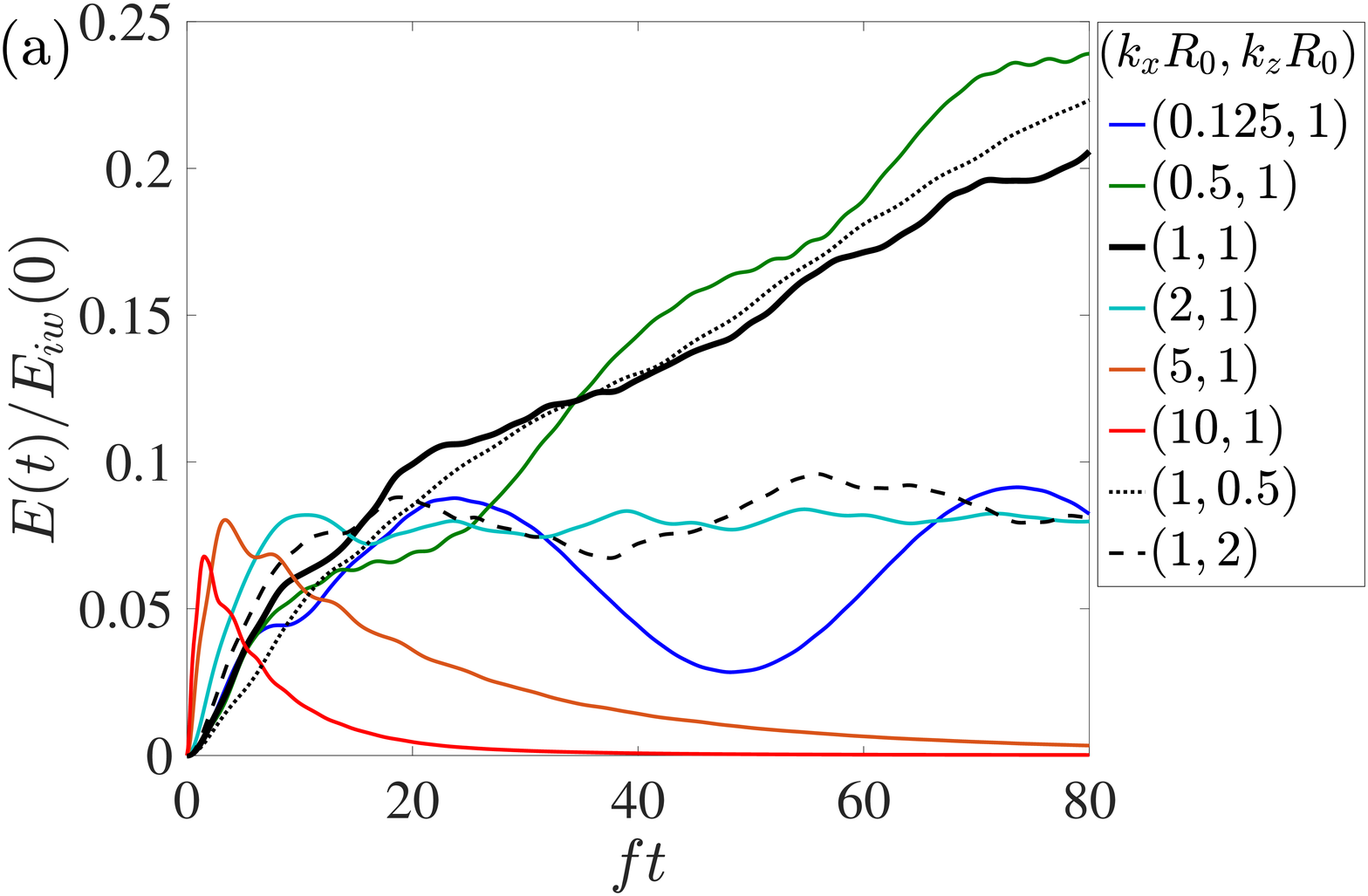}
   \includegraphics[height=5.2cm]{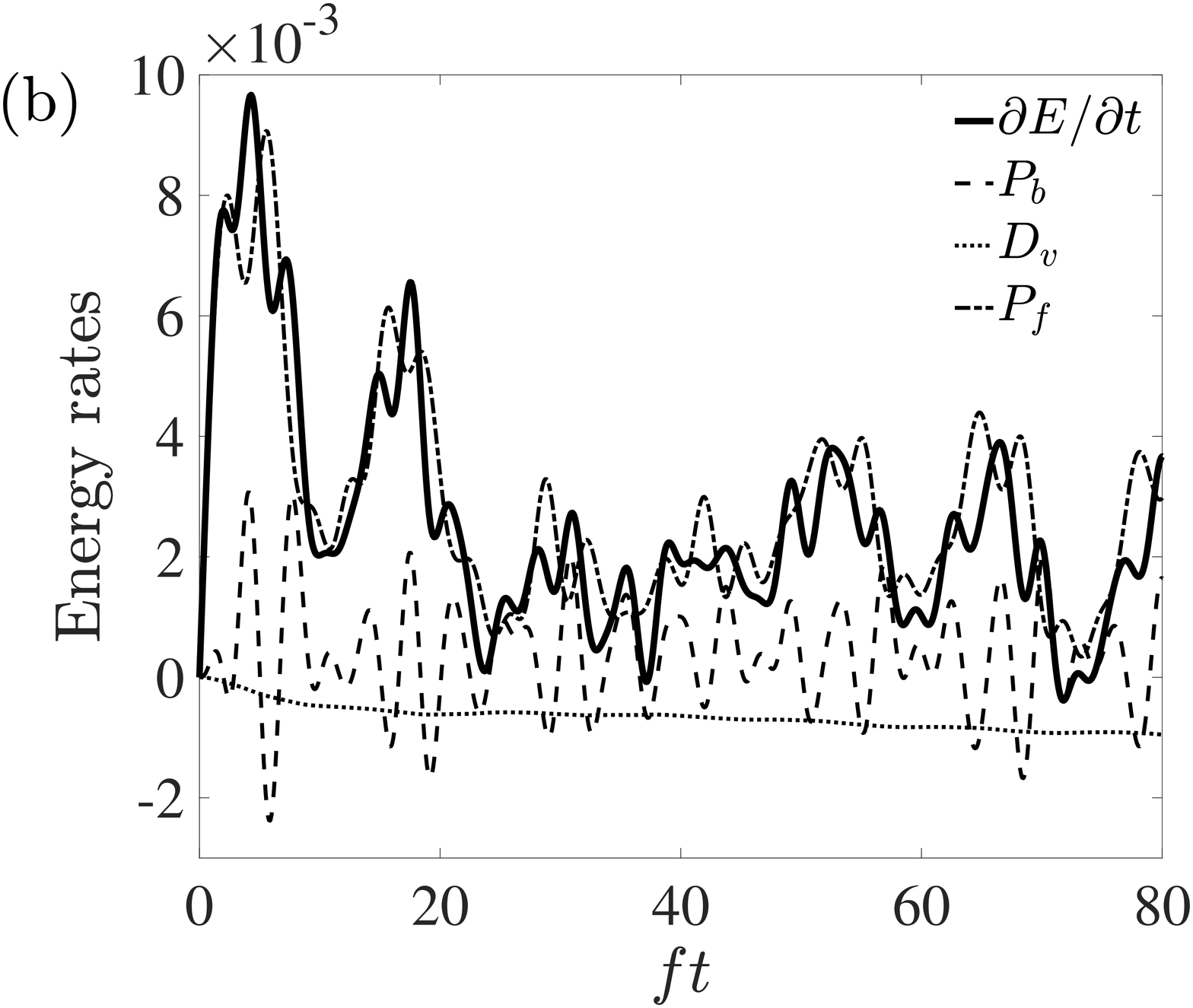}
     \caption{
     (a) Perturbation energy $E(t)$ for the radiating modes created by incoming inertial waves with various wavenumber sets $(k_{x}R_{0},k_{z}R_{0})$. The energy is normalized by the initial inertial-wave energy $E_{iw}(t=0)$ for comparison among different cases. 
     (b) The time rates for perturbation energy $\partial E/\partial t$ (solid), base shear contribution $P_{b}$ (dashed), viscous dissipation $D_{v}$ (dotted) and power injected by the incoming inertial wave-vortex interaction $P_{f}$ (dash-dot) for the case with $(k_{x}R_{0},k_{z}R_{0})=(1,1)$ in (b). The energy rates are normalized by $fE_{iw}(0)$ for non-dimensionalization.}
         \label{Fig_new_rm_energy}
   \end{figure}
The appearance of this radiating mode is also well captured in spatio-temporal diagrams computed by extracting the velocity component $\check{u}$ at $x=0$ as shown in Fig.~\ref{Fig_interaction_spatiotemporal}.
For the case in (a) with $(k_{x}R_{0},k_{z}R_{0})=(0.25,1)$ where the inertial wave has a long wavelength in the $x$-direction and fast phase speed $c_{x}=3.88$, we clearly see the low-velocity ring region around $y/R_{0}\simeq\pm2.7$ and the structure of the radiating mode, i.e., a perturbation confined within the low-velocity ring region and radiating progressive wave in the far field.
For the case in (b) with a larger $k_{x}R_{0}=1$ (the same case in Fig.~\ref{Fig_planarwave_stablevortex}), the interference between the imposed inertial wave and new radiating mode is more apparent.
The shape of perturbation inside the ring region is also more irregular and it seems that the perturbation has a higher azimuthal wavenumber than the mode in (a).  
A similar tendency, i.e., strong mixing and radiation, is observed as $k_{x}$ is further increased to $k_{x}R_{0}=2$ as shown in Fig.~\ref{Fig_interaction_spatiotemporal}(c).

The generation mechanism of radiating modes (denoted by the subscript \guillemotleft rm\guillemotright\ in the following) can be understood mathematically by considering another perturbation $\check{\vec{q}}_{rm}=\check{\vec{q}}-\check{\vec{q}}_{iw}$ that satisfies the following equations:
\begin{equation}
\label{eq:continuity_rm}
\check{\nabla}\cdot\check{\vec{u}}_{rm}=0,
\end{equation}
\begin{equation}
\label{eq:momentum_rm}
\frac{\partial\check{\vec{u}}_{rm}}{\partial t}+\mathcal{L}_{\vec{U}}(\check{\vec{u}}_{rm})=-\nabla\check{p}_{rm}+\check{\vec{f}}_{rm},
\end{equation}
where $\check{\vec{f}}_{rm}$ is the forcing term induced by the interaction between the convective column and inertial wave as
\begin{equation}
\label{eq:forcing_rm}
\check{\vec{f}}_{rm}=-\left(\vec{U}\cdot\check{\nabla}\right)\check{\vec{u}}_{iw}-\left(\check{\vec{u}}_{iw}\cdot\check{\nabla}\right)\vec{U}.
\end{equation}
These equations are obtained by substituting $\check{\vec{q}}=\check{\vec{q}}_{rm}+\check{\vec{q}}_{iw}$ into the linearized perturbation equations (\ref{eq:continuity_cartesian_linear_modal})-(\ref{eq:momentum_z_linear_modal}) and subtracting the equations by the inertial-wave equations (\ref{eq:continuity_cartesian_linear_IW})-(\ref{eq:momentum_z_linear_IW}). 
The interaction term $\check{\vec{f}}_{rm}$ forces the linear system (\ref{eq:momentum_rm}) with a characteristic frequency $\omega$ from the dispersion relation of the incoming inertial wave (\ref{eq:dispersion_IW}).
A stable convective column has neutral or stable modes with frequencies close to $\omega$, thus the forcing can trigger these modes by resonance and new radiating modes appear as the sum of these modes due to the forcing $\check{\vec{f}}_{rm}$ triggered by the vortex-wave interaction. 
An example of a radiating mode forced by this interaction is displayed in Fig.~\ref{Fig_new_rm}(a), which clearly shows the structure of cylindrical radiation.
If we apply the 2D Fourier transform to the velocity field $\check{u}(x,y,t)$ in the spatial directions $x$ and $y$ at a given time $t$, we can obtain, as shown in Fig.~\ref{Fig_new_rm}(b), the wave amplitude $\hat{u}_{rm}(k_{x},k_{y},t)$ in the wavenumber space $(k_{x},k_{y})$ where
\begin{equation}
\hat{u}_{rm}(k_{x},k_{y},t)=\int_{-\infty}^{\infty}\int_{-\infty}^{\infty}\check{u}_{rm}(x,y,t)\exp[-\mathrm{i}(k_{x}x+k_{y}y)]\mathrm{d}x\mathrm{d}y.
\end{equation}
The largest amplitude is obtained at $(k_{x}R_{0},k_{y}R_{0})=(1,0)$, which corresponds to the wavenumber set of the initially-imposed inertial wave. 
We note that other wavenumber components also have amplitudes comparable to this largest one especially around the circle $(k_{x}R_{0})^{2}+(k_{y}R_{0})^{2}=1$ where the most of the energy of the radiating mode lies seemingly. 
This implies that the inertial wave-vortex interaction leads to the energy transfer from the Cartesian wavenumber components $k_{x}$ and $k_{y}$ into the cylindrical wavenumber component $k_{r}=\sqrt{k_{x}^{2}+k_{y}^{2}}$, thus the nature of the planar wave becomes cylindrical as a consequence.
In Fig.~\ref{Fig_new_rm}(c), we also plot the frequency spectrum $\widehat{u}(\omega,x,y)$ for the points at $x=0$ with different $y$ by applying the Fourier transform to $\check{u}$ as
\begin{equation}
\widehat{u}(\omega,x,y)=\int_{-\infty}^{\infty}\check{u}(x,y,t)\exp(-\mathrm{i}\omega t)\mathrm{d}t.    
\end{equation}
We clearly see that the frequency spectra reach their maximum peaks around the frequency $\omega\simeq (1/\sqrt{2})f$, which corresponds to the real part of the inertial-wave frequency $\omega_{iw}$ from Eq.~(\ref{eq:dispersion_IW}). 
This confirms our above statement that the inertial wave triggers stable vortex modes with frequencies close to $\omega_{iw}$ and the radiating mode appears as the sum of these modes.

It is not shown here but we also verified the excitation of radiating modes by an interaction between inertial waves and unstable convective columns. 
However, for unstable convective columns, their most unstable modes eventually become dominant as observed in the previous section on the interaction between unstable convective columns and cylindrical tidally-forced incoming waves and they lead to the destruction of the vortex.

By multiplying with $\check{\vec{u}}_{rm}^{\dag}$ (where $\dag$ denotes the complex conjugate) to the equation (\ref{eq:momentum_rm}) and averaging it over the domain where $x\in\left[-L_{x},L_{x}\right]$ and $y\in\left[-L_{y},L_{y}\right]$, we obtain the following equation for the perturbation energy of a radiating mode: 
\begin{equation}
\frac{\partial E}{\partial t}=P_{b}+D_{v}+P_{f},
\end{equation}
where $E$ is the perturbation kinetic energy of the radiating mode, $P_{b}$ is the energy contribution by the direct interaction between the convective column and the mode, $D_{v}$ denotes the energy loss by viscous dissipation, and $P_{f}$ is the energy production by the forcing induced by the interaction between the convective column and inertial wave, all of which are defined as follows:
\begin{equation}
\begin{aligned}
&E=\frac{1}{2}\int_{-L_{x}}^{L_{x}}\int_{-L_{y}}^{L_{y}}\left(\left|\check{u}_{rm}\right|^{2}+\left|\check{v}_{rm}\right|^{2}+\left|\check{w}_{rm}\right|^{2}\right)\mathrm{d}y\mathrm{d}x,\\
&P_{b}=-\int_{-L_{x}}^{L_{x}}\int_{-L_{y}}^{L_{y}}\left(\frac{\partial U_{x}}{\partial x}|\check{u}_{rm}|^{2}+\frac{\partial U_{x}}{\partial y}\check{u}^{\dag}_{rm}\check{v}_{rm}+\frac{\partial U_{y}}{\partial x}\check{v}^{\dag}_{rm}\check{u}_{rm}\right.\\
&~~~~~~~~~\left.+\frac{\partial U_{y}}{\partial y}|\check{v}_{rm}|^{2}\right)\mathrm{d}y\mathrm{d}x,\\
&D_{v}=-\upnu\int_{-L_{x}}^{L_{x}}\int_{-L_{y}}^{L_{y}}\left[\left|\frac{\partial\check{u}_{rm}}{\partial x}\right|^{2}+\left|\frac{\partial\check{u}_{rm}}{\partial y}\right|^{2}+\left|\frac{\partial\check{v}_{rm}}{\partial x}\right|^{2}+\left|\frac{\partial\check{v}_{rm}}{\partial y}\right|^{2}\right.\\
&~~~~~~~~~\left.+\left|\frac{\partial\check{w}_{rm}}{\partial x}\right|^{2}+\left|\frac{\partial\check{w}_{rm}}{\partial y}\right|^{2}+k_{z}^{2}\left(\left|\check{u}_{rm}\right|^{2}+\left|\check{v}_{rm}\right|^{2}+\left|\check{w}_{rm}\right|^{2}\right)\right]\mathrm{d}y\mathrm{d}x,\\
&P_{f}=-\int_{-L_{x}}^{L_{x}}\int_{-L_{y}}^{L_{y}}\check{\vec{u}}^{\dag}_{rm}\cdot\left[\left(\vec{U}\cdot\check{\nabla}\right)\check{\vec{u}}_{iw}+\left(\check{\vec{u}}_{iw}\cdot\check{\nabla}\right)\vec{U}\right]\mathrm{d}y\mathrm{d}x.
\end{aligned}
\end{equation} 
Figure \ref{Fig_new_rm_energy} show examples of how perturbation energies of different radiating modes change over time. 
In (a), we plot the kinetic energy $E$ normalized by $E_{iw}(0)$, the kinetic energy of the inertial wave $E_{iw}$ at $t=0$, for comparison between cases with different wavenumbers $k_{x}$ and $k_{z}$.
This normalization allows us a quantitative comparison among different inertial waves by considering the same wave energy as an input.
For the cases at $k_{z}=1$, we found that the energy of new radiating modes increases largely as a result of the wave-vortex interaction when the imposed inertial waves have the wavenumber $k_{x}R_{0}\sim O(1)$, the case where the wavelength scale is of the same order of the vortex radius $R_{0}$.
These new radiating modes achieve more than 20$\%$ of the input inertial-wave energy.
The number is non-negligible and thus it highlights the importance of the role of inertial wave-vortex interaction in creating {potential} extra turbulent dissipation in fast-rotating planets or stars.
It is also noteworthy that the new radiating modes gain less energy if the imposed inertial waves have a longer wavelength (e.g. $k_{x}R_{0}=0.125$) or a shorter wavelength (e.g. $k_{x}R_{0}=10$) than the vortex length scale $R_{0}$. 
For the case with $(k_{x}R_{0},k_{z}R_{0})=(1,1)$, we plot in panel (b) different energy-rate contributions: $P_{b}$, $D_{v}$ and $P_{f}$, and we compare with the time rate of perturbation energy $\partial E/\partial t$. 

As expected, the viscous dissipation $D_{v}$ is always negative. Therefore it contributes to the decrease of $E$. However, its magnitude is small compared to other contribution terms. 
The term $P_{b}$ indicating the transfer of momentum between the radiating mode and the convective column is oscillatory around zero. The term $P_{f}$, on the other hand, is always positive for the case with $k_{x}R_{0}=1$, thus the main contribution on the perturbation energy production comes from $P_{f}$, the energy {injection} by the forcing induced by the inertial wave-vortex interaction. 
It is not shown here but we also verified that the forcing term $P_{f}$ is the main source of energy of the new radiating mode for other cases with different wavenumbers.

To summarise the interaction between stable convective columns and planar incoming inertial waves, we verified that the interaction promotes the mixing of momentum around the vortex centre, creates low-velocity ring regions and triggers a new wave-like perturbation radiating in the far field with a frequency close to the characteristic frequency of the primary inertial wave.
The energy of this secondary radiating mode is sourced mainly from the inertial wave-vortex interaction and the amount of the energy of the perturbation is a non-negligible fraction of the input inertial wave energy, especially when the length scale of the inertial wave is comparable to that of the vortex radius. 
This implies that the convective column can act as an additional source of energy to create more efficient turbulent dissipation by generating the secondary radiating mode.

\section{Towards a complete nonlinear picture}
\label{sec:nonlinear}
\begin{figure}
   \centering
   \includegraphics[width=9cm]{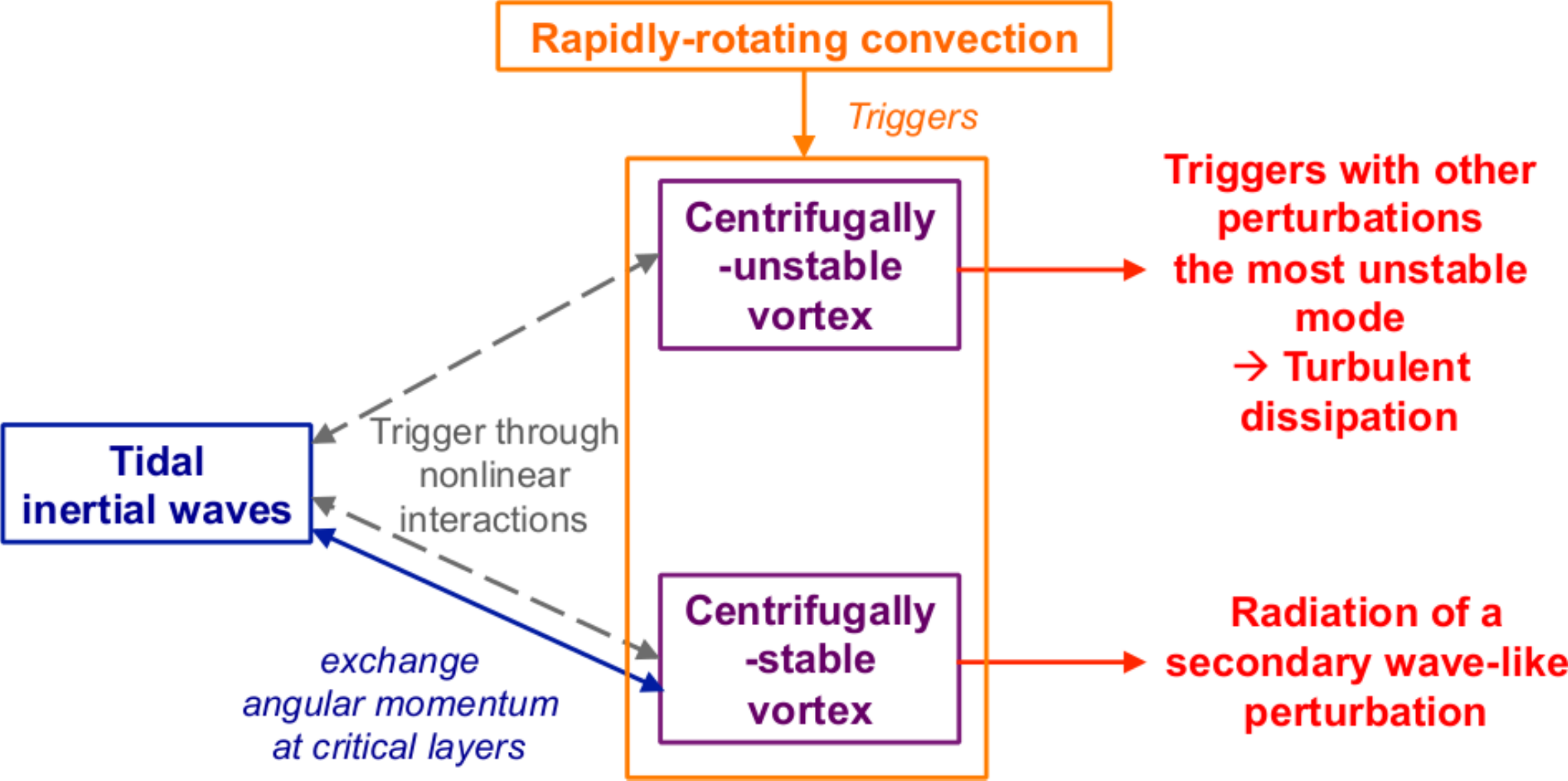}
     \caption{Schematic diagram 
     of possible interactions between tidal inertial waves and convective columnar vortices.
              }
         \label{Fig_Summary}
   \end{figure}

When nonlinear effects are taken into account in the wave-vortex interaction, interesting behaviors are expected. 
In the unstable regime where vortices are intrinsically unstable, tidal waves as any other disturbance (depending on the relative time scales as discussed in Sect.~\ref{sec:linear_cylindrical}) can destabilize them leading to fully turbulent states where energy is dissipated at small scales. 
From the current study, it is not clear which type of perturbation will yield turbulence more effectively. 
Nonlinear simulations can address this question to understand the evolution of an unstable convective vortex along its nonlinear interaction with perturbation such as tidal waves or any noise.
Then, convection or the nonlinear interactions between incoming tidal inertial waves and the vortex \citep{Duran-Matute2013,Boury2021} can trigger new columnar vortices, which can potentially be unstable or not. 
In the unstable case, this can potentially lead to a cycle of creation-destruction of vortices that has been observed in \citet{Barker2013} in their study of the dynamics of the tidal elliptic instability in rotating flows. 
In the stable case, it is interesting to note that tidal inertial waves can transfer the momentum they carry to the vortex at critical layers as identified in Appendix \ref{subsec:critical} but also extract energy from the vortex via an over-reflection/transmission mechanism. These complex behaviors, which we summarise in Fig. \ref{Fig_Summary}, allow us to point out that the modelling of the interactions between tides and rotating convection using effective turbulent friction cannot be applied to describe the interactions between tidal inertial waves and coherent convective vortices in the general case.

The results obtained with the convective column model proposed by \citet{Grooms2010} can also be applied to other vortices triggered by rotating double-diffusive convection \citep[][]{Moll2017} or by the nonlinear interactions of tidal inertial waves \citep{Astoul2022}. The inertial wave-vortex interaction is also crucial in other configurations. For instance, a recent study by \citet{Pizzi2022} investigates the interaction in the presence of precession-turbulence.
They consider a base precessional shear flow which is subject to an instability called the precessional instability \citep[][]{Kerswell1993}.
The instability triggers turbulence and such a turbulent precessional flow contains both vortices as 2D structures and inertial waves as 3D structures. 
They found that the 3D inertial wave becomes stronger as the precession increases and the nonlinear wave-vortex interaction contributes to turbulent dissipation more efficiently. 
They also reported a cyclic behavior of the flow where vortices appear and disappear as a result of the nonlinear interactions \citep[as in][and de Vries et al. in prep., for the case on non-linear evolution of the tidal elliptic instability]{Barker2013}. This promotes our motivation to study in the future the nonlinear interaction between tidally-forced inertial waves and convective columns to better understand the contribution of the wave-vortex interaction to turbulent dissipation along the evolution of rapidly-rotating planets and stars.

\section{Conclusion and discussion}
\label{sec:Conclusion}
In this paper, we investigate how tidally-forced inertial waves interact with a convective columnar vortex in rapidly-rotating planets/stars. In Sect.~\ref{sec:Formulation}, we studied the convective column using a semi-analytical model proposed by \citet{Grooms2010}, and adapted the model without any mathematical singularity around the column center $r=0$ to allow us {to make a complete} theoretical stability analysis. 
It is found that the convective columnar vortex (\ref{eq:base_Dandoy}) is centrifugally stable when $-\Omega_{p}\leq\Omega_{0}\leq3.62\Omega_{p}$ ($\Omega_p$ and $\Omega_{0}$ being the global planetary (stellar) rotation and the rotation of the flow at the center of the vortex, respectively) and unstable otherwise. 

Since the stability is closely linked to the dynamics of the tidal wave-vortex interaction, we studied in great detail in Sect.~\ref{sec:Stability} centrifugally-unstable modes and neutral modes using linear stability analysis. 
We find that the convective column is unstable to axisymmetric $m=0$ perturbations (which can be induced by eccentricity tides) and weakly non-axisymmetric perturbations with azimuthal wavenumbers $m=\left\{1,2\right\}$ (which correspond to obliquity and asynchronous tides, respectively). The WKBJ analysis detailed in Appendix allows us to derive thoroughly explicit expressions of the maximum growth rate at a given ratio $\Omega_{0}/\Omega_{p}$ for unstable modes and the frequency of neutral modes. 
Next, we presented linear simulations of the interaction between convective columns and tidally-forced radial incoming inertial waves. We verified that such an interaction triggers the most unstable mode of an unstable convective column.

In Sect.~\ref{sec:Interaction}, we studied the interactions between stable convective vortices and tidal inertial waves. 
We provided key results by conducting linear simulations of tidally-forced incoming planar inertial waves interacting with a stable vortex.
It is observed that the wave-vortex interaction leads to the mixing of momentum and the creation of low-velocity regions around the vortex core. 
The interaction promotes efficient radiation of a new {wave-like perturbation} in the case where the wavelength of the incoming wave is close to the vortex characteristic length. This secondary wave can be regarded as the sum of neutral or stable modes of the vortex with frequencies close to the characteristic frequency of the primary inertial wave. Since the amplitude of the secondary wave is non-negligible compared to the one of the primary wave it constitutes a supplementary potential source of dissipation. Finally, angular momentum exchanges can occur at critical layers.   

As identified in Sect. ~\ref{sec:nonlinear}, it would be crucial to understand in the near future the role of the nonlinearities. Moreover, it would be necessary to go beyond the simplified traditional polar $f$-plane approximation. 
In particular, the 2D non-separable dynamics of tidal inertial wave attractors should be treated. Finally, if the external convective regions of gaseous giant planets are not the seat of the magnetic field generation through dynamo action, this is not the case of low-mass stars where magnetic fields should thus be taken into account \citep[e.g.][]{Barker2014MNRAS,Lin2018,Astoul2019}.  

\begin{acknowledgements}
{The authors thank the referee for her/his comments, which have allowed us to improve our manuscript.} The authors acknowledge support from the European Research Council through ERC grant SPIRE 647383 and from GOLF and PLATO CNES grants at the Department of Astrophysics of CEA.
J. Park acknowledges support from the Royal Astronomical Society and Office of Astronomy for Development through the RAS-OAD astro4dev grant and from the Engineering and Physical Sciences Research Council (EPSRC) through the EPSRC mathematical sciences small grant (EP/W019558/1). 
A. Astoul acknowledges support from the Science and Technology Facilities Council (STFC) grant ST/S000275/1, as well as the Leverhulme Trust for early career grant.
\end{acknowledgements}
\bibliographystyle{aa} 
\bibliography{aa}
\begin{appendix}
\section{WKBJ approximation for large vertical wavenumber $k_{z}$}
\label{sec:WKBJ}
Previous studies have revealed that the centrifugal instability of rotating flows reaches its maximum growth rate as $k_{z}\rightarrow\infty$ in the inviscid limit \citep[][]{Billant2005,Billant2013,Park2013PoF,Park2017}.
We also see in Sect.~\ref{sec:Stability} that eigenvalues of the convective Taylor column demonstrate asymptotic behaviors in the large-$k_{z}$ limit. 
To understand these behaviors and derive analytic expressions of the dispersion relation, we adopt the WKBJ approximation for large $k_{z}$ by applying it to $\hat{u}$ as
\begin{equation}
\label{eq:WKBJ_u}
\hat{u}(r)\sim\exp\left[\frac{1}{\epsilon}\sum_{n=0}^{\infty}\epsilon^{n}S_{n}(r)\right],
\end{equation}
where $\epsilon$ is a small parameter to be defined. 
Applying (\ref{eq:WKBJ_u}) to the 2nd-order ODE (\ref{eq:2ODE_u}), we find
\begin{equation}
\epsilon=\frac{1}{k_{z}},~~
S_{0}^{'2}=\Delta,~~
S'_{1}=-\frac{1}{2}\left(\frac{1}{r}-\frac{Q'}{Q}\right)-\frac{\Delta'}{4\Delta}.
\end{equation}
The leading-order term $S_{0}$, where $S'_{0}=\pm\sqrt{\Delta}$, determines the exponential behavior of the solution, and it depends on the sign of $\Delta$. 
For instance, if $\Delta>0$, we express $\hat{u}$ as an evanescent solution:
\begin{equation}
\label{eq:WKBJ_u_evanescent}
\hat{u}=\frac{Q^{\frac{1}{2}}}{r^{\frac{1}{2}}\Delta^{\frac{1}{4}}}\left[A_{1}\exp\left(k_{z}\int_{r}\sqrt{\Delta(t)}\mathrm{dt}\right)+A_{2}\exp\left(-k_{z}\int_{r}\sqrt{\Delta(t)}\mathrm{dt}\right)\right],
\end{equation}
or as a wavelike solution if $\Delta<0$:
\begin{equation}
\label{eq:WKBJ_u_wavelike}
\begin{aligned}
\hat{u}=\frac{Q^{\frac{1}{2}}}{r^{\frac{1}{2}}(-\Delta)^{\frac{1}{4}}}&\left[B_{1}\exp\left(\mathrm{i}k_{z}\int_{r}\sqrt{-\Delta(t)}\mathrm{dt}\right)\right.\\
&\left.+B_{2}\exp\left(-\mathrm{i}k_{z}\int_{r}\sqrt{-\Delta(t)}\mathrm{dt}\right)\right].
\end{aligned}
\end{equation}

%
   \begin{figure}
   \centering
   \includegraphics[width=6cm]{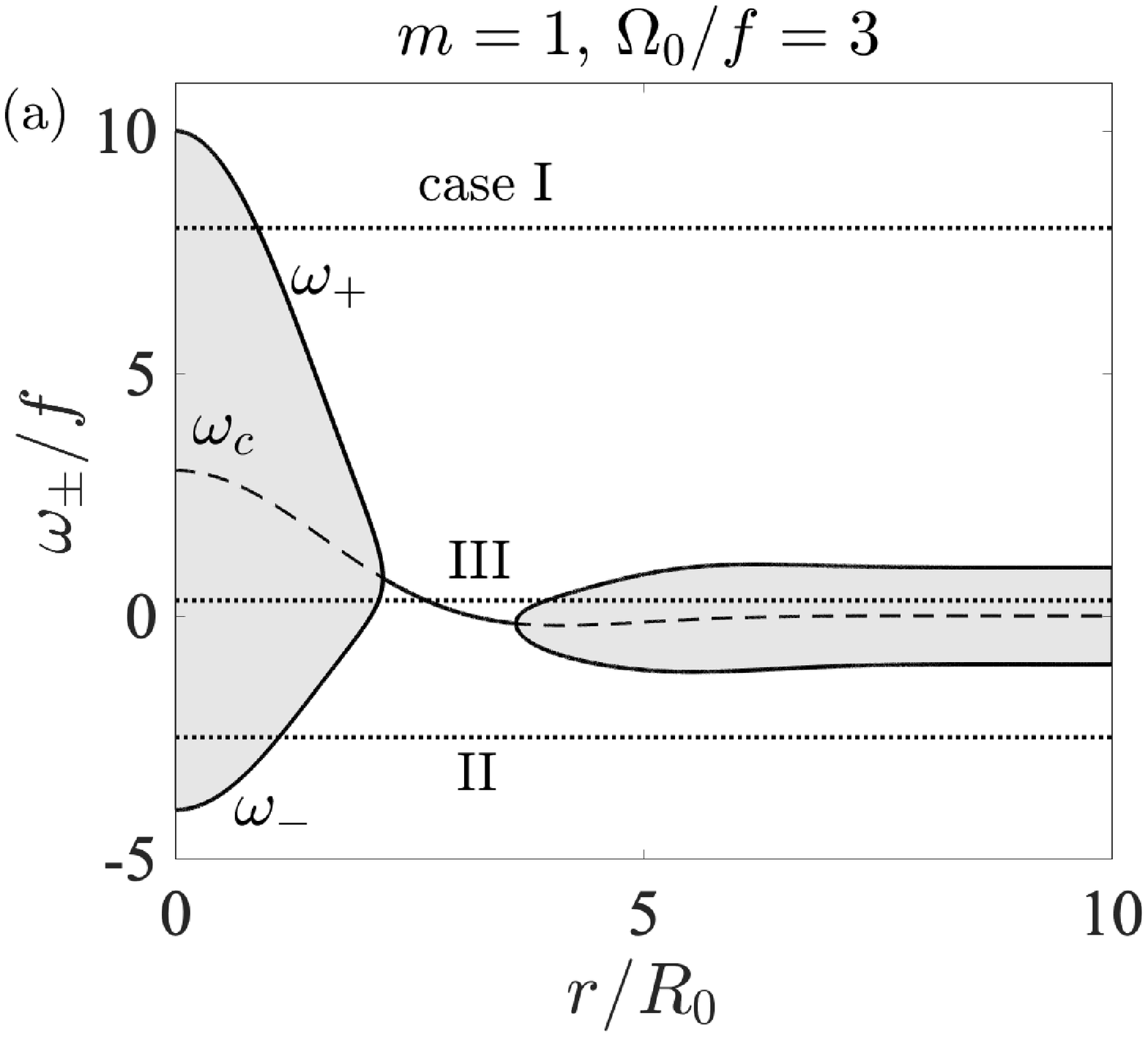}
   \includegraphics[width=6cm]{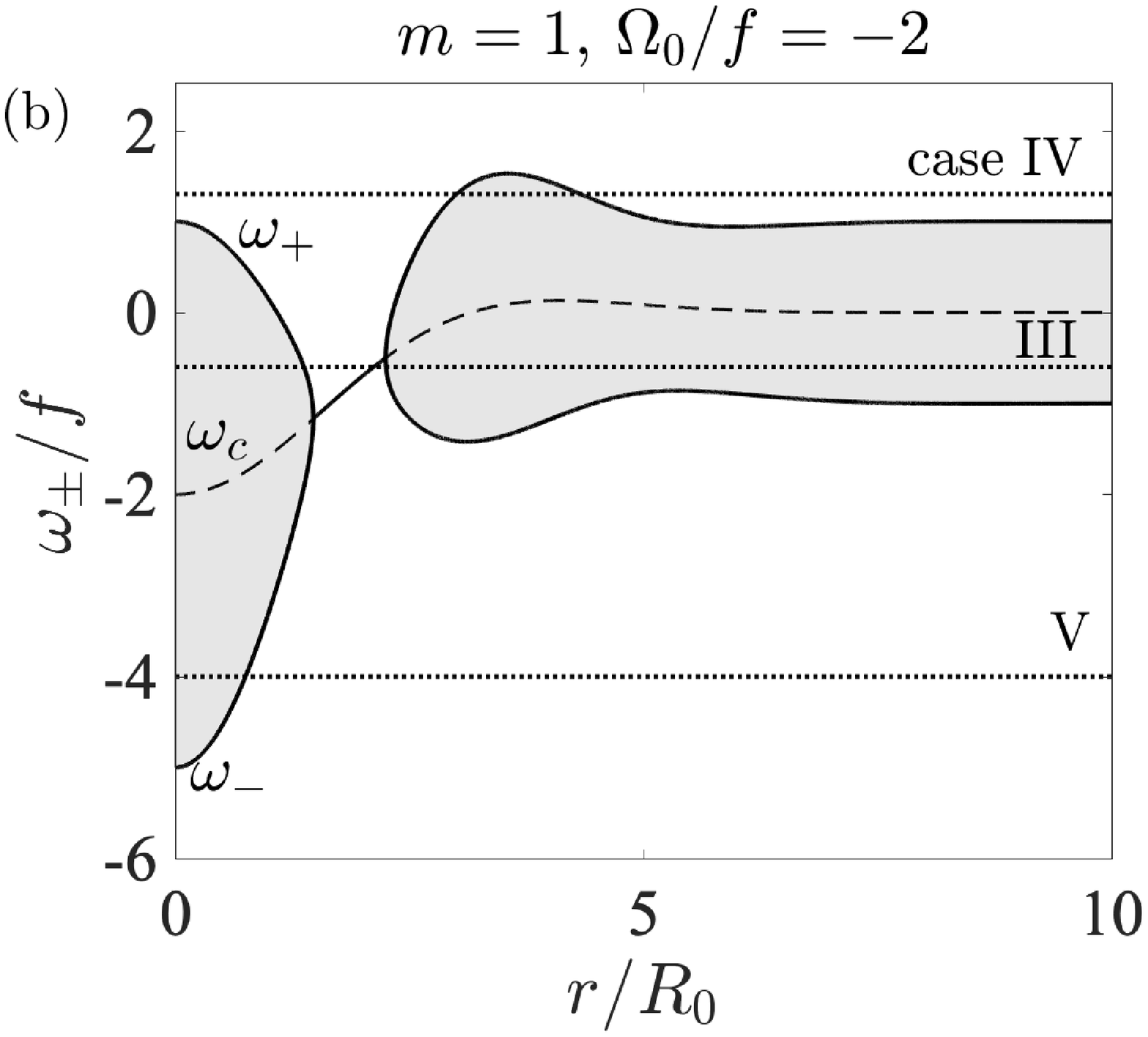}
     \caption{Epicyclic frequencies $\omega_{\pm}$ (solid lines) and critical frequency $\omega_{c}$ (dashed line) for $m=1$ and (a) $\Omega_{0}/f=3$ and (b) $\Omega_{0}/f=-2$.
     Gray area denotes the region where the solution is wavelike (i.e., $\Delta<0$) while white area denotes the evanescent region (i.e., $\Delta>0$). 
     Dotted lines represent frequencies $\omega_{r}$ for neutral modes in the upper (cases I and IV) and lower (cases II and V) branches, and for unstable modes (case III).
              }
         \label{Fig_epi_freq}
   \end{figure}
To comprehend more easily the sign of $\Delta(r)=1-\Phi/\left(-\omega+m\Omega\right)^{2}$, it is convenient to define the radial epicyclic frequencies $\omega_{\pm}(r)$:
\begin{equation}
\omega_{\pm}(r)=m\Omega\pm\mathrm{Re}\left(\sqrt{\Phi}\right),
\end{equation}
\citep[see also,][]{Ledizes2005,Ledizes2008,Park2013}.
We also define the critical frequency $\omega_{c}$ and radius $r_\mathrm{c}$ at which the Doppler-shifted frequency $s$ vanishes:
\begin{equation}
\omega_{c}=m\Omega(r_\mathrm{c}),
\end{equation}
\citep[i.e. the frequency and radius at which the corotation resonance occurs in the inviscid case, see e.g.][]{Astoul2021}.
In Fig.~\ref{Fig_epi_freq}a, we plot $\omega_{\pm}$ and $\omega_{c}$ for $\Omega_{0}/f=3$ and $m=1$. 
The epicyclic frequencies $\omega_{\pm}$ depend on $m$ and $\Omega_{0}$, and their behavior can be upside down for anticyclonic cases with $\Omega_{0}/f<0$ (see e.g., Fig.~\ref{Fig_epi_freq}b for $\Omega_{0}/f=-2$).
The gray-shaded area between $\omega_{+}$ and $\omega_{-}$ denotes the region where the solution is wavelike (i.e., $\Delta<0$), while the white area denotes the region where the solution is evanescent (i.e., $\Delta>0$).
This implies that, for a given frequency $\omega$, the WKBJ solution is wavelike if $\omega$ lies in the range $\omega_{-}<\omega<\omega_{+}$ while the solution is evanescent if $\omega>\omega_{+}$ or $\omega<\omega_{-}$.
The crossings of $\omega_{r}=\omega_{\pm}$ indicate where $\Delta(r)$ is zero \citep[see also,][]{Park2012}. 
This implies that they indicate the location of the turning point $r_{t}$ where $\Delta(r_{t})=0$ and where solution behavior changes from wavelike to evanescent, or vice versa. 
We also display in Fig.~\ref{Fig_epi_freq} typical frequencies of the neutral modes in the upper and lower branches (cases I, II, IV, and V) and unstable modes (case III). 
If the frequency $\omega$ lies in the range $\omega_{c}(0)<\omega<\omega_{+}(0)$ like in the case I, we have one turning point $r_{t}$ and the solution is wavelike in the range $0<r<r_{t}$ while it is evanescent outside the turning point $r>r_{t}$. 
At this frequency, we can construct an eigenfunction as the exponentially decaying solution for $r>r_{t}$ while waves are trapped between the turning point $r_{t}$ and $r=0$ \citep[see also,][]{Ledizes2005}.
For the cases II and V where the frequency lies in the range $\omega_{-}(0)<\omega<\omega_{c}(0)$, we can construct the eigenfunction in the same way.
For the case IV, the frequency $\omega_{r}$ crosses $\omega_{+}$ implying that there are two turning points. 
Between the two turning points, the solution is wavelike while it is evanescent elsewhere. 
This configuration corresponds to the ring mode. 
The case III of the unstable mode is different; first, it has a frequency in the range $0<\omega<\omega_{c}(0)$ implying that there is a critical point $r_{c}$ where $\omega_{c}(r_{c})=0$.
The unstable mode also has a positive growth rate $\omega_{i}>0$ thus $\Delta$ is no longer real but complex on the real $r$-axis.
Division of the WKBJ solutions into wavelike and evanescent solutions around the turning point $r_{t}$ is, therefore, not applicable on the real $r$-axis and we need to investigate the solution behavior of the unstable mode in the complex plane. 

In the following subsections, we will derive in detail the dispersion relations of the frequency for the upper and lower branches of the neutral modes (cases I, II, IV, and V) using the epicyclic frequencies.
And for the case III, we will use the analysis in the complex plane used by \citet{Billant2005} to derive the frequency and growth rate of the unstable modes. 

\subsection{Dispersion relations for neutral modes}
We first investigate the dispersion relations of the neutral modes that have a real frequency $\omega$.
In Fig.~\ref{Fig_epi_freq}, we display possible frequencies where we can construct the neutral mode. 
For instance, in the case I where the frequency lies in the range $\max(\omega_{c})<\omega_{r}<\max(\omega_{+})$, we have one turning point $r_{t}$ and the solution is evanescent outside $r_t$: 
\begin{eqnarray}
\label{eq:WKBJ_neutral_modes_1}
    \hat{u}=\frac{Q^{1/2}}{r^{1/2}\Delta^{1/4}}A_{1}\exp\left(-k_{z}\int_{r_{t}}^{r}\sqrt{\Delta(t)}~\mathrm{d}t\right).
\end{eqnarray}
In order to impose the decaying boundary condition as $r\rightarrow\infty$, $A_{2}=0$ is imposed in Eq.~(\ref{eq:WKBJ_u_evanescent}). 
Besides, by considering the connection around the turning point $r_{t}$ \citep[see also,][]{Olver1974,Billant2005,Ledizes2005}, we obtain the wavelike solution in the range $0<r<r_{t}$ as
\begin{eqnarray}
\label{eq:WKBJ_neutral_modes_2}
\hat{u}=\frac{A_1}{2}\frac{Q^{1/2}}{r^{1/2} (-\Delta)^{1/4}}\bigg[\exp\left(\mathrm{i}k_{z}\int_{r}^{r_{t}}\sqrt{-\Delta(t)}~\mathrm{d}t-\mathrm{i}\frac{\pi}{4}\right)\nonumber\\
+\exp\left(-\mathrm{i}k_{z}\int_{r}^{r_{t}}\sqrt{-\Delta(t)}~\mathrm{d}t+\mathrm{i}\frac{\pi}{4}\right)\bigg].
\end{eqnarray}
This mode is the core mode since the wave is confined between the core $r=0$ and the turning point $r_{t}$ \citep[][]{Ledizes2005}.
Depending on the azimuthal wavenumber $m$, we impose the boundary condition at the center $r=0$ as
\begin{equation}
\label{eq:boundary_condition_core}
\hat{u}=0~~\mathrm{if}~~m\neq1~~~~\mathrm{or}~~~~~~\frac{\mathrm{d}\hat{u}}{\mathrm{d}r}=0~~\mathrm{if}~~m=1,
\end{equation}
based on the asymptotic relations \eqref{eq:solution1_center} and \eqref{eq:solution2_center} \citep[see also,][]{Saffman1992}.
Imposing such conditions to \eqref{eq:WKBJ_neutral_modes_2} leads to the following quantization relations at leading order:
\begin{eqnarray}
    \label{eq:Quantization_neutral_modes_core}
\begin{aligned}
    &k_z \int_{0}^{r_t} {\sqrt{-\Delta(t)}dt} = \left(n-\frac{1}{4}\right)\pi,~~~\mathrm{if}~~m\neq1,\\
    &k_z \int_{0}^{r_t} {\sqrt{-\Delta(t)}dt} = \left(n+\frac{1}{4}\right)\pi,~~~\mathrm{if}~~m=1,
\end{aligned}
\end{eqnarray}
where $n$ is an integer indicating the mode number. 
We see that the right-hand-side terms are always finite thus the integrals on the left-hand side should converge to zero as $k_z\rightarrow\infty$. 
This implies that the turning point migrates to the center $r=0$ in this limit. 
It allows us to expand all the functions around $r=0$:
\begin{eqnarray}
\label{eq:Expansion_neutral_modes}
\begin{aligned}
&\Phi(r) = \Phi_0 + \frac{\Phi''_0}{2} r^2+O(r^3),\\
&\Omega(r) = \Omega_0 + \frac{\Omega''_0}{2} r^2+O(r^3),
\end{aligned}
\end{eqnarray}
where the subscript $0$ represents the fact that the functions are evaluated in $r=0$. 
Applying the expansions (\ref{eq:Expansion_neutral_modes}) to the quantization relations (\ref{eq:Quantization_neutral_modes_core}), we can also expand the frequency in the power of $k_{z}$ as 
\begin{equation}
\label{eq:expansion_frequency}
\omega = \omega_{0} + \frac{\omega_{1}}{k_{z}}+O\left(\frac{1}{k_{z}^{2}}\right). 
\end{equation}
After having inserted the expansion \eqref{eq:expansion_frequency} into the quantization conditions \eqref{eq:Quantization_neutral_modes_core}, we find the frequency for upper branches as
\begin{equation}
    \label{eq:dispersion_neutral_modes_core_upper_appendix}
\begin{aligned}
    &\omega=m\Omega_{0}+\sqrt{\Phi_{0}}-\frac{2}{k_{z}}\left(n-\frac{1}{4}\right) \sqrt{-\frac{\Phi_0''}{2} - m\Omega_0'' \sqrt{\Phi_0}},&\mathrm{if}~m\neq1,\\
    &\omega=m\Omega_{0}+\sqrt{\Phi_{0}}-\frac{2}{k_{z}}\left(n+\frac{1}{4}\right) \sqrt{-\frac{\Phi_0''}{2} - m\Omega_0'' \sqrt{\Phi_0}},&\mathrm{if}~m=1.\\
\end{aligned}
\end{equation}
A good agreement between the asymptotic dispersion relation \eqref{eq:dispersion_neutral_modes_core_upper_appendix} and the numerical solutions can be concluded from Fig. \ref{Fig_eigenvalues_k}a especially for large $k_{z}$.
Note that $\omega_{1}$ is negative for the upper branches thus the frequency $\omega$ increases with $k_{z}$ and asymptotes to $m\Omega_{0}+\sqrt{\Phi_{0}}$ as $k_{z}\rightarrow\infty$. 

The quantization relations \eqref{eq:Quantization_neutral_modes_core} can be similarly applied for lower branches for other core-mode cases II and V in Fig.~\ref{Fig_epi_freq} where the frequency $\omega$ now meets the epicyclic frequency $\omega_{-}$. 
In this case, we found the frequency expansion as follows
\begin{equation}
    \label{eq:dispersion_neutral_modes_core_lower_appendix}
\begin{aligned}
    &\omega=m\Omega_{0}-\sqrt{\Phi_{0}}+\frac{2}{k_{z}}\left(n-\frac{1}{4}\right) \sqrt{-\frac{\Phi_0''}{2} + m\Omega_0'' \sqrt{\Phi_0}},&\mathrm{if}~m\neq1,\\
    &\omega=m\Omega_{0}-\sqrt{\Phi_{0}}+\frac{2}{k_{z}}\left(n+\frac{1}{4}\right) \sqrt{-\frac{\Phi_0''}{2} + m\Omega_0'' \sqrt{\Phi_0}},&\mathrm{if}~m=1.\\
\end{aligned}
\end{equation}
For the lower branches, $\omega_{1}$ is now positive, and the frequency $\omega$ decreases and asymptotes to $m\Omega_{0}-\sqrt{\Phi_{0}}$ as $k_{z}$ increases.
We see in Fig.~\ref{Fig_eigenvalues_k}b that the asymptotic dispersion relations \eqref{eq:dispersion_neutral_modes_core_lower} agree well with numerical results for the lower-branch frequency. 

On the other hand, we also have a mode called the ring mode \citep[][]{Ledizes2005} where the wavelike solution is confined between the two turning points $r_{t1}$ and $r_{t2}$ where $r_{t1}<r_{t2}$ (see e.g., the case IV in Fig.~\ref{Fig_epi_freq}b).
In this case, the wavelike solution (\ref{eq:WKBJ_neutral_modes_2}) is connected around $r_{t1}$ by the evanescent solution in the range $0<r<r_{t1}$:
\begin{eqnarray}
\label{eq:WKBJ_neutral_modes_ring_mode_rt1}
    \hat{u}=\frac{Q^{1/2}}{r^{1/2}\Delta^{1/4}}A_{2}\exp\left(-k_{z}\int_{r}^{r_{t1}}\sqrt{\Delta(t)}~\mathrm{d}t\right).
\end{eqnarray}
Matching the solutions \eqref{eq:WKBJ_neutral_modes_2} and \eqref{eq:WKBJ_neutral_modes_ring_mode_rt1} leads to the following quantization condition:
\begin{eqnarray}
    \label{eq:Quantization_neutral_modes_ring}
   k_z \int_{r_{t1}}^{r_{t2}} {\sqrt{-\Delta(t)}dt} = \left(n-\frac{1}{2}\right)\pi,
\end{eqnarray}
\citep[see also,][]{Billant2005}.
As $k_{z}$ increases, these turning points migrate towards the point $r_{\max+}$ at which $\omega_{+}$ is the maximum and the derivative of $\omega_{+}$ is zero:
\begin{equation}
\label{eq:r_max_plus}
    -2 m\Omega'(r_{\max+})\sqrt{\Phi(r_{\max+})} = \Phi'(r_{\max+}).
\end{equation}
Such a behavior around $r_{\max+}$ leads to the following expression for the frequency:
\begin{equation}
\label{eq:frequency_neutral_modes_ring_upper}
\begin{aligned}
&\omega=\omega_{0}-\frac{\omega_{1}}{k_{z}},~\\
&\omega_{0}=m\Omega(r_{\max+})+\sqrt{\Phi(r_{\max+})},\\
&\omega_{1}=\left.\left(n-\frac{1}{2}\right)\sqrt{-\frac{\Phi''}{2}-3m^{2}\Omega^{'2}+\frac{2m\Omega'\Phi'+m\Omega''\Phi}{-\omega_{0}+m\Omega}}\right|_{r=r_{\max+}}.
\end{aligned}
\end{equation}
The ring mode in the upper branches can be constructed when $\omega_{+}(0)<\omega_{+}(r_{\max+})$, i.e.,
\begin{equation}
\label{eq:ring_condition_plus}
m<\frac{\sqrt{\Phi_{r_{\max+}}}-\sqrt{\Phi_{0}}}{\Omega_{0}-\Omega_{r_{\max+}}}.
\end{equation}
We found numerically that this inequality corresponds to $-1.2<\Omega_{0}/f<0$ for $m=0$, $-2.8<\Omega_{0}/f<0$ for $m=1$, and $\Omega_{0}/f<0$ for $m\geq2$.

For lower branches, we can have the same quantization condition as in Eq.~(\ref{eq:Quantization_neutral_modes_ring}) when $\omega_{-}(0)>\omega_{-}(r_{\min-})$ where
\begin{equation}
    \label{eq:r_min_minus}
    2 m\Omega'(r_{\min-})\sqrt{\Phi(r_{\min-})} = \Phi'(r_{\min-}).
\end{equation}
The inequality $\omega_{-}(0)>\omega_{-}(r_{\min-})$ is equivalent to 
\begin{equation}
\label{eq:ring_condition_minus}
m<\frac{\sqrt{\Phi_{0}}-\sqrt{\Phi_{r_{\min-}}}}{\Omega_{0}-\Omega_{r_{\min-}}},
\end{equation}
and it corresponds to $-1.2<\Omega_{0}/f<0$ for $m=0$, $-0.8<\Omega_{0}/f<0$ for $m=1$, $\Omega_{0}/f>-0.6$ for $m=2$, and $\Omega_{0}/f>0$ for $m\geq3$.
We also have the frequency expression evaluated at $r_{\min-}$ for the ring mode in the lower branches as follows:
\begin{equation}
\label{eq:frequency_neutral_modes_ring_lower}
\begin{aligned}
&\omega=\omega_{0}+\frac{\omega_{1}}{k_{z}},~\\
&\omega_{0}=m\Omega(r_{\min-})-\sqrt{\Phi(r_{\min-})},\\
&\omega_{1}=\left.\left(n-\frac{1}{2}\right)\sqrt{-\frac{\Phi''}{2}-3m^{2}\Omega^{'2}+\frac{2m\Omega'\Phi'+m\Omega''\Phi}{-\omega_{0}+m\Omega}}\right|_{r=r_{\min-}}.
\end{aligned}
\end{equation}
We also checked that the asymptotic frequencies \eqref{eq:frequency_neutral_modes_ring_upper} and \eqref{eq:frequency_neutral_modes_ring_lower} are in good agreement with numerical results.

\subsection{Dispersion relations for unstable modes}
\label{sec:disp_unst}
   \begin{figure}
   \centering
   \includegraphics[width=7cm]{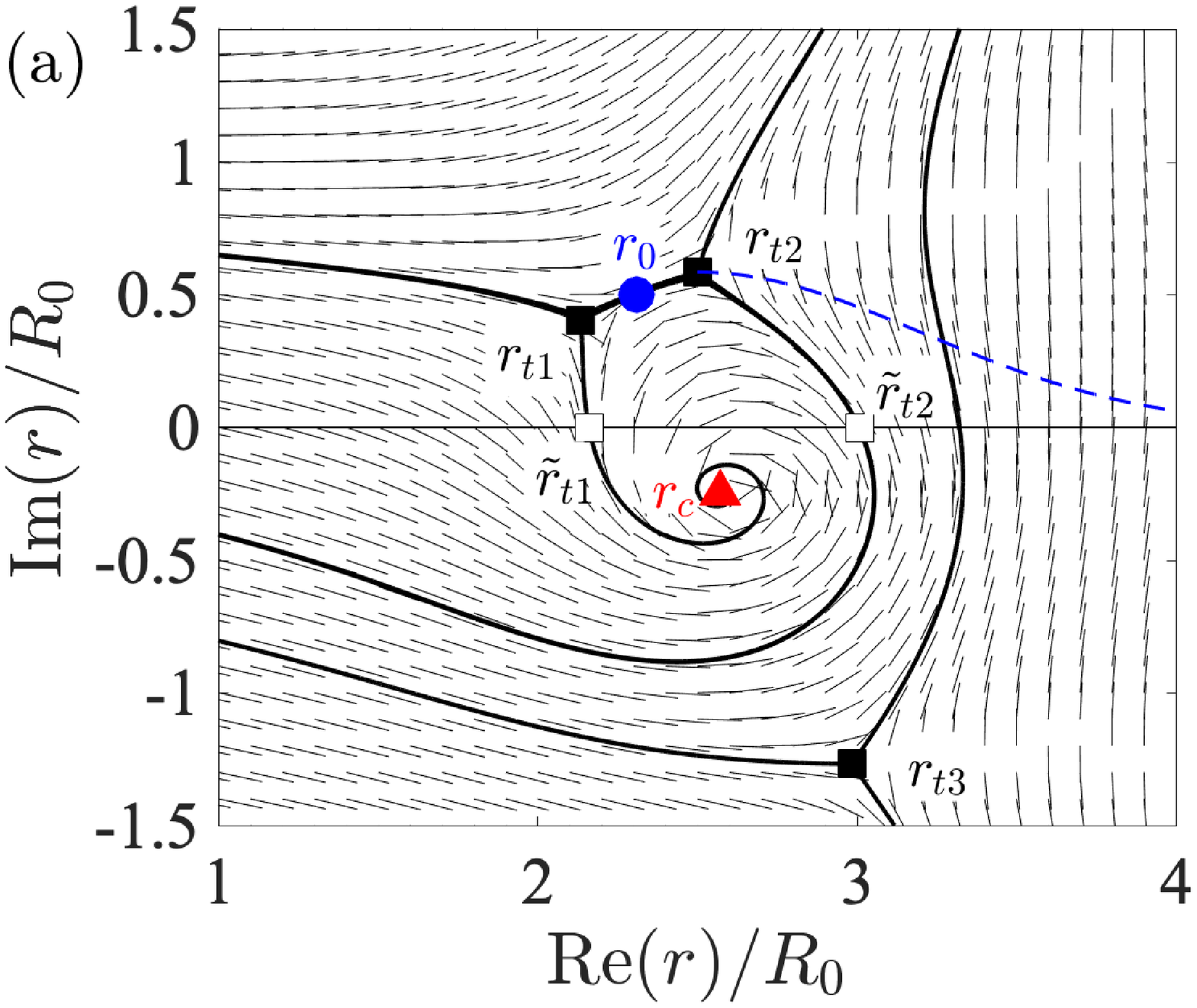}
   \includegraphics[width=7cm]{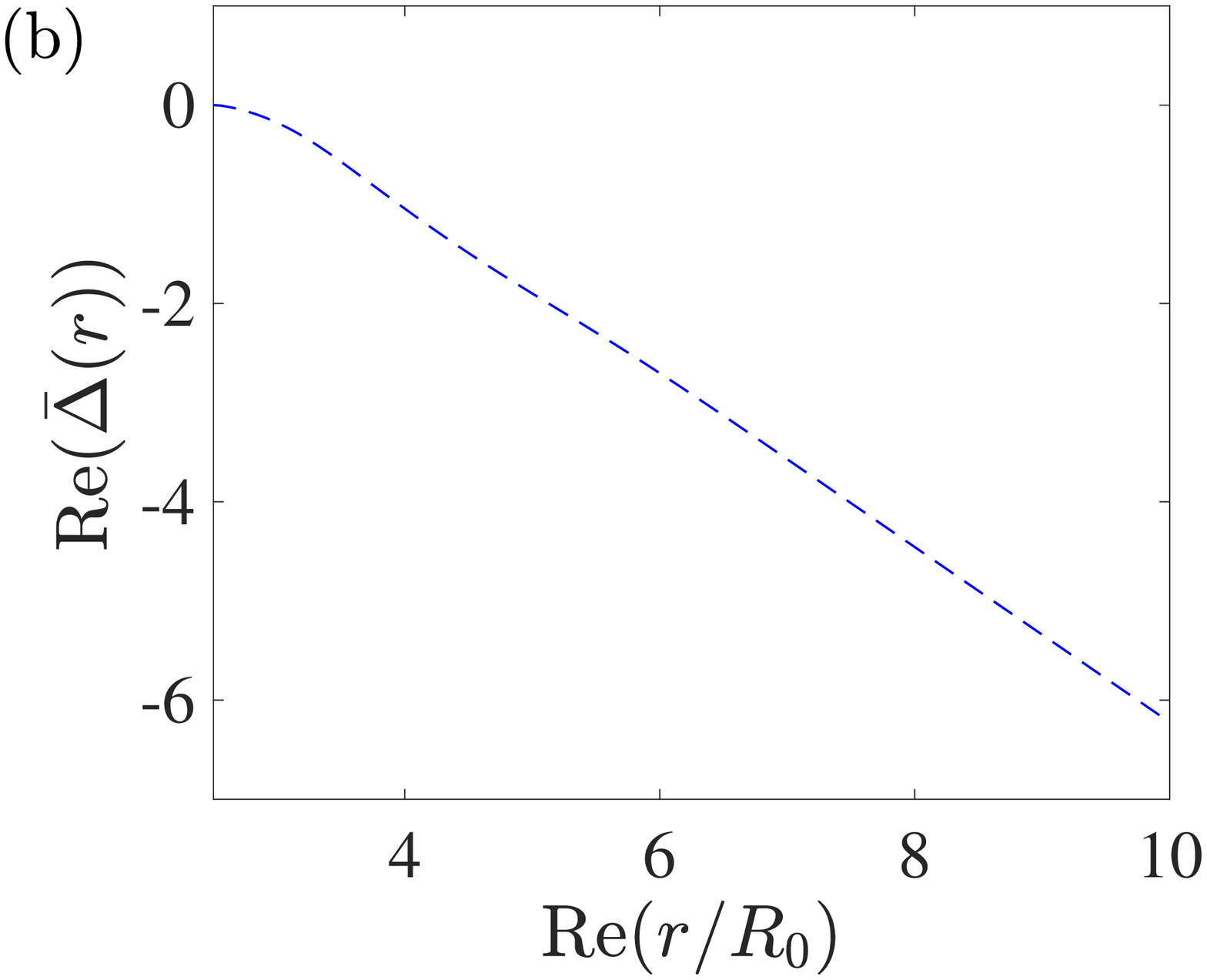}
     \caption{(a) An example of the Stokes lines network for $\omega=0.778+0.46\mathrm{i}$ at $m=2$, $k_{z}R_{0}=30$, and $\Omega_{0}/f=3$.
     Black solid lines denote the Stokes lines, short lines indicate the direction where $\mathrm{Re}(\Delta)$ remains constant, and a blue-dashed line denotes the progressive path. (b) The integral $\bar{\Delta}(r)$ versus the real part of $r$ along the progressive path in (a).
              }
         \label{Fig_stokes_lines}
   \end{figure}
For a general azimuthal wavenumber $m$ and complex frequency $\omega$, the function $\Delta(r)$ is complex on the real $r$-axis and the turning points $r_{t}$ where $\Delta(r_{t})=0$ are in the complex plane. 
Figure \ref{Fig_stokes_lines} shows locations of the turning points in the complex plane for $m=2$ and $\Omega_{0}/f=3$ when the eigenfrequency $\omega=0.778+0.45\mathrm{i}$ at $k_{z}R_{0}=30$ from the numerical computation is considered.
Due to the oscillatory behaviors of the Bessel function $J_{0}$ of the angular velocity profile (\ref{eq:base_Dandoy}) as $|r|\rightarrow\infty$ for $|\arg(r)|<\pi/2$ \citep[see also,][]{Abramowitz}, there are many turning points in the complex plane. 
But while other turning points are far from the real $r$-axis, the two turning points $r_{t1}$ and $r_{t2}$ are somewhat close to the real $r$-axis and they will influence the WKBJ solution effectively. 
To better understand the exponential behavior of the WKBJ solution, we also draw in Fig.~\ref{Fig_stokes_lines} the Stokes lines defined as 
\begin{equation}
\label{eq:def_Stokes}
\mathrm{Re}\left(\Delta(r)\right)=0,
\end{equation}
 \citep[see also,][]{Olver1974,Billant2005}.
The Stokes lines emanate from the turning points and they delimit the regions of the WKBJ solutions in the same behavior. 
The points $\tilde{r}_{t1}$ and $\tilde{r}_{t2}$ where the Stokes lines cross the real $r$-axis also delimit the solutions on the real $r$-axis.
By the definition (\ref{eq:def_Stokes}), the WKBJ solutions on the Stokes line between the two turning points are wavelike. 
Furthermore, it is noticeable that a Stokes line emanated from $r_{t1}$ shows a spiral pattern around the critical point $r_{c}$ since $\Delta$ is singular. 

The Stokes line network in the complex plane in Fig.~\ref{Fig_stokes_lines} is very similar to that of the Carton \& McWilliams vortex studied in \citet{Billant2005}. 
We follow their approach by considering the WKBJ solution that decreases exponentially from $r=r_{t1}$ in the range $0<r<r_{t1}$:
\begin{equation}
\label{eq:WKBJ_Stokes_eq1}
\hat{u}=C_{1}\frac{Q^{1/2}}{r^{1/2}\Delta^{1/4}}\exp\left[-k_{z}\int_{r}^{r_{t1}}\sqrt{\Delta(t)}~\mathrm{d}t\right].
\end{equation}
This solutions matches with the wavelike solution in the range $r_{t1}<r<r_{t2}$:
\begin{equation}
\label{eq:WKBJ_Stokes_eq2}
\hat{u}=2C_{1}\frac{Q^{1/2}}{r^{1/2}(-\Delta)^{1/4}}\sin\left[k_{z}\int_{r_{t1}}^{r}\sqrt{-\Delta(t)}~\mathrm{d}t+\frac{\pi}{4}\right].
\end{equation}
Furthermore, we have the evanescent solution after the turning point $r_{t2}$
\begin{eqnarray}
\label{eq:WKBJ_Stokes_eq3}
\hat{u}=\frac{Q^{1/2}}{r^{1/2}\Delta^{1/4}}\left[D_{1}\exp\left(-k_{z}\int_{r_{t2}}^{r}\sqrt{\Delta(t)}~\mathrm{d}t\right)\right.\nonumber\\
\left.+D_{2}\exp\left(k_{z}\int_{r_{t2}}^{r}\sqrt{\Delta(t)}~\mathrm{d}t\right)\right].
\end{eqnarray}
We note that for $r>r_{t2}$, there are many turning points in the complex plane as a result of the oscillatory behavior of the vortex profile in Eq.~ (\ref{eq:base_Dandoy}).
These turning points, like $r_{t3}$ in Fig.~\ref{Fig_stokes_lines}, are far from the real axis. With this characteristic, we can avoid the multiple-turning-point analysis by choosing a proper progressive path \citep[][]{Ledizes2005}.
More precisely, the idea is to check whether the integral $\bar{\Delta}(r)=\int_{r_{t2}}^{r}\sqrt{\Delta}\mathrm{d}t$ is holomorphic (i.e. complex differentiable) and $\mathrm{Re}(\bar{\Delta})$ is monotonic along the progressive path.
As we can see in Fig.~\ref{Fig_stokes_lines}b, the integral $\bar{\Delta}$ is monotonically decreasing along the progressive path displayed in Fig.~\ref{Fig_stokes_lines}a. 
This implies that we can keep the WKBJ solution (\ref{eq:WKBJ_Stokes_eq3}) along the progressive path as $r\rightarrow\infty$, thus we can impose $D_{2}=0$ to apply the exponentially decreasing boundary condition.

By connecting at $r=r_{t2}$ the solution (\ref{eq:WKBJ_Stokes_eq2}) and the solution (\ref{eq:WKBJ_Stokes_eq3}) with $D_{2}=0$, we obtain the following quantization condition:
\begin{equation}
\label{eq:WKBJ_quantization}
k_{z}\int_{r_{t1}}^{r_{t2}}\sqrt{-\Delta}~\mathrm{d}r=\left(n+\frac{1}{2}\right)\pi,
\end{equation}
where $n$ is the mode number. 
It is identical to the quantization condition obtained by \citet{Billant2005}.
We see from Eq.~(\ref{eq:WKBJ_quantization}) that $\mathrm{Im}\left(\int_{r_{t1}}^{r_{t2}}\sqrt{-\Delta}\mathrm{d}r\right)=0$ (i.e., $\mathrm{Re}\left(\int_{r_{t1}}^{r_{t2}}\sqrt{\Delta}\mathrm{d}r\right)=0$), and this implies that the two turning points should be connected by the Stokes lines as seen in Fig.~\ref{Fig_stokes_lines}.
Also, it is important to note that the right-hand side of Eq.~(\ref{eq:WKBJ_quantization}) is always finite while the integral on the left-hand side should become zero as $k_{z}\rightarrow\infty$.
This implies that $r_{t1}$ and $r_{t2}$ should collapse to a point as $k_{z}\rightarrow\infty$.
From this information, \citet{Billant2005} considered a double turning point $r_{0}$ between the two turning points (see also Fig.~\ref{Fig_stokes_lines}), where the radial derivative of $\omega_{+}=m\Omega(r)+\mathrm{i}\sqrt{-\Phi(r)}$ becomes zero, i.e.,
\begin{equation}
\label{eq:double_turning_point}
2m\Omega'(r_{0})\sqrt{-\Phi(r_{0})}-\mathrm{i}\Phi'(r_{0})=0.
\end{equation}

   \begin{figure*}
   \centering
   \includegraphics[height=4.8cm]{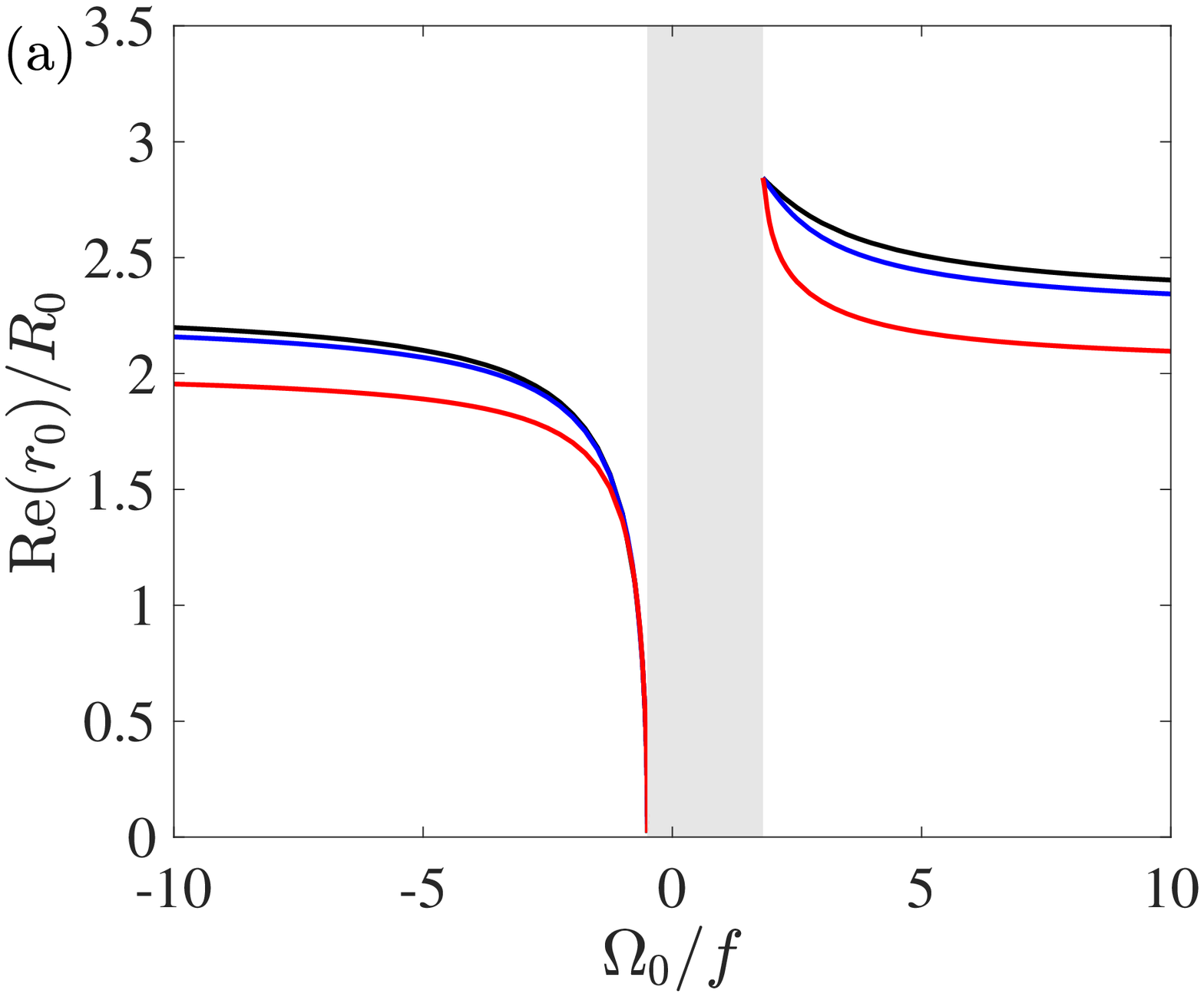}
   \includegraphics[height=4.8cm]{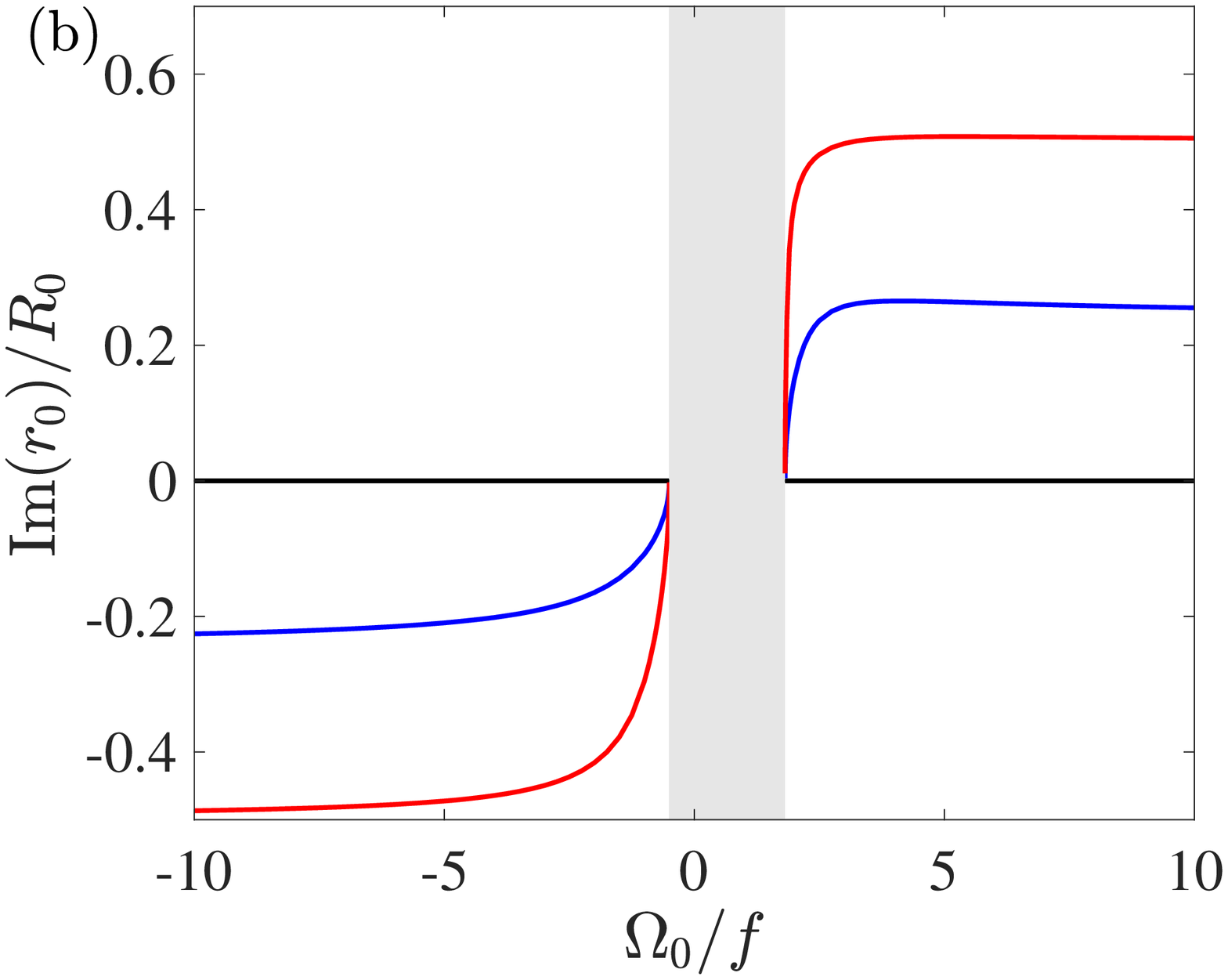}
   \includegraphics[height=4.8cm]{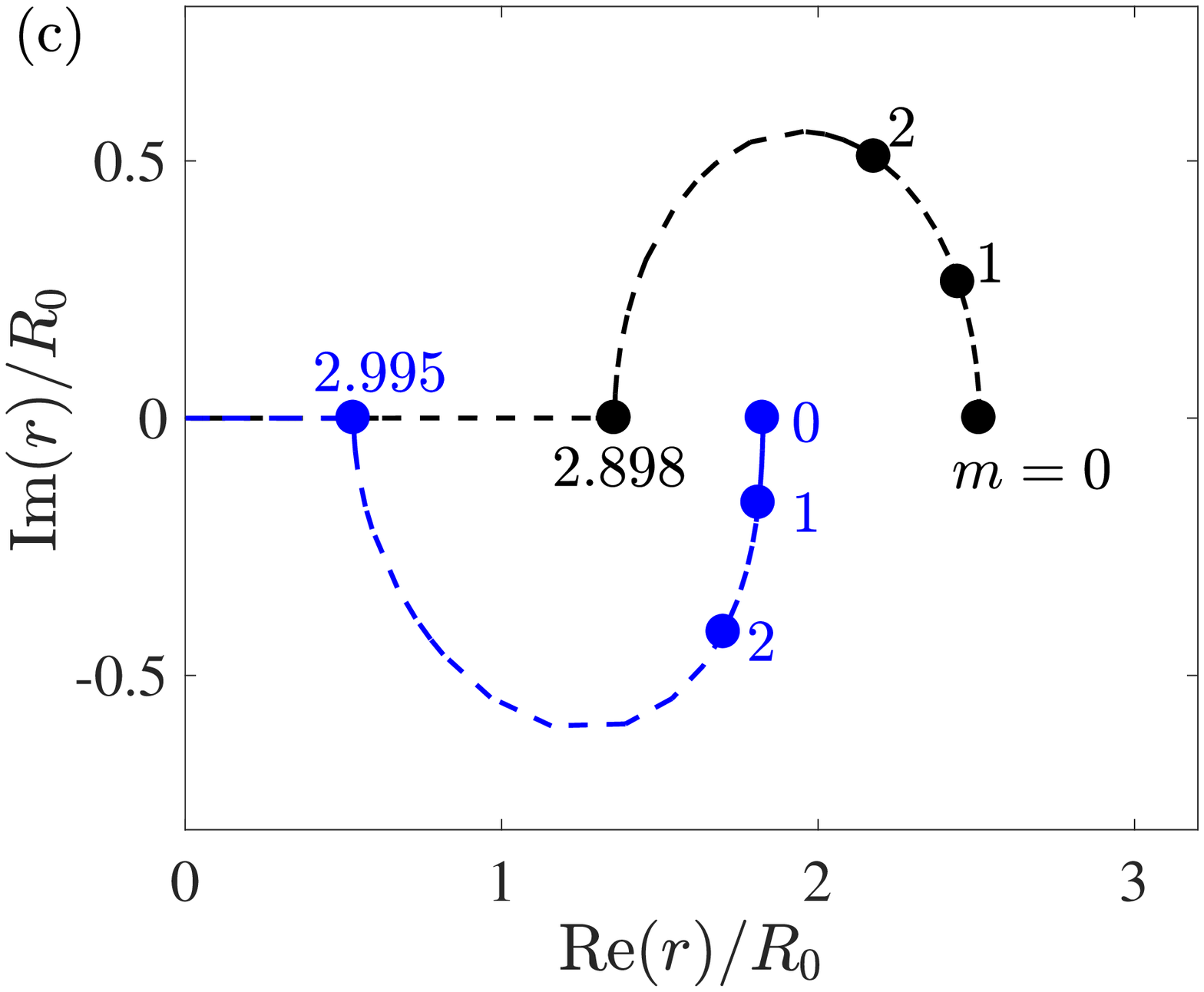}
     \caption{(a) Real and (b) imaginary parts of the double turning point $r_{0}$ versus $\Omega_{0}/f$ for different azimuthal wavenumbers: $m=0$ (black), $m=1$ (blue) and $m=2$ (red). 
     Gray shaded areas represent the centrifugally-stable regime. 
     (c) Movement of the double turning point $r_{0}$ in the complex plane as $m$ changes for $\Omega_{0}/f=5$ (black) and $\Omega_{0}/f=-2$ (blue).
              }
         \label{Fig_dtp}
   \end{figure*}
In Fig.~\ref{Fig_dtp}, we show how the location of the double turning point $r_{0}$ changes in the complex plane for both cyclonic and anticyclonic cases for various azimuthal wavenumbers. 
For cyclonic cases $\Omega_{0}/f>0$, $r_{0}$ departs from the real axis at $m=0$ and lies in the upper complex plane ($\mathrm{Im}\left(r_{0}\right)>0$) for $m>0$, while it lies in the lower complex plane ($\mathrm{Im}\left(r_{0}\right)<0$) for anticyclonic cases $\Omega_{0}/f<0$.
For both cases, we found that the double turning point $r_{0}$ disappears for $m\leq3$. 

Once the location of $r_{0}$ is known, we can apply the Taylor expansion around $r_{0}$ in the limit $k_{z}\rightarrow\infty$ to derive the following dispersion relation as
\begin{equation}
\label{eq:dispersion_relation_complex_w_appendix}
\omega=\omega_{0}+\frac{\omega_{1}}{k_{z}}+O\left(\frac{1}{k_{z}^{2}}\right),
\end{equation}
where
\begin{equation}
\label{eq:dispersion_relation_w0_appendix}
\omega_{0}=m\Omega(r_{0})+\mathrm{i}\sqrt{-\Phi(r_{0})},
\end{equation}
\begin{equation}
\label{eq:dispersion_relation_w1_appendix}
\omega_{1}=\left.\frac{2n+1}{2\sqrt{2}\mathrm{i}}\sqrt{\Phi''-2m^{2}\Omega^{'2}+2\mathrm{i}m\Omega''\sqrt{-\Phi}}~\right|_{r=r_{0}},
\end{equation}
\citep[see also,][]{Billant2005}.
In Fig.~\ref{Fig_eigenvalues_k}(c,d), we see that the WKBJ prediction (\ref{eq:dispersion_relation_complex_w}) is in good agreement with numerical results for the most unstable eigenvalues (i.e., $n=1$) for various azimuthal wavenumbers $m$.
\section{Note on the critical layers}
\label{subsec:critical}
There is a certain range of frequency where neutral, stable, or even unstable modes can possess a critical layer at $r=r_{c}$ where $s=0$.
In the inviscid limit $\nu=0$, the critical radius $r_{c}$ is a pole of the function $\Delta$ and thus the second-order ordinary differential equation Eq.~(\ref{eq:2ODE_u}) becomes singular. 
This can also be seen from the WKBJ solutions such that the divergence of $\Delta$ leads to the divergence of the leading-order term of the WKBJ solution (\ref{eq:WKBJ_u_wavelike}). 
Therefore, a dedicated analysis of Eq.~({\ref{eq:2ODE_u}}) in the vicinity of $r_{c}$ is required to be performed. 
As $r$ goes to $r_\mathrm{c}$, Eq.~({\ref{eq:2ODE_u}}) can be approximated at leading order
\begin{equation}
    \frac{\mathrm{d}^2\hat{u}}{\mathrm{d}r^2}+\frac{\alpha_k^2(1+\mathrm{Ro_c})}{\mathrm{Ro_c}^2(r-r_\mathrm{c})^2}\hat{u}=0,
    \label{eq:corot}
\end{equation}
where $\mathrm{Ro_{c}}$ is the shear Rossby number  and $\alpha_{k}$ is the ratio of vertical to azimuthal wavenumbers defined as:
\begin{equation}
\mathrm{Ro_c}=\frac{r_c\Omega'(r_\mathrm{c})}{f+2\Omega(r_\mathrm{c})}~~\text{ and }~~\alpha_k=\frac{k_zr_\mathrm{c}}{m}.
\end{equation}
Note that Eq. ({\ref{eq:corot}}) is analogous to the differential equation for inertial waves close to a corotation resonance when using a cylindrical differential rotation profile in a local shear box model in Cartesian coordinates \citep[see in particular Sect. 3.4 of][when the box is at the south pole of a spherical shell]{Astoul2021}. As introduced in \cite{Astoul2021} or \cite{Alvan2013} for internal gravity waves approaching a corotation resonance,
we define a new parameter $\Od$ appearing in Eq. (\ref{eq:corot}) as
\begin{equation}
    \Od=\frac{\alpha_k^2(1+\mathrm{Ro_\mathrm{c}})}{\mathrm{Ro_\mathrm{c}^2}}.
\end{equation}
Applying the Frobenius series to solve the equation (\ref{eq:corot}) leads to the following solutions at leading order:
\begin{equation}
\label{eq:corot_Rless}
    \hat{u}\simeq a_{1}(r-r_\mathrm{c})^{\frac{1}{2}+\sqrt{\frac{1}{4}-\Od}}+
    a_{2}(r-r_\mathrm{c})^{\frac{1}{2}-\sqrt{\frac{1}{4}-\Od}},
\end{equation}
if $\Od\leq 1/4$ where $a_{1}$ and $a_{2}$ are constants, or
\begin{equation}
\label{eq:corot_Rmore}
    \hat{u}\simeq b_{1}(r-r_\mathrm{c})^{\frac{1}{2}+\mathrm{i}\sqrt{\Od-\frac{1}{4}}}+
    b_{2}(r-r_\mathrm{c})^{\frac{1}{2}-\mathrm{i}\sqrt{\Od-\frac{1}{4}}},
\end{equation}
if $\Od>1/4$ where $b_{1}$ and $b_{2}$ are constants.

The solutions (\ref{eq:corot_Rmore}) in the regime $\Od>1/4$ can experience a strong attenuation as it crosses $r=r_{c}$.
For instance, if we consider the analytic continuity of Eq. (\ref{eq:corot_Rmore}) for $r>r_{\mathrm{c}}$ by taking the lower half in the complex plane\footnote{Since $\mathrm{Im}\{r-r_\mathrm{c}\}\simeq-\mathrm{Im}\{\omega\}/(m\Omega'_\mathrm{c})$ provided that $m\Omega'_\mathrm{c}>0$, and with the radiation condition $\mathrm{Im}\{\omega\}>0$ where $\mathrm{Im}$ features the imaginary part \citep[for a more detailed description of the mathematical analysis about the connection of solutions]{Miles1961,Booker1967,Astoul2021}.}, we obtain the solution for $r<r_{c}$ as
\begin{equation}
\label{eq:corot_Rmore_below}
    \hat{u}\simeq \aava{-\mathrm{i}}c_{1}|r-r_\mathrm{c}|^{\frac{1}{2}+\mathrm{i}\sqrt{\Od-\frac{1}{4}}}\aava{-\mathrm{i}}
    c_{2}|r-r_\mathrm{c}|^{\frac{1}{2}-\mathrm{i}\sqrt{\Od-\frac{1}{4}}},
\end{equation}
where $c_{1}=b_{1}\exp\left(\pi\sqrt{\Od-1/4}\right)$ and $c_{2}=b_{2}\exp\left(\aava{-}\pi\sqrt{\Od-1/4}\right)$ are the amplitudes of the solution for $r<r_\mathrm{c}$ related to the amplitudes of the solution for $r>r_\mathrm{c}$.
To determine in which directions the waves propagate from the corotation, we introduce the wave action flux $\mathcal{A}$, which is an invariant quantity along the radius $r$ in the inviscid limit. 
This quantity is conserved in the whole domain except at corotation, and is thus a useful tool to determine the direction of the wave energy propagation and quantify the angular momentum flux exchanges between the waves and the mean flow \citep[e.g.][in the context of (inertial-)gravity waves]{Grimshaw1975, Grimshaw1979,Andrews1978}. In the cylindrical coordinates, this invariant can be derived as
\begin{equation}
    \mathcal{A}=\left\langle\frac{rpu}{s}\right\rangle_{\theta,z}=\frac{1}{2}\mathrm{Re}\left\{\frac{r\hat{p}\hat{u}^*}{s}\right\},
    \label{eq:wavAct}
\end{equation}
where $\langle\cdot\rangle_{\theta,z}$ denotes the average in the $z$- and $\theta$-directions and $\mathrm{Re}$ the real part. 
By taking a linear combination of the momentum equations (\ref{eq:continuity_mode}), (\ref{eq:momentum_azimuthal_mode}), and (\ref{eq:momentum_vertical_mode}) in the inviscid limit, one can get an expression for the pressure:
\begin{equation}
    \hat{p}=-\mathrm{i}\frac{s\hat{u}'+\hat{u}/r\left[s-m\left(f+\zeta\right)\right]}{Q}.
\end{equation}
Its substitution into the wave action flux Eq. (\ref{eq:wavAct}) yields:
\begin{equation}
\label{eq:wave_flux}
    \mathcal{A}=\frac{r}{2Q}\mathrm{Im}\{\hat{u}'\hat{u}^*\}.
\end{equation}
Injecting the solutions Eqs. (\ref{eq:corot_Rmore}) and (\ref{eq:corot_Rmore_below}) into Eq.  (\ref{eq:wave_flux}) leads to the following wave action flux above and below the corotation:
\begin{eqnarray}
    \label{eq:wave_flux_corot}
\begin{aligned}
    &\mathcal{A} = \frac{r_{c}}{2Q_{c}}\mu\left(|b_{1}|^{2}-|b_{2}|^{2}\right),&\mathrm{if}~~r>r_{c},\\
    &\mathcal{A} = \frac{r_{c}}{2Q_{c}}\mu\left(|b_{2}|^{2}\exp(-2\pi\mu)-|b_{1}|^{2}\exp(2\pi\mu)\right),&\mathrm{if}~~r<r_{c},
\end{aligned}
\end{eqnarray}
where $\mu=\sqrt{\mathcal{R}-1/4}$, provided again that $m\Omega'_c>0$. 
In this framework, the direction of wave propagation in the region $r>r_{\mathrm{c}}$ is given by the sign of $s\mathcal{A}$ since this quantity is related to the group velocity  \citep[e.g.][]{Grimshaw1975,Bretherton1968}. 
From Eq. (\ref{eq:wave_flux_corot}) and given that $s\gtrless 0$ for $r\gtrless r_\mathrm{c}$, the first wave of amplitude $|b_1|$ is travelling outwards since $s\mathcal{A}>0$ (for both $r>r_{\mathrm{c}}$ and $r<r_{\mathrm{c}}$), while the second wave of amplitude $|b_2|$ is travelling inwards since $s\mathcal{A}<0$. 
This implies that both waves are attenuated as their wave action fluxes are reduced by a factor of $\mathrm{exp}(-2\pi\mu)$ after they get through the corotation, as similarly reported in \cite{Astoul2021} for inertial waves in a shear box model with cylindrical differential rotation when $\Od>1/4$. 
In the regime $\Od\leq 1/4$, the wave action flux analysis cannot determine the wave direction unambiguously and thus a further numerical analysis is required as it has been done in the aforementioned paper.
Depending on the boundary conditions and other internal properties such as wave frequency, wavenumbers or the shear Rossby number, the inertial waves may experience either damping or over-reflection/over-transmission.
The latter can ultimately lead to shear instability if the waves interfere constructively and sustain their growth as successive over-reflection (over-transmission) of the waves occurs \citep[][]{Lindzen1988,Harnik2007,Astoul2022}.
\end{appendix}

\end{document}